\documentclass[12pt,a4paper]{article}
\usepackage{geometry}
\usepackage[utf8]{inputenc}
\usepackage{setspace}

\usepackage{multicol}
\usepackage{diagbox}
\usepackage{caption}
\usepackage{subcaption}
\usepackage{jcappub}
\usepackage{bm}
\usepackage{epsfig}
\usepackage{bbold}
\usepackage{graphicx}
\usepackage{amsmath}
\usepackage{amssymb}
\usepackage{amsbsy}
\usepackage{color}
\usepackage{slashed}
\usepackage{afterpage}
\usepackage{psfrag}
\usepackage[noabbrev]{cleveref}
\usepackage{dsfont}
\usepackage{enumerate}
\usepackage[shortlabels]{enumitem}
\usepackage[utf8]{inputenc}
\usepackage{amsmath}
\usepackage[dvipsnames]{xcolor}
\usepackage{graphicx}
\usepackage{braket}
\usepackage[most]{tcolorbox}
\usepackage{empheq}
\usepackage{soul}
\usepackage{float}

\title{\begin{center}
From Cusps to Swallowtails: Domain Wall singularities in 2+1 dimensions
\end{center}}

\author[a,b,c]{Jose J. Blanco-Pillado,}
\author[d]{Daniel Jiménez-Aguilar,}
\author{and}
\author[e]{Oriol Pujolàs}

\affiliation[a]{IKERBASQUE, Basque Foundation for Science, 48011, Bilbao, Spain}
\affiliation[b]{EHU Quantum Center, University of the Basque Country, UPV/EHU, Bilbao, Spain}
\affiliation[c]{Department of Physics, University of the Basque
Country, UPV/EHU, 48080, Bilbao, Spain}
\affiliation[d]{Institute of Cosmology, Department of Physics and Astronomy, Tufts University, Medford, MA 02155, USA}
\affiliation[e]{Institut de Física d’Altes Energies (IFAE) and Barcelona Institute of Science and Technology (BIST), Campus UAB, 08193 Bellaterra, Barcelona, Spain}

\emailAdd{josejuan.blanco@ehu.eus}
\emailAdd{Daniel.Jimenez\_Aguilar@tufts.edu}
\emailAdd{pujolas@ifae.es}

\makeatletter
\gdef\@fpheader{}
\makeatother

\abstract{The effective Nambu-Goto description of $(2+1)$-dimensional domain walls predicts singular behavior of its worldsheet resulting in swallowtail bifurcations. This phenomenon is intimately related to the formation of cusps, which emerge in different forms that we identify and classify. We describe in detail how swallowtail bifurcations generically arise in the collision of wiggles on straight domain wall strings, as well as in the collapse of closed loops, even for smooth initial conditions. Remarkably, by means of accurate lattice simulations, we find that these distinctive swallowtail features are reproduced in the field theory evolution of sufficiently thin walls, typically emitting a significant fraction of their initial energy in the process. 
These results suggest that such singular evolutions could potentially have important implications for the observable signatures associated with the collapse of domain wall networks in (3+1) dimensions in the early universe.}

\begin{document}

\maketitle

\newpage

%%%%%%%%%%%%%%%%%%%%%%%%%%%%
\section{Introduction}
%%%%%%%%%%%%%%%%%%%%%%%%%%%%

Solitonic solutions play a fundamental role across various areas of physics, ranging from
condensed matter to string theory and cosmology (see for example \cite{rajaraman1982solitons,BISHOP19801,Johnson:2003cvf,Manton:2004tk,Vilenkin:2000jqa}). Understanding the dynamics of solitons
is fundamental for determining their observational signatures. In cosmological contexts, an 
essential aspect of soliton evolution is their potential decay through the emission of radiation, 
as this process determines both their natural lifetime and the expected observational imprints 
associated with the emitted radiation across different channels.

Topological defect networks such as domain walls, cosmic strings, monopoles, etc. are expected to form in the early universe \cite{Kibble:1976sj}. To assess their observational consequences, it is necessary to characterize their evolution throughout cosmic history. This, in turn,
requires large-scale numerical simulations capable of capturing the dynamics of these networks
over extended time scales.

One of the most prominent examples of solitonic structures in cosmology is cosmic strings \cite{Vilenkin:2000jqa}. The evolution of string networks has been investigated extensively through a variety of numerical approaches. A widely used method is based on the Nambu-Goto (NG) approximation \cite{nambu1970lectures,Goto:1971ce}, in which strings are modeled as infinitely thin, one-dimensional objects, thereby neglecting physical effects associated with their finite core structure (see \cite{Albrecht:1984xv, Bennett:1989ak,
  Allen:1990tv, Vanchurin:2005pa,Vanchurin:2005yb, Martins:2005es, Ringeval:2005kr,Olum:2006ix,
  Blanco-Pillado:2011egf,Blanco-Pillado:2013qja} and references therein). An alternative approach involves lattice field theory simulations, where the full non-linear dynamics of the underlying field configurations are evolved on a discretized lattice, and the behavior of the string network is extracted by identifying and tracking the localized core regions corresponding to the string solutions (see for example \cite{Vincent:1997cx,Moore:2001px,Hindmarsh:2017qff,Helfer:2018qgv,Correia:2020yqg,Correia:2021tok}).

Both approximate methods are expected to yield consistent results in regimes where the soliton remains smooth and, in
its rest frame, is well described by its lowest-energy configuration. However, discrepancies may arise in
situations involving soliton excitations, interactions between solitons or the development of regions 
with high curvature. The emergence of such states from smooth initial configurations can in some cases be predicted using the NG dynamics,  it is clear
that, in those cases the NG approximation breaks down, as it does not account for the soliton's finite thickness, and one should use lattice field theory simulations to understand the correct dynamics.

In fact, high-curvature regions play a crucial role in the emission of radiation from solitonic configurations. This radiation consists of excitations of the fundamental fields underlying the soliton and is particularly prominent in regions where the soliton undergoes rapid, nonlinear dynamics. Accurately characterizing the associated energy loss and spectral properties of the emitted radiation requires detailed analysis through lattice field theory simulations.

A well-known example of this type of dynamics occurs in the context of cosmic strings in 
3+1 dimensions, where cusps —localized regions in which the string momentarily bends back on itself and reaches the speed of light— naturally arise in the evolution of closed string loops \cite{Turok:1984cn}. Solutions to the Nambu-Goto equations of motion predict the formation of such cusps, and field theory simulations have confirmed that the evolution of the solitonic string closely tracks the Nambu-Goto prediction up to the point where string segments begin to overlap \cite{Olum:1998ag}. These simulations have provided valuable insights into the role of nonlinear field interactions at the point of overlap, enabling a quantitative understanding of the backreaction effects and an estimation of the energy radiated during these high-curvature events.

In this work, we investigate the dynamics of domain walls in $2+1$ (domain wall strings) dimensions, which exhibit notable similarities to both cosmic strings and domain walls in $3+1$ dimensions and offers obvious computational advantages. However, there are also important differences. The reduced co-dimension in the 2+1 case gives rise to distinctive geometric features, such as caustics and swallowtail bifurcations, which, as we will show here, emerge generically during the evolution of these objects.

\begin{figure}[h!]
    \centering
\includegraphics[width=0.45\linewidth]{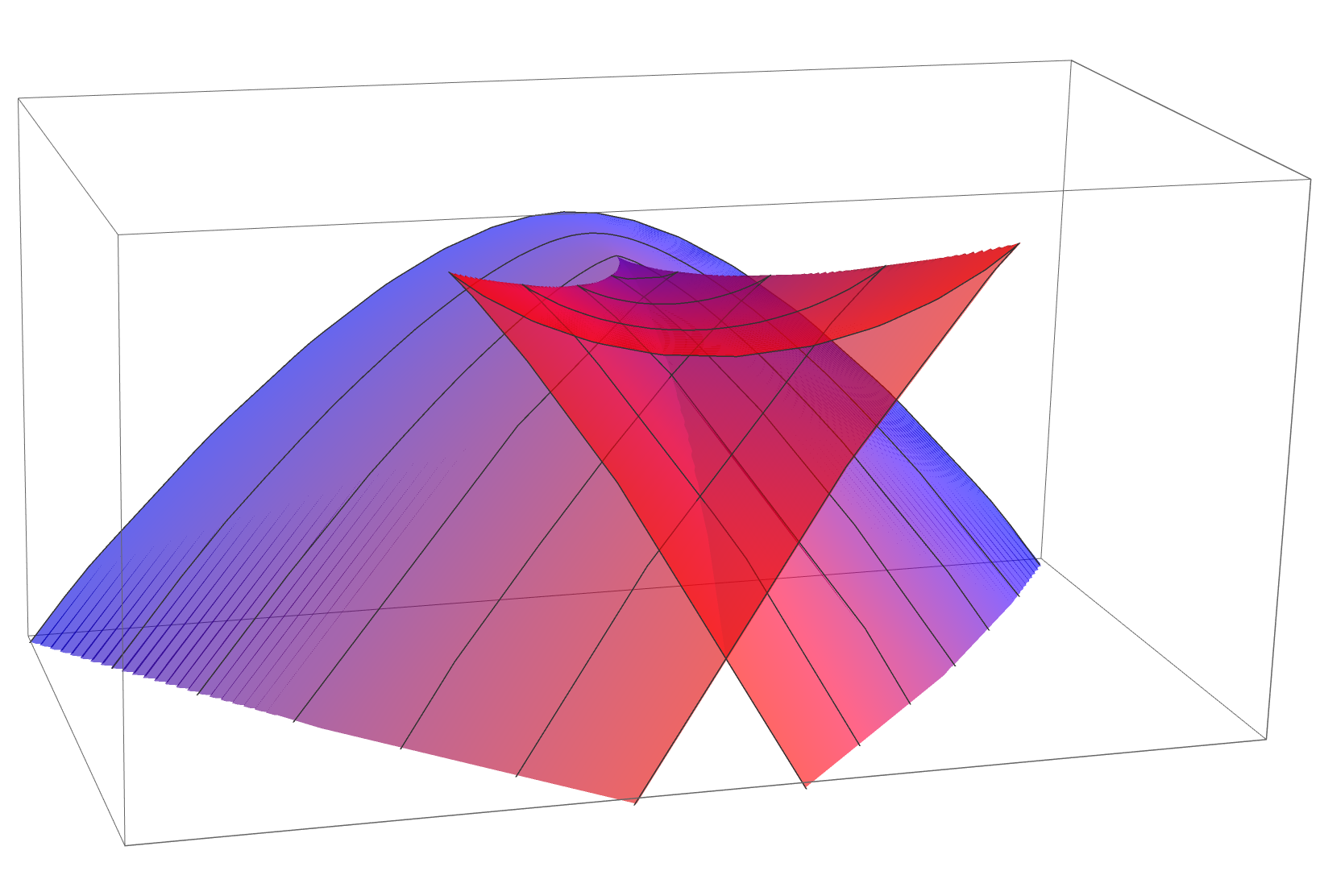}
\quad
\includegraphics[width=0.45\linewidth]{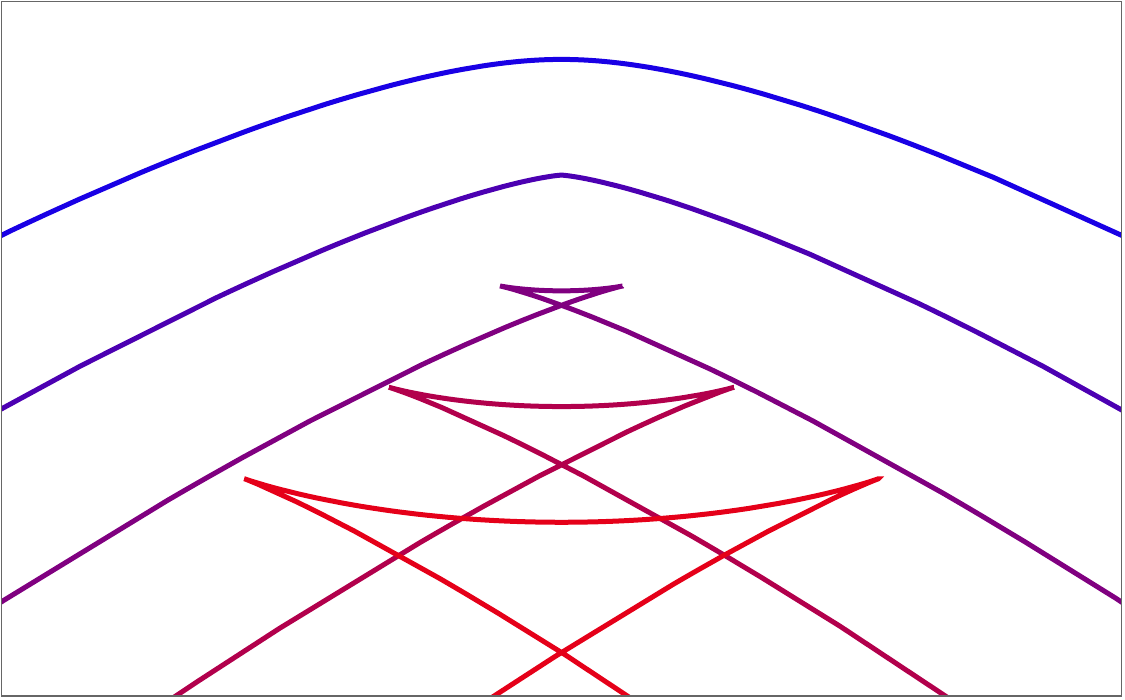}
    \caption{Left: The swallowtail singular surface \eqref{st} in the $\{a,b,c\}$ parameter space, with $b$ ($a$) the horizontal (vertical) axis. Right: Cuts at constant $c$ slices, with a linear offset in $c$ added for visualization. 
    The caustic develops at $c=0$, second line from above. In these axes, the curve can be described by the following parametrization: $b\propto |a|^{4/3}$. 
    As we will show, almost identical singularities arise in the domain wall worldsheet in $2+1$ dimensions with $c$ playing the role of time.}
    \label{fig:ST} 
\end{figure}

The swallowtail bifurcation is a well-known singular surface that arises in a variety of mathematical and physical contexts \cite{arnold1981}. These surfaces can be characterized as the set of points in a parameter space $\{a,b,c\}$ that separates regions with differing numbers of real roots for a given polynomial. In particular, the swallowtail singularity is associated with the quartic polynomial $p\left(x\right)=a+b\,x+ c\,x^2+x^4$, which can possess up to four real roots. The swallowtail surface corresponds to configurations with an odd number of real roots and can be parametrically described by
\begin{equation}\label{st}
a = c~\sigma^2 + 3 ~\sigma^4, \quad b = 2 ~c~ \sigma + 4~ \sigma^3,
\end{equation}
where $\sigma$ is a free parameter. This surface is illustrated in Fig.~\ref{fig:ST}.

The existence of such singular configurations within the Nambu-Goto framework was anticipated in \cite{Hoppe:1995sn,Eggers:2009zz}, and in this work we will identify and analyze the specific initial conditions necessary for their development. 
However, it is important to realize that the existence of such features in the NG framework does not necessarily imply that solitons in field theory will follow the same trajectory, as doing so would require traversing regions of high curvature or domain wall overlap. One might therefore expect these configurations to be suppressed in field theory. Here, we demonstrate that this is not the case; rather, 
field theory simulations of these initial conditions lead to configurations that closely resemble NG predictions, even when the assumptions underlying the NG approximation are significantly violated.

For illustrative purposes, we mainly focus on a case study that is particularly simple and well posed, consisting in the collisions of wiggles traveling on an infinitely long wall, see Fig.~\ref{fig:initial condition wiggles}. Left- or right- moving wiggles correspond to Vachaspati waves \cite{Vachaspati:1990sk} in the field theory, and can have an arbitrary profile. 
It is straightforward then to prepare a smooth initial state for the collision amenable for the comparison between field theory and NG evolutions, since  for well-localized wiggles the initial overlap can be reduced arbitrarily by increasing the initial separation. This guarantees good agreement between NG and field theory evolutions all the way until the collision. As we will see, for large enough amplitude wiggles, swallowtail bifurcations appear during the collision both in NG and in field theory simulations. We also present results for `loops’, walls with closed topology.

An additional motivation for the present work arises from Domain Wall (DW) networks, which are receiving a revived interest recently due to the substantial impact on cosmology mainly in the form of Gravitational Waves and Primordial Black Holes. Compared to Cosmic Strings, the dynamics of DWs remain substantially less well understood. Largely, this stems from the fact that, in contrast with those of the Cosmic String case, the DW equations of motion in the NG approximation are not integrable. Questions like what types of singularities can be formed even in the NG approximation and the conditions for their formation are not only interesting per se but also key aspects in the evolution of general DW network models. 

The remarkable progress until now in this direction \cite{Hoppe:1995sn,Eggers:2009zz,Eggers-etal} leads to a clear picture: in the NG approximation, closed DWs in 3+1 indeed develop generically singularities of a particular `swallowtail bifurcation’ type. The persistence of this phenonemon at the level of field theory DWs is the focus of the present and of a companion article \cite{3+1}. 

The remainder of the paper is structured as follows. In Sec. 2, we introduce the field theory model that serves as the basis for our analysis and outline the procedure for deriving the corresponding effective action governing the soliton dynamics in $2+1$d. In Sec. 3, we examine the integrability of the equations of motion derived from the Nambu-Goto effective action, both in the conformal gauge, commonly adopted in the cosmic string literature, and in the static gauge. Sec. 4 focuses on the emergence of cusp-like features in these 2+1-dimensional solitons evolving according to NG dynamics, highlighting the distinctive characteristics of such events in these lower-dimensional configurations. In Sec. 5, we analyze the formation of swallowtail bifurcations and their connection to cusp formation. In Sec. 6 we discuss the emergence of these swallowtail configurations from smooth initial conditions involving two localized wiggles on an otherwise straight domain wall string. In Sec. 7, we demonstrate that the singular behavior predicted by the NG approximation is indeed realized in full lattice field theory simulations. Finally, in Sec. 8 we write the main conclusions of this work and present a brief discussion on the potential extension of these phenomena to domain walls in 3+1 dimensions as well as the broader implications of these findings for a cosmological scenario. For clarity, technical derivations and supplementary material supporting key results in the main text are presented in the Appendices.

%%%%%%%%%%%%%%%%%%%%%%%%%%%%%%%%%%%%%%%%%%%%%%%%%%%%%%%%
\section{Domain Wall Strings in Field Theory}\label{sec:FTintro}
%%%%%%%%%%%%%%%%%%%%%%%%%%%%%%%%%%%%%%%%%%%%%%%%%%%%%%%%

In this work, we examine the dynamics of domain wall strings arising in a scalar field theory governed by the action
 \begin{equation}
S=\int dx~dy~dt\left[\frac{1}{2}\eta^{\mu\nu}\partial_{\mu}\phi\partial_{\nu}\phi-\frac{\lambda}{4}\left(\phi^{2}-\eta^{2}\right)^{2}\right]\,,
\label{eq:FT action}
\end{equation}
where $\phi$ denotes a real scalar field living in a $2+1$ dimensional flat spacetime.
As one of the simplest solitonic solutions, this domain wall configuration has been extensively studied in the literature. 
Here, we provide a brief overview of its key characteristics, emphasizing aspects relevant to our subsequent discussion, while referring the reader to comprehensive reviews for a more detailed analysis (see for example \cite{Vachaspati:2006zz}).

The static soliton field configuration, corresponding to a solution of the equations of motion derived from the previous action, is given by
\begin{equation}
\phi_k\left(x^{\mu}\right)= \eta \tanh\left[\sqrt{\frac{\lambda}{2}} \eta x\right]~.
\end{equation}

In (2+1)-dimensional spacetime, this soliton solution extends along a single spatial dimension, effectively behaving as a domain wall string.

Beyond the static solution, the spectrum of perturbations around the soliton can be analyzed to identify 
different types of excitations. The lowest-energy excitation corresponds to the zero mode, which is massless 
and arises as a Goldstone mode due to the breaking of translational symmetry by the soliton solution. This mode describes transverse fluctuations of the domain wall string, commonly referred to as wiggles, propagating along its 
extended spatial direction. At higher energies, the soliton also supports a massive bound state, which corresponds 
to excitations associated with variations in the soliton's width. From the perspective of an effective worldsheet theory, this mode has a mass comparable to that of field excitations in the bulk vacuum\footnote{It is also worth noting that these modes are transient in nature, as they gradually decay into massive bulk radiation as a result of the non-linear interactions inherent in the theory \cite{Blanco-Pillado:2020smt}.}. Additionally, the spectrum includes an infinite set of massive states corresponding to the scattering states of the bulk field, which are slightly modified due to the presence of the soliton. Despite its relative simplicity, this model captures many essential features of higher-dimensional solitonic objects, making it a valuable system for detailed investigation.

While the full dynamics of this field theory configuration can be studied using lattice field theory simulations, it is often desirable to develop an effective description that captures the soliton's 
low-energy behavior with a reduced number of degrees of freedom. This approach involves integrating 
out the massive modes to derive an effective action for the domain wall string's zero mode. It is well known 
that, at leading order, this effective action is described by the Nambu-Goto action \cite{Forster:1974ga,1980ZPhyB..36..329D}. Recent studies \cite{Blanco-Pillado:2022rad} have explicitly demonstrated the validity of this approximation by comparing the 
results of full field theory simulations with NG dynamics in this model. These comparisons indicate 
that the NG action accurately reproduces the motion of the domain wall, except for potential deviations 
arising from radiation emission.

Interestingly, recent results reported in \cite{Blanco-Pillado:2024bev} indicate that the leading-order corrections to the Nambu-Goto dynamics arising from domain wall string curvature predominantly couple to the bound state mode of the soliton. As a result, these corrections are substantially more suppressed than previously expected. This observation has important implications: in the absence of excitations of the bound mode, higher curvature corrections to the NG evolution are effectively absent in this model, suggesting that any singularities predicted by the NG dynamics are not regulated by such corrections.

Motivated by this, we begin our analysis by considering the NG dynamics as an effective description of domain wall strings in 
2+1 dimensions, treating it as a complete effective action in the regime where bound mode excitations are negligible\footnote{It is important to emphasize that these structures are ultimately field theory solitons, and as such, one expects radiation to be emitted during singular events on the worldsheet. A detailed investigation of this radiation will be presented in a subsequent section, where we numerically analyze the full field-theoretic dynamics.}.

%%%%%%%%%%%%%%%%%%%%%%%%%%%%%%%%%%%%%%%%%%%%%%%%%%%%%%%%
\section{Nambu-Goto Dynamics}
%%%%%%%%%%%%%%%%%%%%%%%%%%%%%%%%%%%%%%%%%%%%%%%%%%%%%%%%
As previously discussed, the description of domain wall strings within the framework of the Nambu-Goto action relies on the thin-wall approximation, in which the soliton is assumed to have negligible thickness. Consequently, its dynamics are entirely characterized by the spacetime region traversed by the wall as it evolves—its worldsheet. Mathematically, the worldsheet can be described as a general parametrization of a two-dimensional surface embedded in spacetime, given by
\begin{equation}
X^{\alpha}=X^{\alpha}\left(\zeta^{i}\right)\,,
\label{eq:X zeta}
\end{equation}
where $\zeta^{0},\zeta^{1}$ are the two worldsheet coordinates. The Nambu-Goto action is given by \cite{nambu1970lectures,Goto:1971ce}
\begin{equation}
S_{\text{NG}}=-\mu\int d^{2}\zeta\sqrt{-\gamma}\,,
\label{eq:NG action}
\end{equation}
where $\mu$ is the energy per unit length of the string and $\gamma$ is the determinant of the worldsheet metric given by,
\begin{equation}
\gamma_{ij}=\eta_{\alpha\beta}\partial_{i}X^{\alpha}\partial_{j}X^{\beta}\,.
\label{eq:worldsheet metric}
\end{equation}
Here, $\eta_{\alpha\beta}$ is the Minkowski metric in 2+1 dimensions and the partial derivatives denote differentiation with respect to the worldsheet coordinates.

Physically, the Nambu-Goto action is directly proportional to the area of the worldsheet swept out by the solitonic object as it evolves through spacetime. As a consequence, the theory possesses reparametrization invariance, allowing for the freedom to choose arbitrary coordinate systems on the worldsheet to describe its dynamics. This gauge freedom permits the adoption of various coordinate choices, commonly referred to as different gauges. In the following, we describe two gauge choices that will play a central role in our analysis of domain wall string solutions in the subsequent sections.

%%%%%%%%%%%%%%%%%%%%%%%%%%%%%%%%%%%%%%%%%%%%%%%%%%%%%%%%
\subsection{Conformal and Static gauges}
%%%%%%%%%%%%%%%%%%%%%%%%%%%%%%%%%%%%%%%%%%%%%%%%%%%%%%%%

There are two common gauges used to fix the redundancy in the parameterization \eqref{eq:X zeta}.
In the conformal gauge, in which $\gamma_{ij}=\sqrt{-\gamma}\eta_{ij}$, and  $X^{0}=\zeta^{0}=t$, the equations of motion for the position of the string, $\vec{X}$, reduce to
\begin{equation}
\ddot{\vec{X}}-\vec{X}\,''= \vec{0}~,
\label{eq:NG equation in conformal gauge}
\end{equation}
with the gauge condition imposing the following constraints:
\begin{equation}
\dot{\vec{X}}\cdot\vec{X}'=0\,,
\label{eq:gauge constraint 1}
\end{equation}
\begin{equation}
\dot{\vec{X}}^{2}+\vec{X}'^{\,2}=1\,.
\label{eq:gauge constraint 2}
\end{equation}
In these equations, dots and primes respectively denote derivatives with respect to the timelike coordinate, $\zeta^{0}=t$, and the spacelike coordinate $\zeta^{1}=\sigma$. Using \eqref{eq:gauge constraint 2}, one can easily check that the coordinate $\sigma$ is proportional to the string energy measured from some arbitrary point on the string, namely, we have the relation
\begin{equation}
d\sigma=\frac{dl}{\sqrt{1-\dot{\vec{X}}^2}}\,,
\label{eq:dsigma}
\end{equation}
where $dl$ is the physical length of an infinitesimal string segment. Moreover, the first gauge condition
indicates that the velocity is always perpendicular to the tangent vector to the wall, $\vec{X}'$.

The general solution of (\ref{eq:NG equation in conformal gauge}) can be written in terms of the functions $\vec{a}(\sigma)$ and $\vec{b}(\sigma)$, namely,
\begin{equation}
\vec{X}\left(t,\sigma\right)=\frac{1}{2}\left[\vec{a}\left(\sigma-t\right)+\vec{b}\left(\sigma+t\right)\right]\,,
\label{eq:x lm rm}
\end{equation}
and the gauge conditions give the constraints
\begin{equation}
|\vec{a}\,'|=|\vec{b}\,'|=1\, ,
\label{eq:constraint ab prime}
\end{equation}
where the primes here denote the derivative of the function with respect to its argument.

The initial conditions $\vec{X}(0,\sigma)$ and $\dot{\vec X}(0,\sigma)$ fully determine the functions  $\vec a(\sigma)$ and $\vec b(\sigma)$. Consequently, equation \eqref{eq:x lm rm} explicitly provides the complete time evolution of the system, thereby illustrating the integrability of the equations of motion.\\

The other gauge that will be relevant for us is the {\em static} gauge. In this gauge, one aligns one of the intrinsic parameters of the worldsheet with one of the coordinates on the background spacetime, namely  $\zeta^1=X^2\equiv y$ and we also choose $\zeta^0=X^0\equiv t$, so all the information about the dynamics of the string will be stored in a single function that parametrizes the position of the soliton in the other coordinate, namely, $X^1=\psi(y,t)$. This parametrization is particularly useful for deformations of a straight string as the ones we will discuss in the subsequent sections of this paper. Furthermore, this will also be useful in comparing these NG configurations with our field theory simulations. 

In this gauge, the NG action takes the form 
\begin{equation}
S_{\text{NG}}=-\mu\int dy~dt \sqrt{1-(\partial_t\psi)^2+(\partial_y\psi)^2}\,~,
\label{eq:NG action static}
\end{equation}
which is often referred to as Dirac-Born-Infeld (DBI)  field theory. In $(1+1)$ dimensions this is an integrable model \cite{Barbashov:1966frq,Barbashov:1966nvq,whitham,Brunelli:1997kh}, see also \cite{Dubovsky:2012wk}. It is possible to write down generic solutions of the equation of motion (with $\mu$ running over $\{t,y\}$)
\begin{equation}\label{eom static}
\partial_\mu\left(\frac{\partial^\mu \psi}{\sqrt{1-(\partial\psi)^2}}\right)=0
\end{equation}
directly in this gauge, as done first in \cite{Barbashov:1966frq,Barbashov:1966nvq} writing them in parametric form. In light of the relation with the conformal gauge, it is clear that generic solutions (containing both left- and right- movers) must  be expressible as
\begin{align}\label{eq:static g soln}
     \psi(\sigma,t) &= \frac12 \big[a_x(\sigma-t)+b_x(\sigma+t)\big]  \\[2mm]\label{eq:static g soln2}
     y(\sigma,t)&=\frac12 \big[a_y(\sigma-t)+b_y(\sigma+t)\big]  
\end{align}
with $\vec{a}$, $\vec{b}$ satisfying \eqref{eq:constraint ab prime}. It is straightforward to check after some algebra that indeed (\ref{eq:static g soln} - \ref{eq:static g soln2})  satisfies \eqref{eom static}, using $\partial_y |_{t=const} = \frac{1}{\partial_\sigma y}\,\partial_\sigma $ and 
$\partial_t |_{y=const} =\partial_t -\frac{\partial_t y}{\partial_\sigma y}\, \partial_\sigma$.

Despite being integrable in the sense of (\ref{eq:static g soln} - \ref{eq:static g soln2}), the DBI model (NG in static gauge) is known to develop caustic singularities \cite{Barbashov:1966nvq,Barbashov:1966frq,whitham,Felder:2002sv,Babichev:2016hys,Mukohyama:2016ipl}. 
The origin of singularities in this gauge can be traced to the form of the DBI Lagrangian \eqref{eq:NG action static}, which leads to a built-in `speed limit' condition 
\begin{equation}
(\partial\psi)^2=(\partial_t \psi)^2-(\partial_y \psi)^2 < 1 ~, 
\end{equation}
which is not guaranteed to hold everywhere during the time evolution. Points on the worldsheet where speed is exceeded correspond to caustic singularities and the subsequent time evolution breaks down, in this gauge.  

For practical purposes, the conformal gauge is more convenient to obtain the time evolution. In Appendix~\ref{appendix wiggles} we derive the full mapping between conformal and static gauges. In particular, the form of left- and right-mover functions $\vec{a},\,\vec{b}$ in terms of the initial condition in static gauge (see Eqs. (\ref{eq:ax prime psi} - \ref{eq:sigma of y}) and (\ref{eq:ax} - \ref{eq:by})), which is important for the case study of collisions of wiggles presented in Sec.\ref{sec:wiggle collisions} and \ref{sec:FT}.

As we shall show explicitly, it is easy to construct  solutions representing the collisions of two wiggles traveling in opposite directions that lead  to caustics (cusps) starting completely smooth initial conditions. 
The evolution past the caustic event in static gauge is ill defined as it would require formally a multivalued $\psi$ \cite{Barbashov:1966nvq,Barbashov:1966frq,whitham,Felder:2002sv}. The description in conformal gauge instead does not break down in any sense (in the NG approximation) and generically gives rise to swallowtail bifurcations.

%%%%%%%%%%%%%%%%%%%%%%%%%%%%%%%%%%%%%%%%%
\section{Cusps in the NG approximation}
%%%%%%%%%%%%%%%%%%%%%%%%%%%%%%%%%%%%%%%%%

Cusps were first discussed in the NG dynamics of 
cosmic strings in $3+1$ dimensions \cite{Turok:1984cn}.
In this context, cusps are spacetime events located at specific points on the string worldsheet with coordinates $\left(t_{*},\sigma_{*}\right)$ where the string moves at the speed of light, i.e. at a cusp,  
\begin{equation}
    \dot{\vec{X}}\,\Big|_*=1\,,
\end{equation}
which is equivalent to
\begin{equation}
    \vec{X'}\,\Big|_*=0
\end{equation}
by the constraint \eqref{eq:gauge constraint 2}. 
In other words, cusps are points on the string that momentarily move at the speed of light and have vanishing tangent vector. 

Taking the time derivative of (\ref{eq:x lm rm}), one can obtain the velocity of the point on the string labeled by $\sigma$ at time $t$:
\begin{equation}
\dot{\vec{X}}\left(t,\sigma\right)=\frac{1}{2}\left[\vec{b}\,'\left(\sigma+t\right)-\vec{a}\,'\left(\sigma-t\right)\right]\,.
\label{eq:velocity}
\end{equation}
On the other hand, the tangent vector at each point is proportional to
\begin{equation}
\vec{X}'\left(t,\sigma\right)=\frac{1}{2}\left[\vec{a}\,'\left(\sigma-t\right)+\vec{b}\,'\left(\sigma+t\right)\right]\,.
\label{eq:r prime}
\end{equation}
Therefore, in general, the condition for cusp formation is
\begin{equation}
\vec{a}\,'\left(\sigma_{*}-t_{*}\right)=-\vec{b}\,'\left(\sigma_{*}+t_{*}\right)~,
\label{eq:general condition for cusp formation}
\end{equation}
for some point $\left(t_{*},\sigma_{*}\right)$ on the worldsheet. Given that the functions $\vec{a}\,'(\sigma)$ and 
$\vec{b}\,'(\sigma)$  are constrained to lie on the unit sphere due to the imposed normalization condition, these points can be identified as the intersections of their respective trajectories on this so-called Kibble-Turok sphere. This means that, generically, cusps are isolated events; they occur at a point in spacetime.

Furthermore, by imposing these conditions alongside the constraint equations of the Nambu-Goto solution, one can show that the generic local structure of such events exhibits a characteristic behavior of the form $X\propto Y^{3/2}$, where $X$ and $Y$ denote coordinates in the plane of the string in the vicinity of the singularity. This non-analytic behavior defines the geometry of a cusp and justifies the terminology used to describe such events.

%%%%%%%%%%%%%%%%%%%%%%%%%%%%%%%%%%%%%%%%%
\subsection{Cusps are non-isolated in 2+1}
%%%%%%%%%%%%%%%%%%%%%%%%%%%%%%%%%%%%%%%%%
As we argued above, in spacetime dimensions higher than $2+1$, cusps are generically isolated events, meaning that a single point on the worldsheet satisfies the cusp conditions. In contrast, in $2+1$ dimensions, cusps are typically not isolated. Moreover, there exist two distinct types of cusp configurations. In the following, we will explicitly demonstrate both of these claims by analyzing the cusp condition in general.

To see this, it is enough to expand near a cusp event 
\begin{equation}
    (t,\sigma)=(t_*,\sigma_*) + (\delta t,\delta\sigma)
\end{equation}
and see whether the cusp condition can hold (and how generically) in nearby points. It is implicit here that $\delta t,\delta\sigma$ are infinitesimally small quantities.

At lowest order in $\delta t,\delta\sigma$ we have 
$$
\vec{X}' = \vec{X}''_* \,\delta\sigma+\dot{\vec{X}}'_* \, \delta t~,
$$
hence the condition on $\delta t$ and $\delta \sigma$ for the cusp to be non-isolated is
\begin{equation}\label{delta sigma}
    \vec{X}''_* \,\delta\sigma + \dot{\vec{X}}'_* \, \delta t =0~.
\end{equation}

Differentiating the constraints with respect to $\sigma$ and using $\vec{X'}_*=\vec{0}$, one quickly sees that both vectors $\vec{X''}_*$ and $\dot{\vec{X'}}_*$ are orthogonal to $\dot{\vec{X}}_*$. This result is true in any number of dimensions. However, it has very different implications in $2+1$ or in higher dimensions. In higher dimensions $\vec{X''}_*, \dot{\vec{X'}}_*$ belong to an orthogonal plane and generically are not parallel.
In $2+1$ instead $\vec{X''}_*, \dot{\vec{X'}}_*$ must be parallel. This implies that there is a 1-parameter family of solutions to \eqref{delta sigma}. Assuming  $\vec{X''}_* \neq\vec{0}$, the solution is of the form 
$$
\delta\sigma =  C \,  \delta t ~.
$$
with $C$ a constant.

This implies that such a cusp exists and propagates both before and after the point $(t_*,\sigma_*)$, indicating that cusps in this context are not isolated events in spacetime but instead persist over a finite time interval. We refer to such configurations as {\it semi-persistent cusps}. An extreme case arises when the cusp persists indefinitely, as observed in certain rigidly rotating solutions previously identified in the literature \cite{Burden:1984xk}. These permanent or eternal cusps represent highly pathological cases, which we will not address any further in this work. However, as we will argue in the following sections, semi-persistent cusps can generically arise from smooth initial conditions in string dynamics in $2+1$ dimensions.

An interesting situation arises in the case where
$$
\vec{X}''_*=\vec{0}.
$$
The condition for the cusp to be non-isolated becomes
\begin{equation}\label{delta sigma 2}
     \frac12 \,\vec{X}'''_* \,(\delta \sigma)^2 + \dot{\vec{X}}'_* \, \delta t =\vec{0}~.
\end{equation}
Again, using the constraints and $\vec{X}'_*=\vec{X}''_*=\vec{0}$  one finds that (only) in $2+1$ dimensions $\vec{X}'''_*$ must be parallel to $\dot{\vec{X}}'_*$ and so there is a 1-parameter family of solutions. In this case, they are of the form 
$$
\delta \sigma_\pm = \pm D  \sqrt{\delta t}~,
$$
with $D$ a constant. Now, there is a pair of solutions only for $\delta t>0$, that is after $(t_*,\sigma_*)$. Moreover, one can see that each of them is a `standard' cusp of semi-persistent type ($\vec{X}'=\vec{0}$ but $\vec{X}''\neq\vec{0}$ on the cusps at $\delta t>0$). The $\vec{X}''_*=\vec{0}$ can occur (if at all) only at one space-time point. So in a sense it is an instanteneous type of cusp -- however, one that is attached to a pair of semi-persistent cusps.

Physically, what this represents is the creation of a pair of semi-persistent cusps at this moment. We will refer to this type of cusps as the {\it bifurcation cusps}.

%%%%%%%%%%%%%%%%%%%%%%%%%%%%%%%%%%%%%
\section{Swallowtail Singularities}
\label{sec:cusps}
%%%%%%%%%%%%%%%%%%%%%%%%%%%%%%%%%%%%%

As discussed in the previous section, the emergence of a bifurcation cusp signifies the onset of two semi-persistent cusps. In this section, we investigate the local behavior of the string in the vicinity of the bifurcation point and demonstrate that the resulting worldsheet geometry evolves into a configuration that closely corresponds to the canonical swallowtail structure. In doing so, we also examine the distinct local characteristics that differentiate bifurcation cusps from semi-persistent cusps.

%%%%%%%%%%%%%%%%%%%%%%%%%%%%%%%%%%%%%%%%%%%%%
\subsection{Near cusp singularity expansion}
%%%%%%%%%%%%%%%%%%%%%%%%%%%%%%%%%%%%%%%%%%%%%

Let us start by studying a generic cusp in $2+1$ dimensions. As we have already mentioned, the fact that the cusp moves at the speed of light 
implies $\vec{a}\,'_{*}=-\vec{b}\,'_{*}$. Here and henceforth, the subscript $*$ denotes evaluation 
at the cusp: $t=t_{*}$, $\sigma=\sigma_{*}$~.  
Without loss of generality, we may choose the $x$ axis to be aligned with the direction of the cusp's velocity. Taking this into account, we have
\begin{equation}
\vec{a}\,'_{*}=\left(-1,0\right)\,\,\,,\,\,\,\vec{b}\,'_{*}=\left(1,0\right)\,.
\label{eq:ab prime}
\end{equation} 

Our goal here is to obtain a fourth-order expansion of the worldsheet in both $t$ and $\sigma$ in the vicinity of the cusp. To this end, we require the derivatives of the functions $\vec{a}$ and $\vec{b}$ up to fourth order. Following an approach analogous to that employed in the 3+1-dimensional case \cite{Vilenkin:2000jqa}, we make use of the constraints imposed by the conformal gauge. Specifically, we use the condition given in Eq.~(\ref{eq:constraint ab prime}), which is equivalent to $\vec{u}\,'\cdot\vec{u}\,'=1$, where $\vec{u}$ generically denotes $\vec{a}$ or $\vec{b}$.

Taking the derivative of this expression, we get
\begin{equation}
\vec{u}\,''\cdot\vec{u}\,'=0\,.
\label{eq:constraint derived once}
\end{equation}
Using (\ref{eq:ab prime}), we immediately conclude that $\vec{a}\,''_{*}$ and $\vec{b}\,''_{*}$ lie on the $y$ axis (perpendicular to the velocity of the cusp):
\begin{equation}
\vec{a}\,''_{*}=\left(0,\alpha_{2}\right)\,\,\,,\,\,\,\vec{b}\,''_{*}=\left(0,\beta_{2}\right)\,,
\label{eq:ab prime prime}
\end{equation}
where we leave $\alpha_{2}$ and $\beta_{2}$ unspecified for now. 

The derivative of (\ref{eq:constraint derived once}) is
\begin{equation}
\vec{u}\,'''\cdot\vec{u}\,'+|\vec{u}\,''|^{2}=0\,,
\label{eq:constraint derived twice}
\end{equation}
which combined with (\ref{eq:ab prime}) and (\ref{eq:ab prime prime}) yields
\begin{equation}
\vec{a}\,'''_{*}=\left(\alpha_{2}^{2},\alpha_{3}\right)\,\,\,,\,\,\,\vec{b}\,'''_{*}=\left(-\beta_{2}^{2},\beta_{3}\right)
\label{eq:ab prime prime prime}
\end{equation}
for some unspecified $\alpha_{3}$ and $\beta_{3}$. Finally, the derivative of (\ref{eq:constraint derived twice}) reads
\begin{equation}
\vec{u}\,^{IV}\cdot\vec{u}\,'+3\,\vec{u}\,'''\cdot\vec{u}\,''=0\,.
\label{eq:constraint derived three times}
\end{equation}
Using (\ref{eq:ab prime}), (\ref{eq:ab prime prime}) and (\ref{eq:ab prime prime prime}), we find
\begin{equation}
\vec{a}\,^{IV}_{*}=\left(3\alpha_{2}\alpha_{3},\alpha_{4}\right)\,\,\,,\,\,\,\vec{b}\,^{IV}_{*}=\left(-3\beta_{2}\beta_{3},\beta_{4}\right)\,,
\label{eq:ab prime prime prime prime}
\end{equation}
where $\alpha_{4}$ and $\beta_{4}$ are not known. Now we expand (\ref{eq:x lm rm}) to fourth order in both $t$ and $\sigma$ near $t=t_{*}$ and $\sigma=\sigma_{*}$. Plugging in equations (\ref{eq:ab prime}), (\ref{eq:ab prime prime}), (\ref{eq:ab prime prime prime}) and (\ref{eq:ab prime prime prime prime}), and defining $\bar{X}\left(t,\sigma\right)\equiv X\left(t,\sigma\right)-X_{*}$, $\bar{Y}\left(t,\sigma\right)\equiv Y\left(t,\sigma\right)-Y_{*}$, $\bar{\sigma}\equiv\sigma-\sigma_{*}$, $\bar{t}\equiv t-t_{*}$, where $X$ and $Y$ denote the $x$ and $y$ components of $\vec{X}$, we obtain
\begin{equation}
\begin{split}
\bar{X}\left(\bar{t},\bar{\sigma}\right)&=\bar{t}-\frac{\alpha_{2}^{2}+\beta_{2}^{2}}{12}\,\bar{t}^{3}+\frac{\alpha_{2}\alpha_{3}-\beta_{2}\beta_{3}}{16}\,\bar{t}^{4}+\left(\frac{\alpha_{2}^{2}-\beta_{2}^{2}}{4}\,\bar{t}^{2}-\frac{\alpha_{2}\alpha_{3}+\beta_{2}\beta_{3}}{4}\,\bar{t}^{3}\right)\bar{\sigma}+\\\\
&+\left[-\frac{\alpha_{2}^{2}+\beta_{2}^{2}}{4}\,\bar{t}+\frac{3\left(\alpha_{2}\alpha_{3}-\beta_{2}\beta_{3}\right)}{8}\,\bar{t}^{2}\right]\bar{\sigma}^{2}+\left(\frac{\alpha_{2}^{2}-\beta_{2}^{2}}{12}-\frac{\alpha_{2}\alpha_{3}+\beta_{2}\beta_{3}}{4}\,\bar{t}\right)\bar{\sigma}^{3}+\\\\
&+\frac{\alpha_{2}\alpha_{3}-\beta_{2}\beta_{3}}{16}\,\bar{\sigma}^{4}\,,
\end{split}
\label{eq:expanded X general}
\end{equation}
\begin{equation}
\begin{split}
\bar{Y}\left(\bar{t},\bar{\sigma}\right)&=\frac{\alpha_{2}+\beta_{2}}{4}\,\bar{t}^{2}+\frac{\beta_{3}-\alpha_{3}}{12}\,\bar{t}^{3}+\frac{\alpha_{4}+\beta_{4}}{48}\,\bar{t}^{4}+\\\\
&+\left(\frac{\beta_{2}-\alpha_{2}}{2}\,\bar{t}+\frac{\alpha_{3}+\beta_{3}}{4}\,\bar{t}^{2}+\frac{\beta_{4}-\alpha_{4}}{12}\,\bar{t}^{3}\right)\bar{\sigma}+\\\\
&+\left(\frac{\alpha_{2}+\beta_{2}}{4}+\frac{\beta_{3}-\alpha_{3}}{4}\,\bar{t}+\frac{\alpha_{4}+\beta_{4}}{8}\,\bar{t}^{2}\right)\bar{\sigma}^{2}+\\\\
&+\left(\frac{\alpha_{3}+\beta_{3}}{12}+\frac{\beta_{4}-\alpha_{4}}{12}\,\bar{t}\right)\bar{\sigma}^{3}+\frac{\alpha_{4}+\beta_{4}}{48}\,\bar{\sigma}^{4}\,.
\end{split}
\label{eq:expanded Y general}
\end{equation}

It is instructive at this stage to examine the configuration of the string precisely at the moment of cusp formation, i.e., at $\bar t=0$. In this case, one observes that
\begin{equation}
\begin{split}
\bar{X}\left(\bar{t},\bar{\sigma}\right)&=\left(\frac{\alpha_{2}^{2}-\beta_{2}^{2}}{12}\right)\bar{\sigma}^{3}
+\left(\frac{\alpha_{2}\alpha_{3}-\beta_{2}\beta_{3}}{16}\right)\,\bar{\sigma}^{4}\,,\\
\bar{Y}\left(\bar{t},\bar{\sigma}\right)&=
\left(\frac{\alpha_{2}+\beta_{2}}{4}\right)\bar{\sigma}^{2}
+\left(\frac{\alpha_{3}+\beta_{3}}{12}\right)\bar{\sigma}^{3}+\left(\frac{\alpha_{4}+\beta_{4}}{48}\right)\,\bar{\sigma}^{4}\,.
\end{split}
\end{equation}

From the expression for the string derived above, it is evident that the local behavior of the cusp undergoes a qualitative change when $\alpha_2 = - \beta_2$. This condition is equivalent to requiring that the second derivative of the string vanishes at the cusp, i.e., $|\vec{X}''_*|=0$. According to our classification, this scenario corresponds to a bifurcation cusp. In this case, the local geometry of the string near the cusp is characterized by the scaling relation $\bar X \propto \bar Y^{4/3}$. In contrast, in the generic case where $|\vec{X}''_*|>0$, the cusp exhibits a different scaling behavior, namely 
$\bar X \propto \bar Y^{3/2}$. This corresponds to the class of semi-persistent cusps in
$(2+1)$dimensions, as previously discussed.

Finally, we note, in passing, the existence of another distinct class of cusps that arise when 
$\alpha_2=\beta_2$ . In this case, the cusp exhibits a pathological behavior, as this condition implies a local relation of the form $\bar X \propto \bar Y^{2}$ in the vicinity of the cusp. Consequently, the two branches of the string overlap over a significantly extended region compared to the generic case in which $|\alpha_2|\ne |\beta_2|$.

In the following section, we demonstrate that all the aforementioned types of cusps can, in fact, arise from the collision of localized perturbations on an extended, straight string in $2+1$ dimensions.

%%%%%%%%%%%%%%%%%%%%%%%%%%%%%%%%%%%%%%%%%%%%%%%%%%%%%%%%%%%%%%
\subsection{Swallowtail Geometry near the bifurcation cusp}
%%%%%%%%%%%%%%%%%%%%%%%%%%%%%%%%%%%%%%%%%%%%%%%%%%%%%%%%%%%%%%

We now return to the description of the worldsheet and examine the structure of the hypersurface in the vicinity of the bifurcation cusp. In this case, the condition $\alpha_2=-\beta_2$ holds, and the expansion of the worldsheet (\ref{eq:expanded X general}) and (\ref{eq:expanded Y general}) reduce in this regime to the following expressions:

\begin{equation}
\begin{split}
\bar{X}\left(\bar{t},\bar{\sigma}\right)&=\bar{t}-\frac{\alpha_{2}^{2}}{6}\,\bar{t}^{3}+\frac{\alpha_{2}\left(\alpha_{3}+\beta_{3}\right)}{16}\,\bar{t}^{4}-\frac{\alpha_{2}\left(\alpha_{3}-\beta_{3}\right)}{4}\,\bar{t}^{3}\,\bar{\sigma}+\\\\
&+\left[-\frac{\alpha_{2}^{2}}{2}\,\bar{t}+\frac{3\alpha_{2}\left(\alpha_{3}+\beta_{3}\right)}{8}\,\bar{t}^{2}\right]\bar{\sigma}^{2}-\frac{\alpha_{2}\left(\alpha_{3}-\beta_{3}\right)}{4}\,\bar{t}\,\bar{\sigma}^{3}+\frac{\alpha_{2}\left(\alpha_{3}+\beta_{3}\right)}{16}\,\bar{\sigma}^{4}\,,
\end{split}
\label{eq:expanded X}
\end{equation}
\begin{equation}
\begin{split}
\bar{Y}\left(\bar{t},\bar{\sigma}\right)=\frac{\beta_{3}-\alpha_{3}}{12}\,\bar{t}^{3}+\frac{\alpha_{4}+\beta_{4}}{48}\,\bar{t}^{4}+\left(-\alpha_{2}\,\bar{t}+\frac{\alpha_{3}+\beta_{3}}{4}\,\bar{t}^{2}+\frac{\beta_{4}-\alpha_{4}}{12}\,\bar{t}^{3}\right)\bar{\sigma}+\\\\
+\left(\frac{\beta_{3}-\alpha_{3}}{4}\,\bar{t}+\frac{\alpha_{4}+\beta_{4}}{8}\,\bar{t}^{2}\right)\bar{\sigma}^{2}+\left(\frac{\alpha_{3}+\beta_{3}}{12}+\frac{\beta_{4}-\alpha_{4}}{12}\,\bar{t}\right)\bar{\sigma}^{3}+\frac{\alpha_{4}+\beta_{4}}{48}\,\bar{\sigma}^{4}\,.
\end{split}
\label{eq:expanded Y}
\end{equation} 
If we consider times sufficiently close to the event of the cusp, $\bar{t}\sim\bar{\sigma}^{2}$, and define the following dimensionless variables\footnote{Note that the dimensions of $\alpha_{n}$ are energy to the power $n-1$.},
\begin{equation}
\tilde{X}\equiv-\frac{48\alpha_{2}^{3}}{\alpha_{3}+\beta_{3}}\,\bar{X}\,\,\,,\,\,\,\tilde{Y}\equiv-\frac{48\alpha_{2}^{3}}{\alpha_{3}+\beta_{3}}\,\bar{Y}\,,
\label{eq:dimensionless X Y}
\end{equation}
\begin{equation}
\tilde{t}\equiv-\frac{24\alpha_{2}^{3}}{\alpha_{3}+\beta_{3}}\,\bar{t}\,\,\,,\,\,\,\tilde{\sigma}\equiv\alpha_{2}\,\bar{\sigma}\,,
\label{eq:dimensionless t and sigma}
\end{equation}
equations (\ref{eq:expanded X}) and (\ref{eq:expanded Y}) can be approximated as
\begin{equation}
\tilde{Z}\left(\tilde{t},\tilde{\sigma}\right)\equiv2\tilde{t}-\tilde{X}\left(\tilde{t},\tilde{\sigma}\right)=\tilde{t}\tilde{\sigma}^{2}+3\tilde{\sigma}^{4}\,,
\label{eq:X swallowtail}
\end{equation}
\begin{equation}
-\tilde{Y}\left(\tilde{t},\tilde{\sigma}\right)=2\tilde{t}\tilde{\sigma}+4\tilde{\sigma}^{3}\,,
\label{eq:Y swallowtail}
\end{equation} 
which completely match the swallowtail parametrization given by (\ref{st}) under the identification $\tilde{t}\leftrightarrow c$, 
$\tilde{Y}\leftrightarrow -b$, $\tilde{Z}\leftrightarrow a$.\\

Furthermore, one can demonstrate from the expansion of the worldsheet that immediately following the formation of the bifurcation cusp, two distinct semi-persistent cusps emerge from this point. It can be verified that the local behavior of these resulting cusps is described by the generic scaling relation $\bar X \propto \bar Y^{3/2}$, consistent with the characteristic structure of semi-persistent cusps.

%%%%%%%%%%%%%%%%%%%%%%%%%%%%%%%%%%%%%%%%%%%%%%%%%%%%%%%%%
\section{Swallowtail Singularities in Wiggle Collisions}\label{sec:wiggle collisions}
%%%%%%%%%%%%%%%%%%%%%%%%%%%%%%%%%%%%%%%%%%%%%%%%%%%%%%%
In the preceding sections, we examined the emergence of swallowtail-type singularities consistent with NG dynamics. While their mathematical existence is well established, it does not immediately follow that such structures arise generically from smooth and physically reasonable initial conditions. Indeed, the conditions identified earlier may appear highly fine-tuned (for example the condition that $\alpha_2 = - \beta_2$ at the initial cusp), suggesting that these configurations could be exceedingly rare. In this section, we show that such singular behavior can in fact occur from smooth initial conditions, studying the especially well-defined problem described by the  collision of localized wiggles on an extended, straight domain wall string in $2+1$ dimensions. We further identify the precise conditions required for these singularities to develop in the collision of wiggles of a Gaussian profile and present a systematic procedure for computing the expansion coefficients of the worldsheet embedding in the vicinity of the bifurcation cusp.

%%%%%%%%%%%%%%%%%%%%%%%%%%%%%%%%%%%%%%%%%%%%%%%%%%%%%%%%%
\subsection{Initial Conditions}
%%%%%%%%%%%%%%%%%%%%%%%%%%%%%%%%%%%%%%%%%%%%%%%%%%%%%%%
We begin by considering a traveling wave of arbitrary shape propagating at the speed of light along a straight domain wall string. It is straightforward to verify that such perturbations constitute exact solutions to the Nambu–Goto equations of motion in the static gauge\footnote{Note that, at the level of the NG description, there is no restriction on the functional form of the wave for it to satisfy the equations of motion. In this work, we restrict our attention to initial conditions for which the transverse displacement of the domain wall string is initially described by a single-valued function but there are otherwise no further restrictions in the parameters of the wiggles.}. Specifically, a valid solution is obtained by parametrizing the worldsheet embedding as 
\begin{equation}
X^{\mu}(t,y) = \left(t, ~\psi(t,y\pm t),~ y\right)~,
\end{equation}
where $\psi(t,y)$ describes the transverse displacement of the domain wall string with
respect to the straight line at $x=0$.

In what follows, we focus on localized wave packets characterized by an amplitude $A$ and a width $w$ of the form

\begin{equation}
\psi_{\text{wiggle}} (y)= A \exp{\left[-\frac{y^2}{2w^2}\right]}~.
\label{eq:Gaussian wiggles}
\end{equation}

To build a configuration that remains arbitrarily close to an exact solution of the NG equations, we consider a superposition of two identical, initially well-separated wiggles propagating in opposite directions along the domain wall string. This yields initial conditions of the form

\begin{equation}
X^{\mu}(t,y) = \left(t, ~\psi_{+}(t,y)+ \psi_{-}(t,y),~ y\right)\,,
\end{equation}
where
\begin{equation}
\psi_{\pm}(t,y) = \psi_{\rm wiggle}(y\pm t \mp d)\,.
\end{equation}
\begin{figure}[h!]
\begin{center}
\includegraphics[scale=0.9]{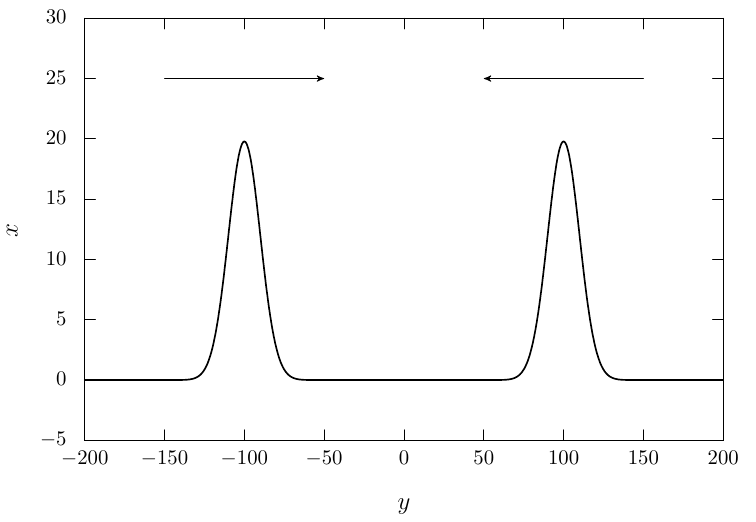}
\caption{Initial configuration for the collision of two wave packets travelling at the speed of light on a straight string.}
\label{fig:initial condition wiggles}
\end{center}
\end{figure}

A snapshot of this configuration is shown in Fig. \ref{fig:initial condition wiggles}.
While the static gauge offers an intuitive visualization — allowing a direct interpretation of the string profile as a function of the spatial coordinate $y$ — it becomes less practical when analyzing the time evolution of the system. In particular, the static gauge description breaks down in regions where the transverse field 
$\psi$ becomes multivalued, which corresponds precisely to the formation of the nontrivial structures we aim to study.

To address this, we recast the initial configuration into the conformal gauge, which is better suited for describing the evolution beyond such singular points. In this gauge, the subsequent dynamics can be computed consistently, without reference to the potential multivaluedness of the embedding.

The translation of our initial data into the conformal gauge requires a nontrivial  transformation, which we perform explicitly. The resulting expressions are presented below, while the full details of the transformation procedure are provided in Appendix \ref{appendix wiggles} for the interested reader.

Recall that the NG description of the string position as a function of time can be obtained in terms of two vectorial functions $\vec{a}(\sigma)$ and $\vec{b}(\sigma)$, namely
\begin{equation}
\begin{split}
X^0 &=t ~,\\
X^1 &=X\left(t,\sigma\right)=\frac{1}{2}\left[a_{x}\left(\sigma-t\right)+b_{x}\left(\sigma+t\right)\right]\,,\\
X^2 &= Y\left(t,\sigma\right)=\frac{1}{2}\left[a_{y}\left(\sigma-t\right)+b_{y}\left(\sigma+t\right)\right]\,.
\end{split}
\end{equation}

Using the expression of the initial conditions
for the domain wall string in the static gauge, the function $\psi(t,y)$ described above, we can obtain the form of the components of each of the relevant functions in the conformal gauge to be
\begin{equation}
a_{x}\left(\sigma\right)=\psi\left(t=0,y\left(\sigma\right)\right)-\int_{-l/2}^{y\left(\sigma\right)} dy\frac{\partial_t \psi}{\sqrt{1+ (\partial_y \psi)^{\,2}-(\partial_t \psi)^{2}}}\,\,\Bigg\rvert_{t=0}\,,
\label{eq:ax}
\end{equation}
\begin{equation}
b_{x}\left(\sigma\right)=\psi\left(t=0,y\left(\sigma\right)\right)+\int_{-l/2}^{y\left(\sigma\right)} dy\frac{\partial_t \psi}{\sqrt{1+ (\partial_y \psi)^{\,2}-(\partial_t \psi)^{2}}}\,\,\Bigg\rvert_{t=0}\,,
\label{eq:bx}
\end{equation}
\begin{equation}
a_{y}\left(\sigma\right)=y\left(\sigma\right)+\int_{-l/2}^{y\left(\sigma\right)} dy\frac{(\partial_y \psi) (\partial_t \psi)}{\sqrt{1+ (\partial_y \psi)^{\,2}-(\partial_t \psi)^{2}}}\,\,\Bigg\rvert_{t=0}\,,
\label{eq:ay}
\end{equation}
\begin{equation}
b_{y}\left(\sigma\right)=y\left(\sigma\right)-\int_{-l/2}^{y\left(\sigma\right)} dy\frac{(\partial_y \psi) (\partial_t \psi)}{\sqrt{1+ (\partial_y \psi)^{\,2}-(\partial_t \psi)^{2}}}\,\,\Bigg\rvert_{t=0}\,.
\label{eq:by}
\end{equation}
\\
In these expressions, $l$ is the size of the box in the $y$ direction, namely, the length of
the string in the absence of wave packets. Furthermore, all the expression should be written in terms of the parameter $\sigma$, the energy per unit length of the configuration, which is related to the spatial coordinate $y$ through the relation

\begin{equation}
\sigma(y)=\int_{-l/2}^{y}dy\frac{1+(\partial_{y}\psi)^{2}}{\sqrt{1+(\partial_{y}\psi)^{2}-(\partial_{t}\psi)^{2}}}\,\,\Bigg\rvert_{t=0}\,,
\label{eq:sigma of y}
\end{equation}
which can be inverted numerically to get $y(\sigma)$.

%%%%%%%%%%%%%%%%%%%%%%%%%%%%%%%%%%%%%%%%%%%%%%%%%%%%%%%%%
\subsection{Nambu-Goto evolution of the colliding wiggles}
%%%%%%%%%%%%%%%%%%%%%%%%%%%%%%%%%%%%%%%%%%%%%%%%%%%%%%%

Once the functional form of $\vec{a}(\sigma)$ and $\vec{b}(\sigma)$ 
 has been determined, the complete evolution of the Nambu–Goto domain wall string can be reconstructed. In this section, we demonstrate that for the class of initial perturbations considered, there exists a critical threshold in the amplitude beyond which the formation of swallowtail-type singularities becomes inevitable.

To establish this result, we begin by showing that, under such initial conditions, the collision of the two traveling waves necessarily leads to the formation of a cusp. Several different approaches could be used to show this; here, we adopt a geometric visualization based on the evolution of the tangent vectors $\vec{a}\,'(\sigma)$ and $-\vec{b}\,'(\sigma)$ .
 Recall that the condition for the appearance of a cusp is the existence of a point $(\sigma_*, t_*)$ on the worldsheet for which $\vec{a}\,'(\sigma_*-t_*) = -\vec{b}\,'(\sigma_*+t_*)$. Furthermore, the conformal gauge constraints require that both  $\vec{a}\,'$ and   $-\vec{b}\,'$
 trace out curves on the unit circle. It is therefore sufficient to show that these curves intersect at some point in spacetime.

In the $2+1$-dimensional case considered here, it is convenient to represent these curves as functions defined over a cylindrical surface at a particular time, where the angular coordinate corresponds to the angle in the $(x,y)$ plane of the functions $\vec{a}\,'$ and $-\vec{b}\,'$ and the axial coordinate spans the domain of $\sigma$.

Plotting $\vec{a}\,'$ and $-\vec{b}\,'$ on this cylinder for various time slices allows us to identify possible intersections. In the asymptotic regions of a straight domain wall, the tangent vectors approach constant values, with 
$\vec{a}\,'(\sigma_{\infty},t) =(0,1)$ and 
$-\vec{b}\,'(\sigma_{\infty},t) =(0,-1)$.  However, within the localized perturbation, both curves undergo a transient excursion in the transverse direction (the $x$ component), returning to their asymptotic values far from the wiggle. For perturbations of sufficiently small amplitude, these deviations are minimal, and the curves never intersect, precluding cusp formation. Conversely, a sufficiently large deviation results in an intersection, indicating the presence of a cusp.

In Appendix \ref{appendix wiggles}, \ref{appendix angles} and \ref{appendix identical-wiggles}, we examine these configurations in detail and derive a necessary condition for the formation of these cusps based on the properties of the initial perturbation on a straight domain wall string. Specifically, we show that a sufficient criterion for cusp formation is that the spatial derivative of the function characterizing the wiggles satisfies, at least, one of these two inequalities:
\begin{equation}
|\partial_y \psi(0,y_{1})|\times|\partial_y \psi(0,y_{4})|>1\,,
\end{equation}
\begin{equation}
|\partial_y \psi(0,y_{2})|\times|\partial_y \psi(0,y_{3})|>1,
\end{equation}
where $y_{1,4}$ and $y_{2,3}$ are, respectively, the ``exterior'' and ``interior'' inflection points in the initial configuration. For the initial state shown in Fig. \ref{fig:initial condition wiggles}, for instance, $y_{1}\approx -110$ and $y_{4}\approx +110$ are the exterior inflection points, and $y_{2}\approx -90$ and $y_{3}\approx +90$ are the interior inflection points. For more details, see the derivation of Eqs. (\ref{eq:condition swallowtail formation 1}) and (\ref{eq:condition swallowtail formation 2}) in Appendix \ref{appendix wiggles} or the comments at the end of Appendix \ref{appendix identical-wiggles}.

\begin{figure}[h!]
\begin{center}
\includegraphics[scale=1.2]{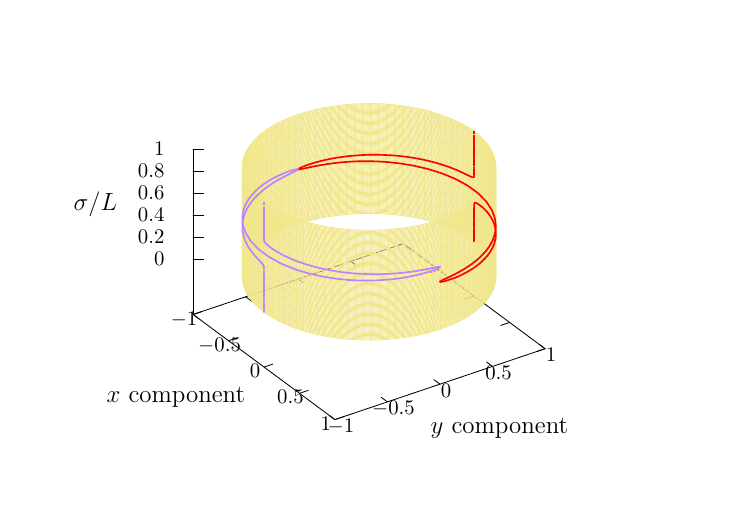}
\caption{Vectors $\vec{a}\,'\left(\sigma-t\right)$ (red) and $-\vec{b}\,'\left(\sigma+t\right)$ (purple) on the surface of the unit-radius cylinder for Gaussian wiggles with the critical amplitude $A=A_{\rm threshold}$. $L$ denotes the invariant length (total amount of $\sigma$) of the string.}
\label{fig:cylinder critical}
\end{center}
\end{figure}

In the case of symmetric wiggles given by the Gaussian functions specified earlier, this condition translates into a threshold amplitude of the form $A= A_{\text{threshold}} = \sqrt{e}\,w$. In Fig. \ref{fig:cylinder critical}, we illustrate the behavior of the tangent vectors at the critical amplitude and at the moment when the unique cusp forms.

\begin{figure}[h!]
\begin{center}
\includegraphics[scale=0.61]{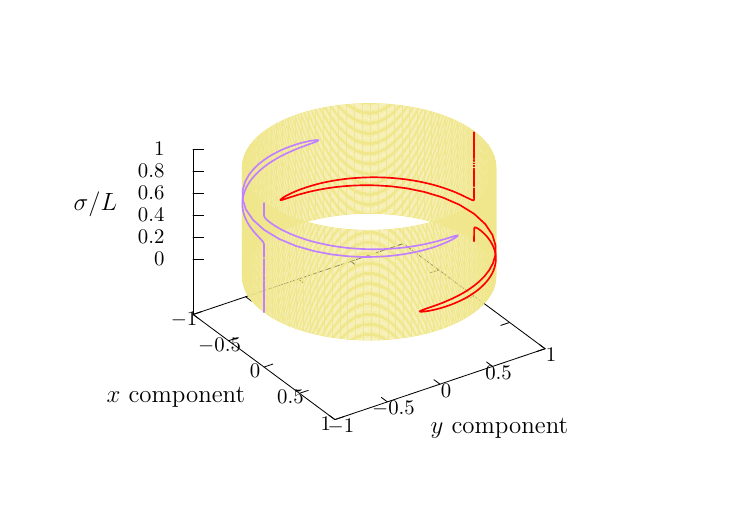}
\includegraphics[scale=0.61]{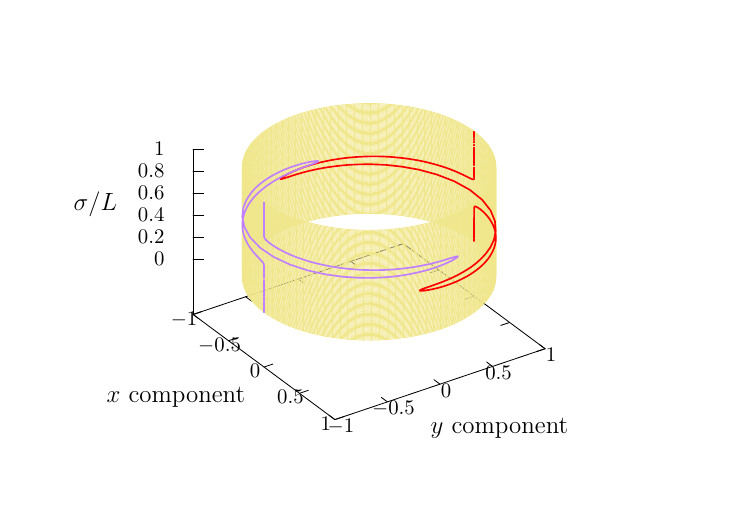}
\includegraphics[scale=0.61]{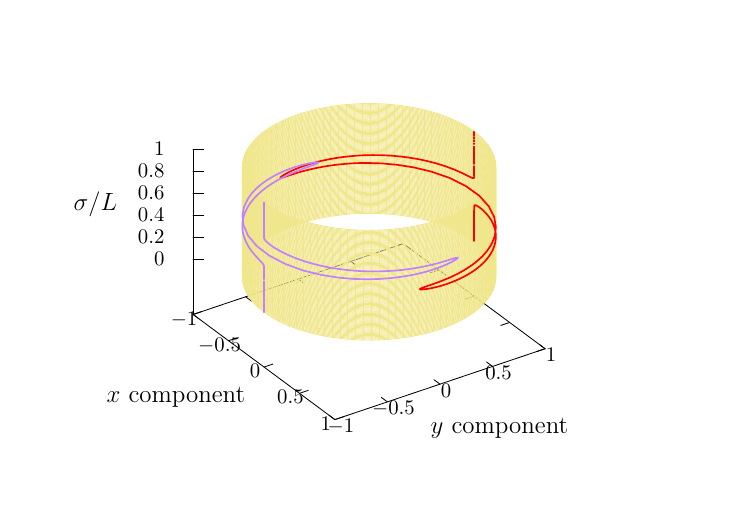}
\includegraphics[scale=0.61]{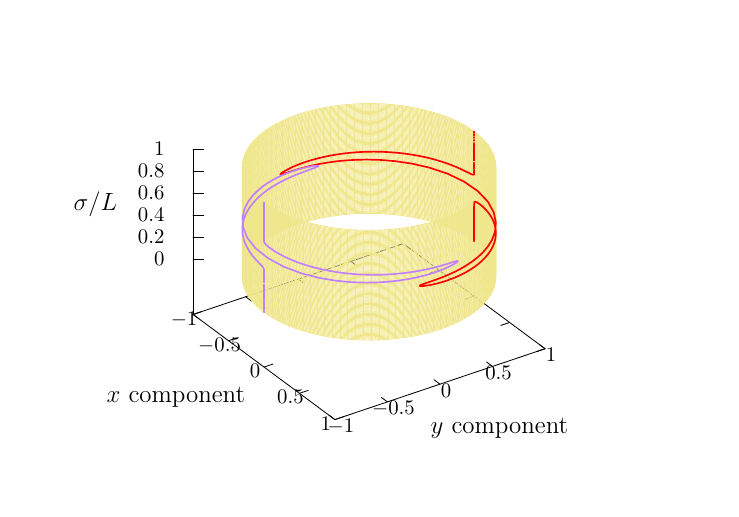}
\caption{Vectors $\vec{a}\,'\left(\sigma-t\right)$ (red) and $-\vec{b}\,'\left(\sigma+t\right)$ (purple) on the surface of the unit-radius cylinder at different times. $L$ denotes the invariant length (total amount of $\sigma$) of the string. Top left: initially, the curves do not overlap. Top right: the curves intersect at a single point and the bifurcation cusp emerges. Bottom left: shortly after, the curves intersect at two points for a finite range of time, which corresponds to the presence of two semi-persistent cusps. Bottom right: the curves intersect again at a single point, meaning that the semi-persistent cusps have annihilated into a bifurcation cusp.}
\label{fig:cylinders}
\end{center}
\end{figure}

We now use the geometric representation of the functions $\vec{a}\,'$ and $-\vec{b}\,'$
on the cylindrical surface to infer the dynamical behavior of the system beyond the
threshold amplitude. In Fig. \ref{fig:cylinders}, we present a representative example with amplitude 
$A = 1.2 A_{\text{threshold}}$. At early times, shortly after the initial conditions are set,
the trajectories of the two functions remain disjoint. However, as time evolves, a moment
arises when the two curves intersect for the first time. This initial intersection signifies
the onset of the bifurcation cusp formation. As the evolution proceeds, there appear two intersections,
signaling the emergence of a pair of semi-persistent cusps. These persist for a finite duration
before annihilating at a later time, giving rise to another bifurcation cusp. All these different situations in the domain wall evolution can be seen in Fig. \ref{fig:cylinders}.

This framework provides an intuitive way to visualize the formation and evolution of cusp
structures on the worldsheet via the behavior of these vector functions on the cylinder. However,
a crucial aspect is not yet captured by this geometric visualization. As previously discussed,
bifurcation cusps are special in that they satisfy an additional condition beyond the standard
cusp criterion. This supplementary condition appears, at first glance, to be fine-tuned and
non-generic. Nonetheless, as we demonstrate in Appendix \ref{appendix angles}, such bifurcation cusps naturally
arise in our class of colliding wiggle configurations and, in fact, they correspond to the initial
intersection between the $\vec{a}\,'$ and $-\vec{b}\,'$. In particular, we show that the new
condition in the second derivatives of these functions implies that $\vec{X}''=\vec{0}$ at the bifurcation
point. This result is established by re-expressing the tangent vectors in terms of their angular
trajectories on the transverse cross-section of the cylinder. This result is further illustrated in Appendix~\ref{appendix identical-wiggles}, where we explicitly compute the coefficients of the bifurcation cusp for the specific case of the collision of two identical traveling waves.

\begin{figure}[h!]
\centering
\includegraphics[scale=0.82]{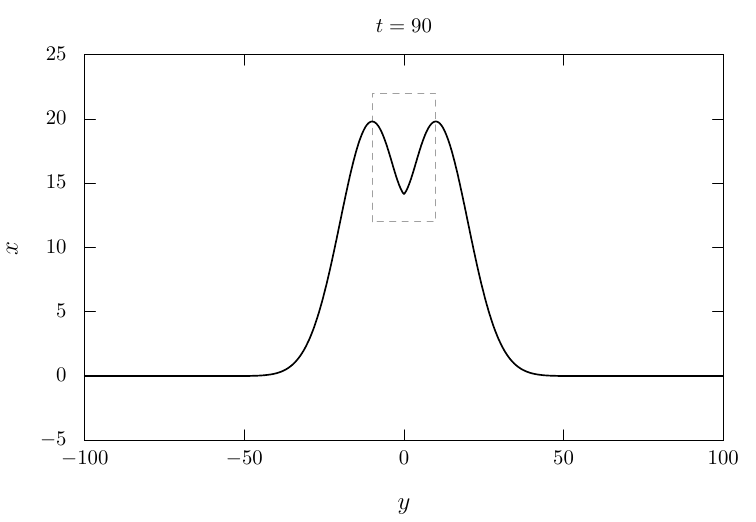}
\includegraphics[scale=0.82]{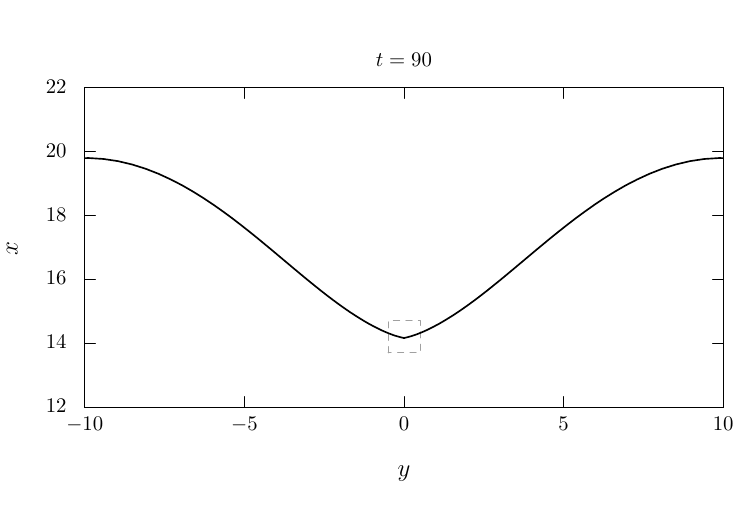}
\begin{minipage}[c]{0.42\textwidth}
    \centering
\includegraphics[scale=0.22]{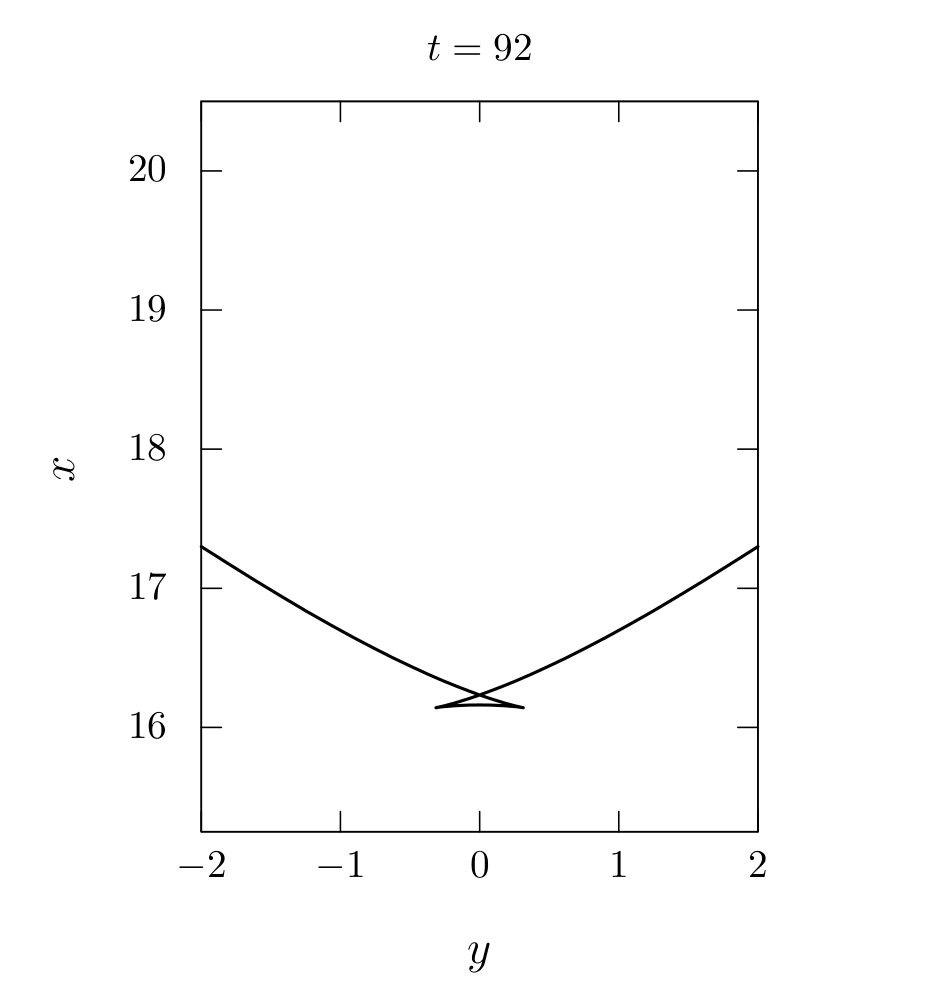}
\end{minipage}
\begin{minipage}[c]{0.42\textwidth}
    \centering
\includegraphics[scale=0.22]{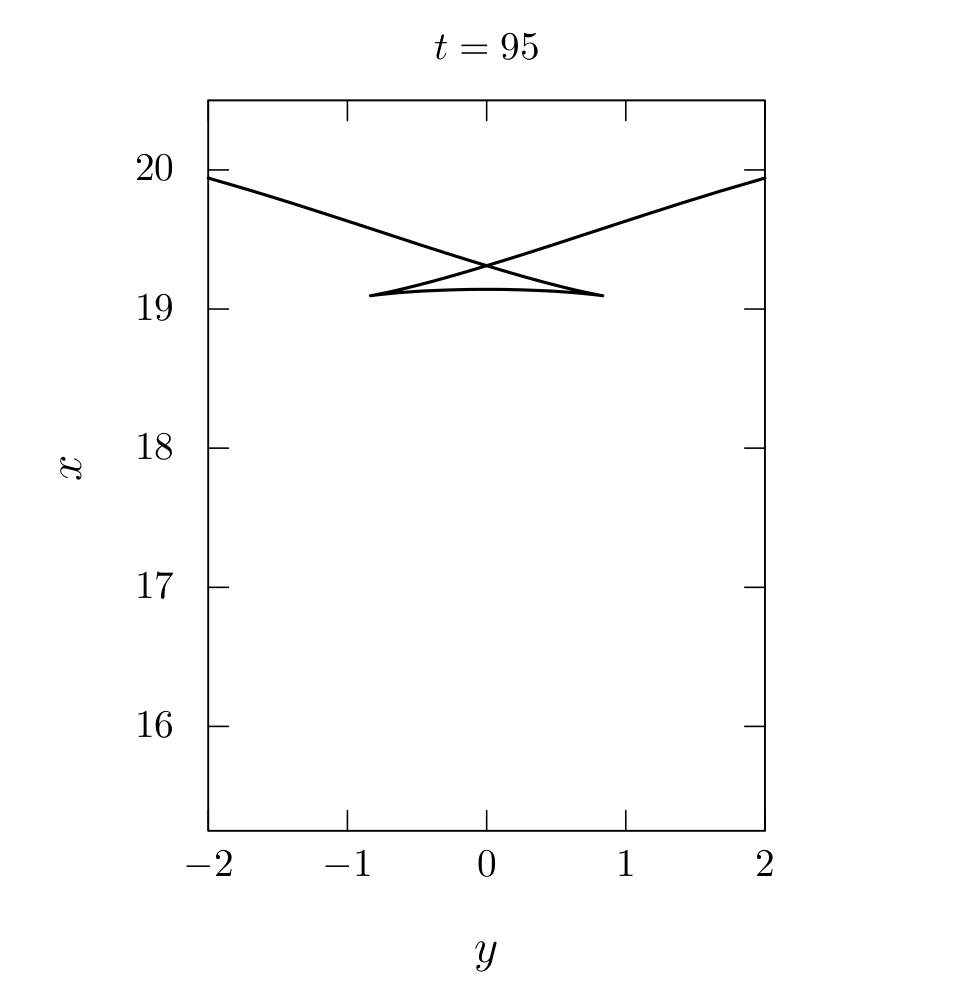}
\end{minipage}
\caption{Formation of the bifurcation cusp and its subsequent evolution in NG. The second snapshot is the amplification of the rectangular region indicated in the first snapshot. Similarly, the last two are zoomed-in views of the squared central region in the second snapshot, but at two later times.}
\label{fig:NG-wiggles}
\end{figure}

In Fig. \ref{fig:NG-wiggles}, we show the NG evolution for our initial wiggles with $A = 1.2 A_{\text{threshold}}$
at different stages of their evolution. We first
depict the formation of the bifurcation cusp. The domain wall string profile at the moment of
cusp formation is well-approximated by the local scaling relation $\bar X \propto \bar Y^{4/3}$, , consistent with the general theory of swallowtail singularities described earlier in the manuscript. Immediately following this event, two additional cusps appear, which we identify as semi-persistent cusps characterized locally by the scaling $\bar X \propto \bar Y^{3/2}$.

\begin{figure}[h!]
\centering
\begin{minipage}[c]{0.49\textwidth}
    \centering
\includegraphics[width=\textwidth]{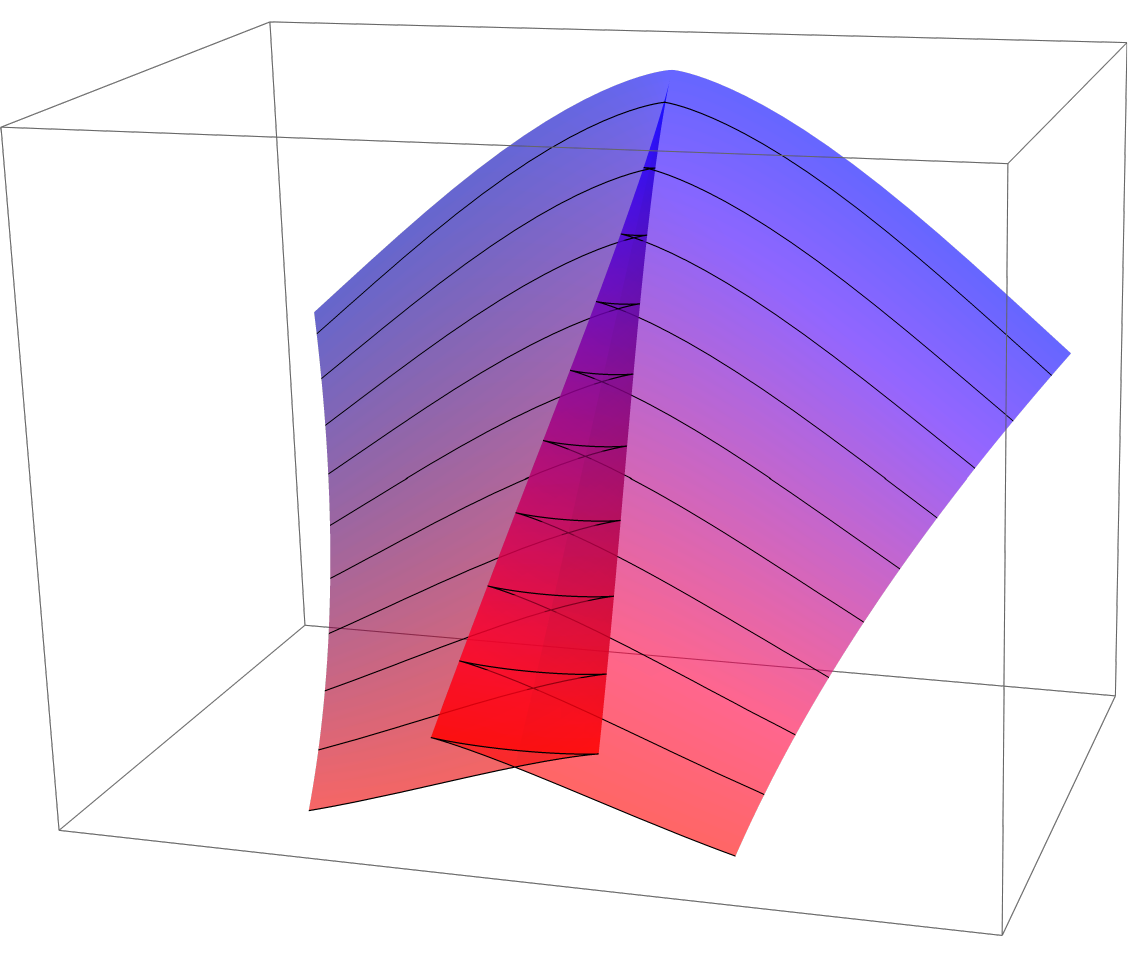}
\end{minipage}
\begin{minipage}[c]{0.49\textwidth}
    \centering
\includegraphics[width=\textwidth]{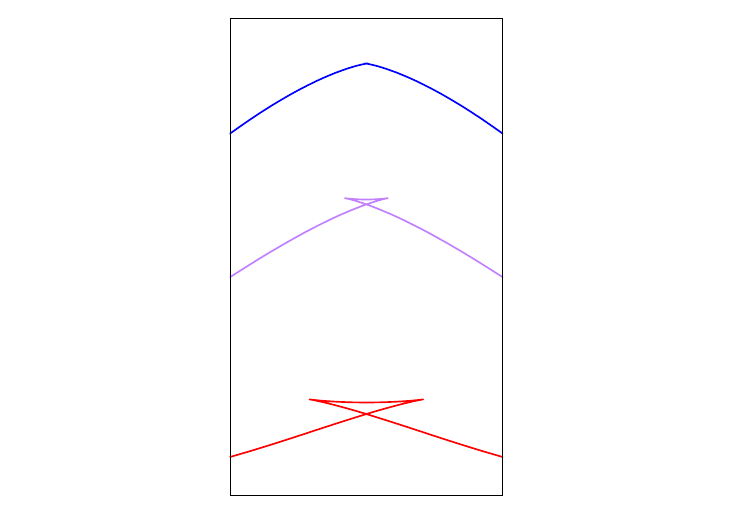}
\end{minipage}
\caption{Left: actual string worldsheet in NG with $y$ as the horizontal axis, $x$ as the vertical axis (inverted) and $t$ as the third axis. Earlier times are shown in blue, and later times in red. Right: constant-time slices of the left panel.}
\label{fig:worldsheet Gaussian}
\end{figure}

\begin{figure}[h!]
\centering
\begin{minipage}[c]{0.49\textwidth}
    \centering
\includegraphics[width=\textwidth,height=5 cm]{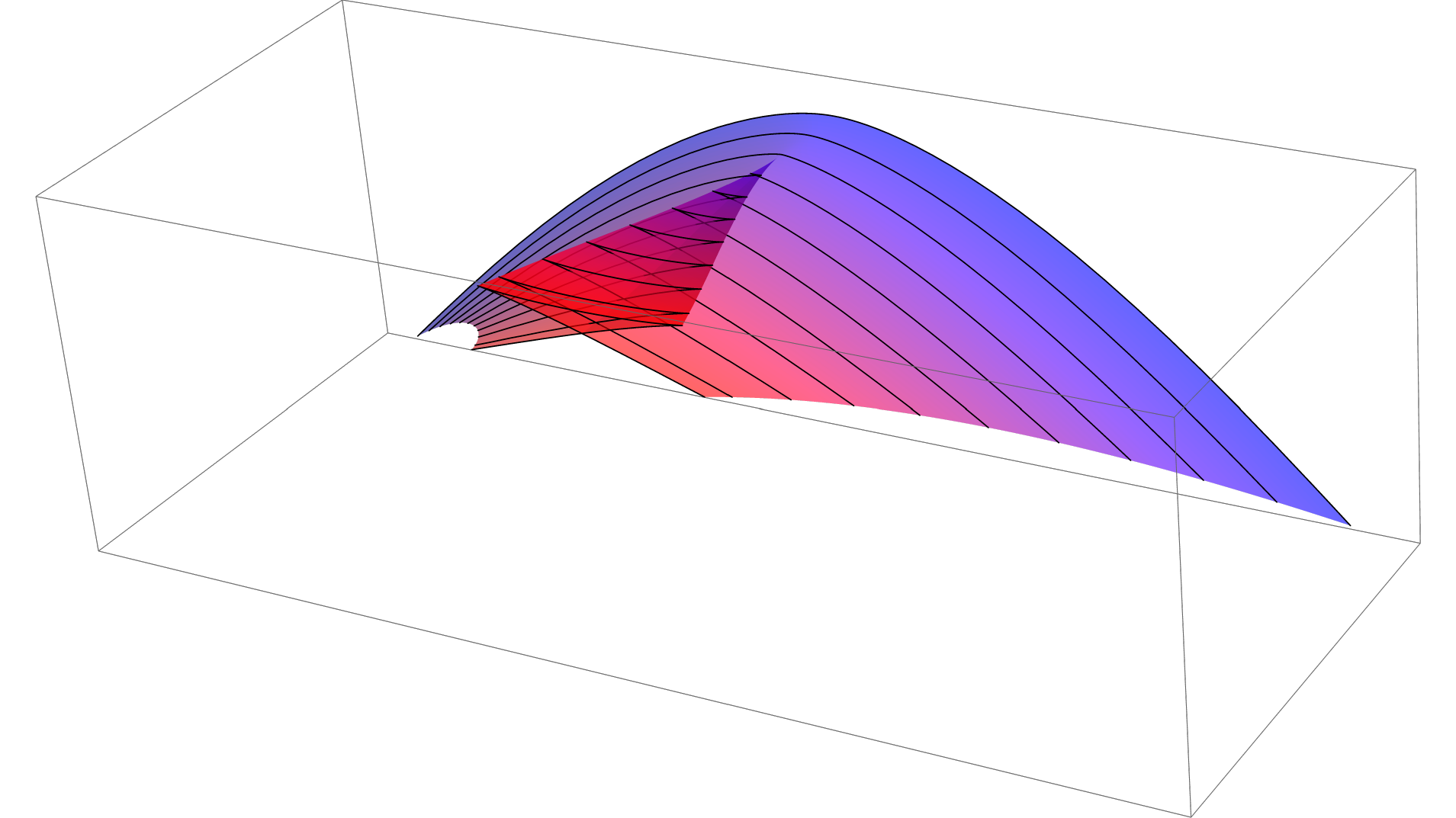}
\end{minipage}
\begin{minipage}[c]{0.49\textwidth}
    \centering
\includegraphics[width=\textwidth,height=4 cm]{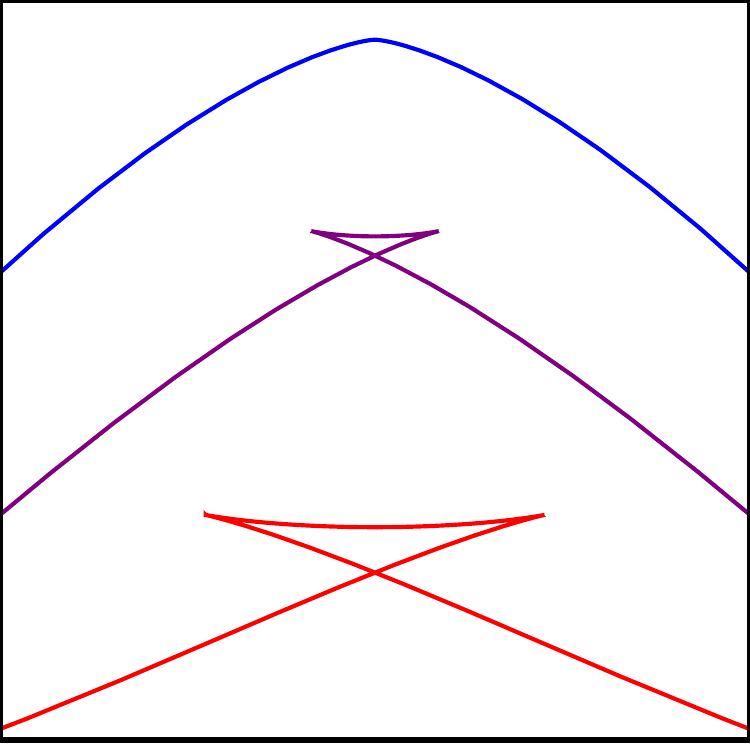}
\end{minipage}
\caption{Left: swallowtail surface as defined by Eqs. (\ref{eq:X swallowtail}) and (\ref{eq:Y swallowtail}) with $-\tilde{Y}$ as the horizontal axis, $\tilde{Z}$ as the vertical axis and $\tilde{t}$ as the third axis, all inverted. Right: constant-time slices of the left panel.}
\label{fig:swallowtail Gaussian}
\end{figure}

Moreover, the full worldsheet geometry of these events confirms that the singular structures conform to the expected classification of swallowtail-type singularities. An illustrative rendering of the worldsheet evolution is shown in Fig. \ref{fig:worldsheet Gaussian}. Additionally, in Fig. \ref{fig:swallowtail Gaussian} we show the parametric swallowtail surface defined by (\ref{eq:X swallowtail}) and (\ref{eq:Y swallowtail}).

Finally, it is evident from the figures that the formation of the bifurcation cusp corresponds,
in the static gauge, to the emergence of multivaluedness in the field $\psi$ or, equivalently, a
self-intersection of the domain wall string. As we will demonstrate in the following sections, this
feature plays a crucial role in the field theory evolution of the system and has significant
physical implications.


\clearpage

%%%%%%%%%%%%%%%%%%%%%%%%%%%%%%%%%%%%%%%%%%%%%%
\section{Field theory evolution}\label{sec:FT}
%%%%%%%%%%%%%%%%%%%%%%%%%%%%%%%%%%%%%%%%%%%%%%

The swallowtail bifurcation is a phenomenon predicted by the Nambu-Goto action. However, as we will show in this section, this catastrophic behavior is closely reproduced in field theory. In the following, we present the results of lattice field theory simulations in the $\lambda\phi^{4}$ model. The equation of motion coming from the field theory action (\ref{eq:FT action}) is

\begin{equation}
\frac{\partial^{2}\phi}{\partial t^{2}}-\frac{\partial^{2}\phi}{\partial x^{2}}-\frac{\partial^{2}\phi}{\partial y^{2}}+\lambda\left(\phi^{2}-\eta^{2}\right)\phi=0\,.
\label{eq:FT eom}
\end{equation}

\subsection{Numerical setup}
We begin by introducing the following dimensionless variables: $\tilde{\phi}=\phi/\eta$, $\tilde{\lambda}=\lambda/\eta^{2}$, $\tilde{t}=\eta^{2}t$, $\tilde{x}=\eta^{2}x$, $\tilde{y}=\eta^{2}y$. In terms of these variables, the equation of motion \eqref{eq:FT eom} reads
\begin{equation}
\frac{\partial^{2}\tilde{\phi}}{\partial \tilde{t}^{2}}-\frac{\partial^{2}\tilde{\phi}}{\partial \tilde{x}^{2}}-\frac{\partial^{2}\tilde{\phi}}{\partial \tilde{y}^{2}}+\tilde{\lambda}\left(\tilde{\phi}^{2}-1\right)\tilde{\phi}=0\,.
\label{eq:dimensionless FT eom}
\end{equation}
In the following, all variables will be dimensionless but we will drop the tildes from them for the sake of simplicity in the notation. In particular, time and length scales are made dimensionless by a multiplicative factor of $\eta^{2}$.

Building on the description of swallowtail singularity formation in the Nambu–Goto framework, we now simulate the collision of two identical perturbations on an otherwise straight domain wall 
configuration. The profile of these perturbations is taken to be the same as that used in the NG analysis, denoted by 
$\psi_{\pm}(t,y)$.

This choice of initial conditions proves particularly useful. It has been shown \cite{Vachaspati:1990sk} that the exact solution to the full non-linear field theory equations for a single arbitrary wiggle propagating at the speed of light along a straight domain wall can be constructed from the static kink solution, $\phi_k(x)$,  and the functional form of the wiggle $\psi_{\pm}(t,y)$ via the composition
\begin{equation}
\phi(t,x,y) = \phi_k\left[x-\psi_{\pm}(t\pm y)\right]~.
\end{equation}

This result enables the construction of an initial configuration comprising two well-separated, finite-length wiggles, symmetrically joined at the center. Such an initial condition can be made arbitrarily close to an exact solution of the full field theory equations of motion, with deviations confined to the narrow region where the two profiles are joined\footnote{This technique was first introduced in the context of cosmic string simulations for cusp formation in \cite{Olum:1998ag}, and has been recently employed in \cite{Blanco-Pillado:2022rad,Drew:2023ptp} to investigate radiation from similar collisional processes.}. Specifically, we will take the initial conditions for the scalar field of the form
\begin{equation}
\phi\left(t=0,x,y\right)=\tanh\left[\frac{x-\psi_{-}\left(t=0,y\right)-\psi_{+}\left(t=0,y\right)}{\sqrt{2/\lambda}}\right]\,,
\label{eq:initial conditions phi}
\end{equation}
with
\begin{equation}
\begin{split}
\dot{\phi}\left(t=0,x,y\right)=\frac{1}{w^{2}}\sqrt{\frac{\lambda}{2}}\,\left[\left(y-d\right)\psi_{+}\left(t=0,y\right)-\left(y+d\right)\psi_{-}\left(t=0,y\right)\right]\times\\\\
\times\text{sech}^{2}\left[\frac{x-\psi_{-}\left(t=0,y\right)-\psi_{+}\left(t=0,y\right)}{\sqrt{2/\lambda}}\right]\,.
\end{split}
\label{eq:initial conditions phi dot}
\end{equation}
\\
The equation of motion \eqref{eq:dimensionless FT eom}, with the initial conditions \eqref{eq:initial conditions phi}-\eqref{eq:initial conditions phi dot}, was solved using the staggered leapfrog method with nearest-neighbor discretization for the spatial derivatives. We perform these simulations in a 2+1 dimensional box with sides of length $L_{x}=200$ and $L_{y}=400$ in the $x$ and $y$ directions, respectively. The domain wall string extends along the $y$ axis, and the lattice spacing and time step are $\Delta x=0.1$ and $\Delta t=0.05$.

Furthermore, we parallelized our code in order to handle the huge amount of points involved in the simulations\footnote{For more information about the implementation of this numerical simulation, see \cite{Blanco-Pillado:2022rad}.}.

%%%%%%%%%%%%%%%%%%%%%%%%%%
\subsection{Results}
%%%%%%%%%%%%%%%%%%%%%%%%%%

\begin{figure}[h!]
\centering
\begin{minipage}[c]{0.48\textwidth}
    \centering
    \includegraphics[width=\textwidth]{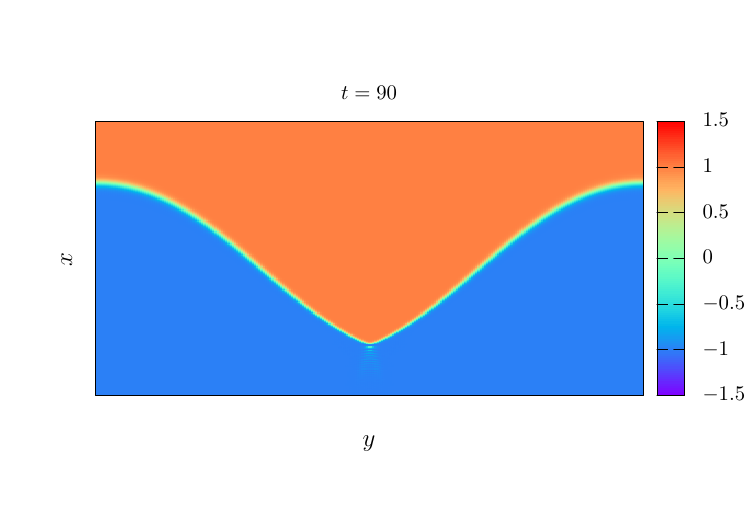}
\end{minipage}
\hfill
\begin{minipage}[c]{0.48\textwidth}
    \centering
    \includegraphics[width=\textwidth]{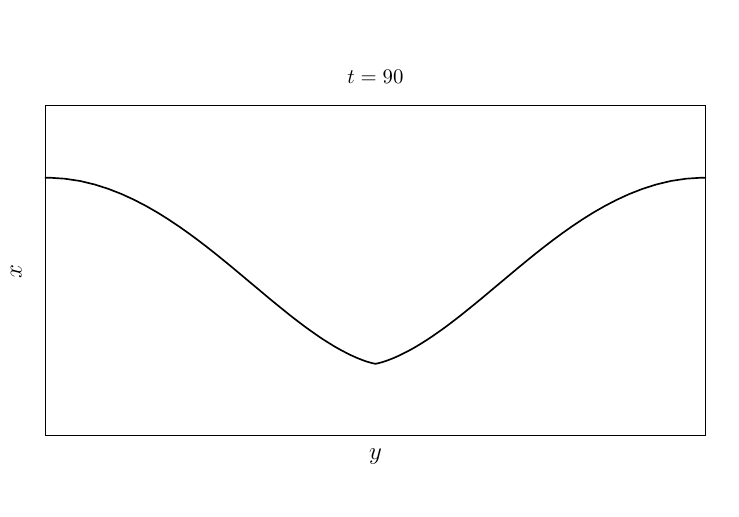}
\end{minipage}
%%%%
\centering
\begin{minipage}[c]{0.42\textwidth}
    \centering
    \includegraphics[width=\textwidth]{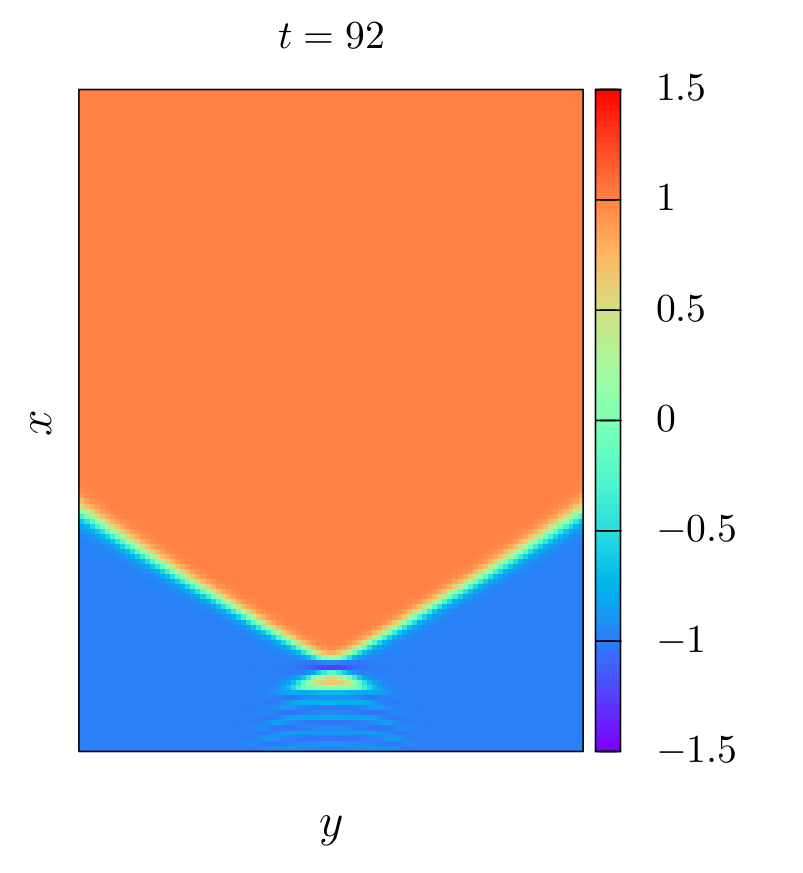}
\end{minipage}
%\hfill
\begin{minipage}[c]{0.42\textwidth}
    \centering
    \includegraphics[width=\textwidth,height=6.6 cm]{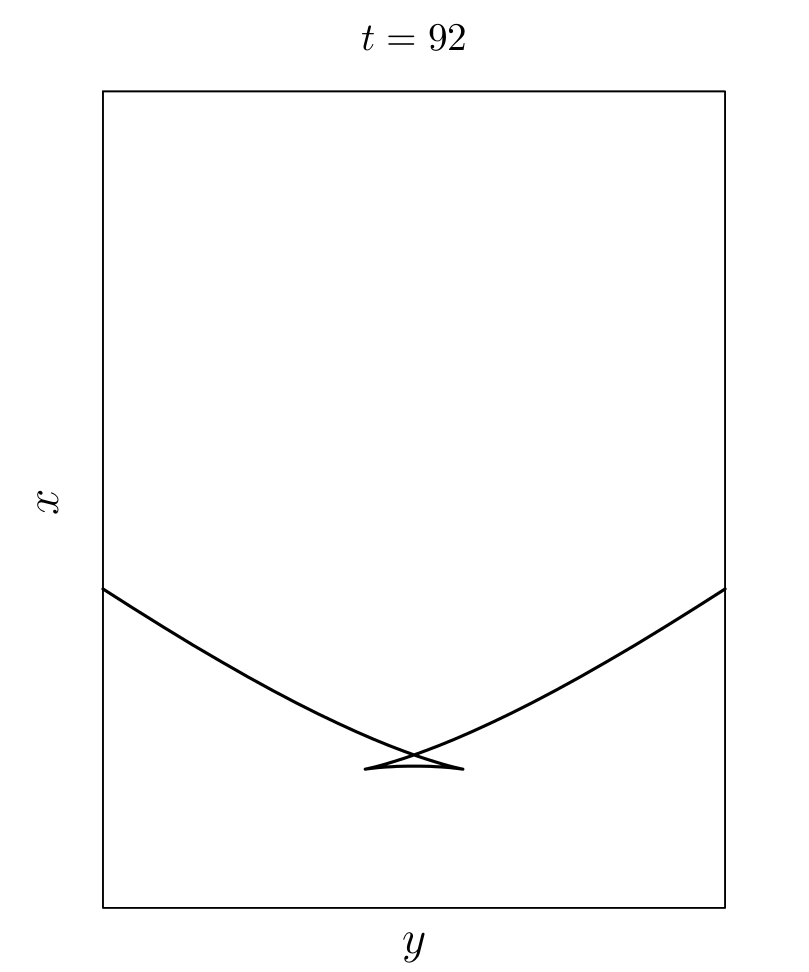}
\end{minipage}
%%%%
\centering
\begin{minipage}[c]{0.41\textwidth}
    \centering
    \includegraphics[width=\textwidth]{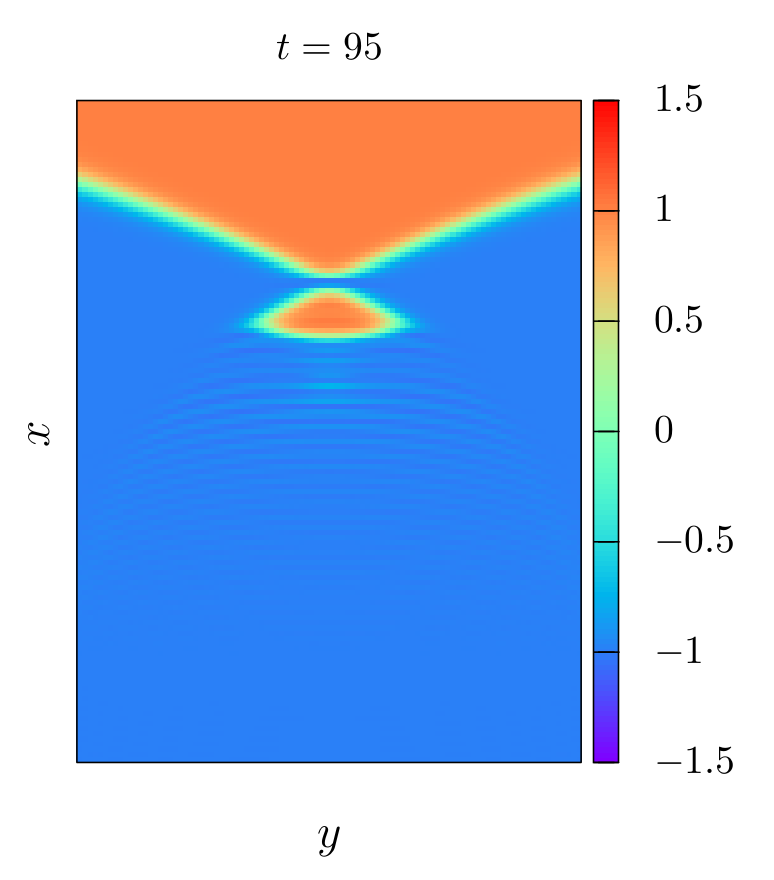}
\end{minipage}
%\hfill
\begin{minipage}[c]{0.41\textwidth}
    \centering
    \includegraphics[width=\textwidth,height=6.6 cm]{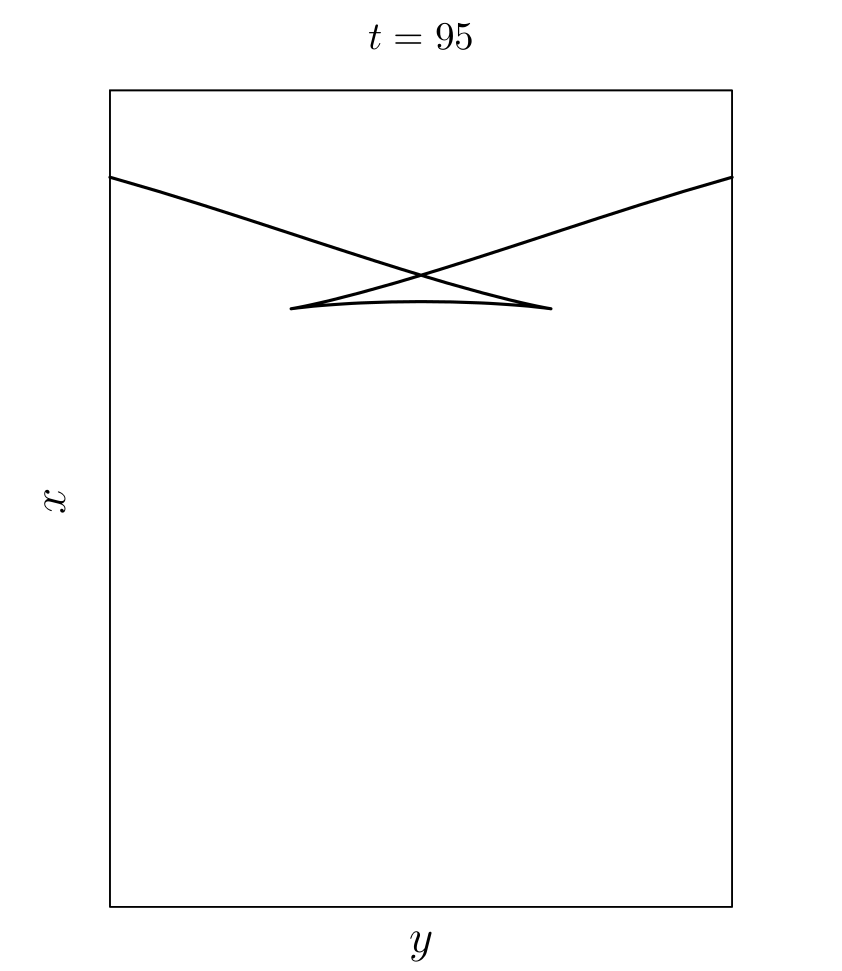}
\end{minipage}
%%%%
\caption{Formation and growth of the swallowtail in field theory (left column) and Nambu-Goto (right column). The initial amplitude of the Gaussian wiggles is $A=1.2A_{\rm threshold}$. The color palette indicates the value of the scalar field.}
\label{fig:amplitude_12}
\end{figure}

We now present the results of our numerical simulations of the full non-linear evolution of the scalar field theory. For low-amplitude initial perturbations, the two incoming wiggles traverse each other with almost no distortion. As shown in \cite{Blanco-Pillado:2022rad}, the primary effect of the interaction in the static gauge is a brief deceleration during the collision phase, while the overall shapes of the outgoing wiggles remain nearly identical to their incoming profiles. This behavior is consistent with the predictions of the Nambu–Goto approximation, confirming that in the small-amplitude regime, the field theory evolution reproduces NG dynamics with high fidelity.

As the amplitude is gradually increased, the overlap region between the colliding wiggles becomes increasingly nonlinear, leading to enhanced curvature and the onset of radiation. Nevertheless, after the interaction, the domain wall string configuration once again relaxes toward a state well described by NG dynamics.

A qualitatively different regime is entered when the wiggle amplitude exceeds a critical threshold, $A_{\text{threshold}}$, as identified in the preceding sections. In this case, the energy localized in the high-curvature region is sufficient to produce a significant burst of radiation. This signals the formation of the bifurcation cusp\footnote{We believe this explains the different behavior of the
collisions encountered in a similar situation in global strings in \cite{Drew:2023ptp}.}.

Of particular interest is the regime well above the threshold. As previously discussed, such configurations give rise to the bifurcation cusp, which subsequently evolves into two symmetric, semi-persistent cusp structures and the appearance of the swallowtail singularity structure. An illustrative example, obtained for an amplitude $A=1.2 A_{\text{threshold}}$, is shown in Fig. \ref{fig:amplitude_12}. A side-by-side comparison of the NG solution and the full field theory simulation reveals remarkable agreement, both qualitatively and quantitatively. The snapshots confirm that the nonlinear evolution of the field closely tracks the NG prediction during the cusp formation and subsequent propagation.

\begin{figure}[h!]
\centering
\includegraphics[scale=0.25]{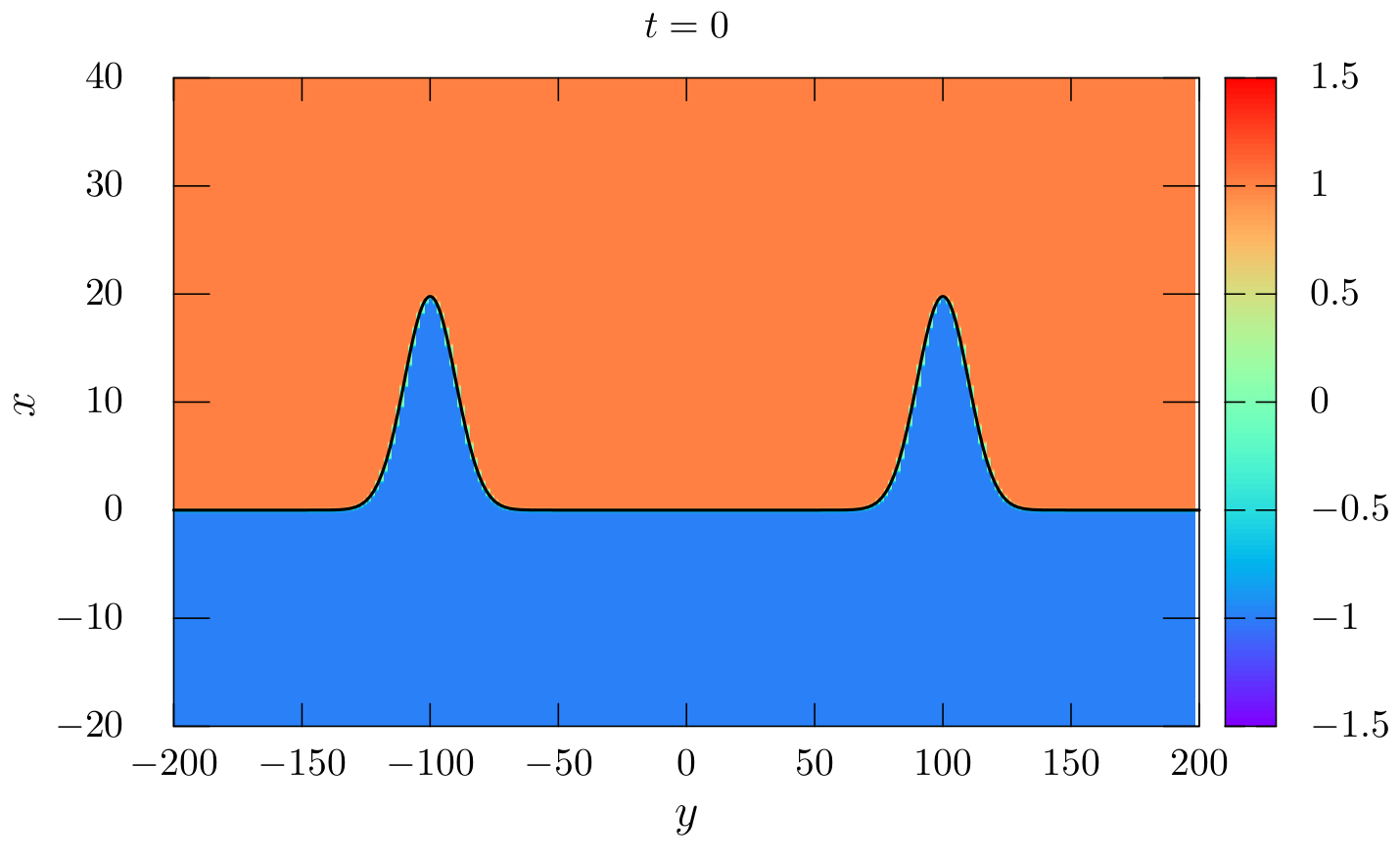}

\vspace{0.5 cm}

\includegraphics[scale=0.25]{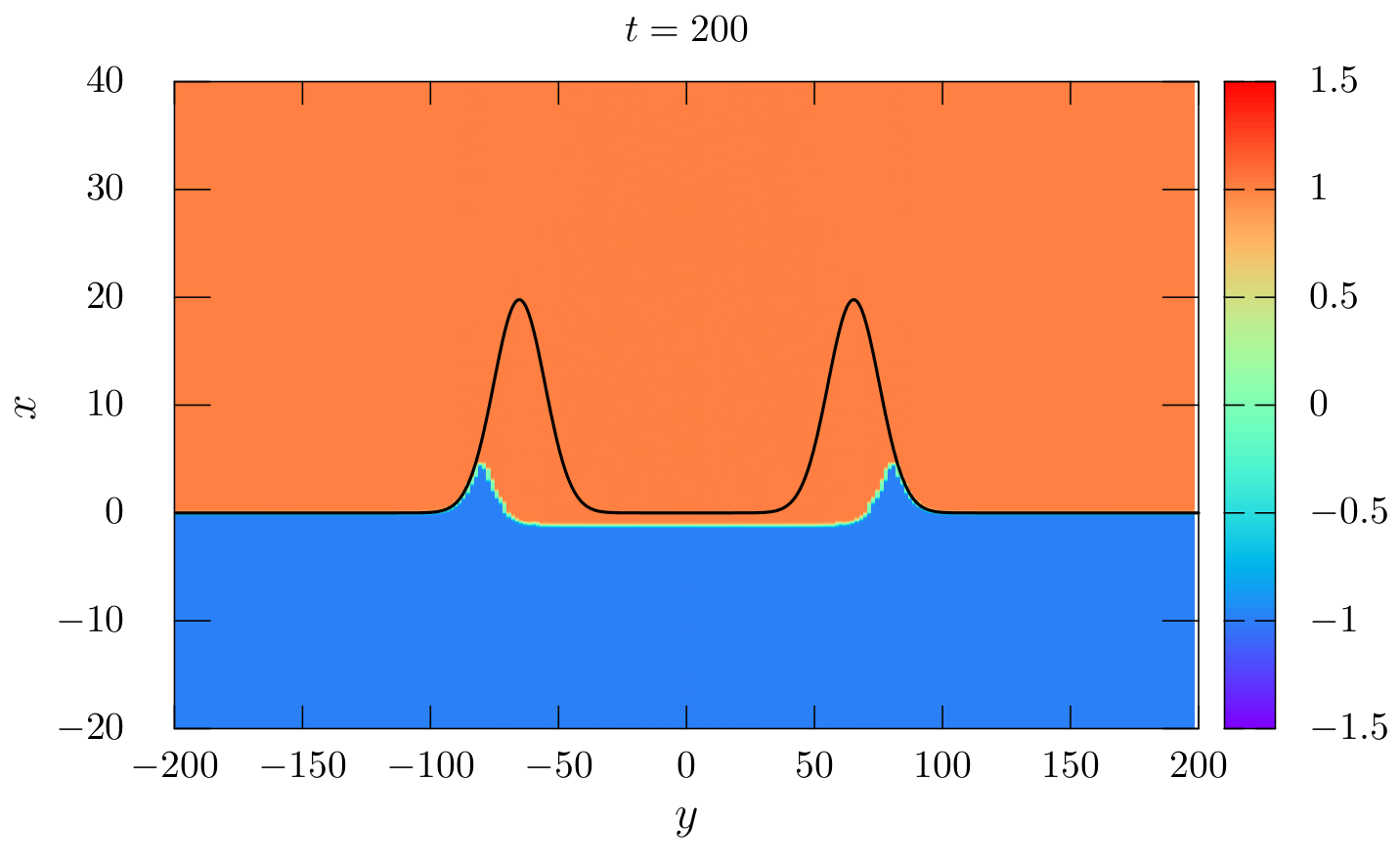}
\caption{Field theory and Nambu-Goto evolutions (the latter in black) of the collision of two Gaussian packets with amplitude $A=1.2A_{\rm threshold}$. From top to bottom: initial state and final configuration after the collision. Note that the wiggles lose a large fraction of their initial length.}
\label{fig:amplitude 12}
\end{figure}

These high-amplitude collisions are also accompanied by substantial radiation emission. A significant fraction of the wiggle energy is transferred to radiative modes, resulting in a post-collision field configuration that differs markedly from the initial conditions (see Fig. \ref{fig:amplitude 12}).

\begin{figure}[h!]
\centering
\hspace*{-1.5cm}
\includegraphics[scale=0.72]{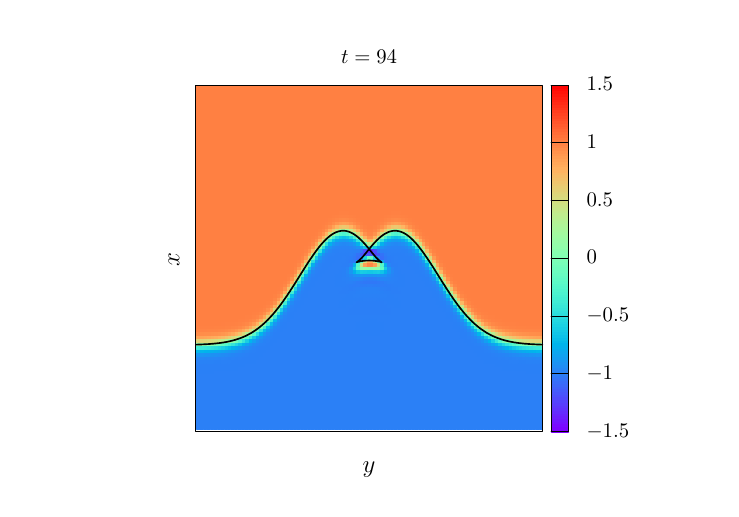}
\hspace*{-1.7cm}
\includegraphics[scale=0.72]{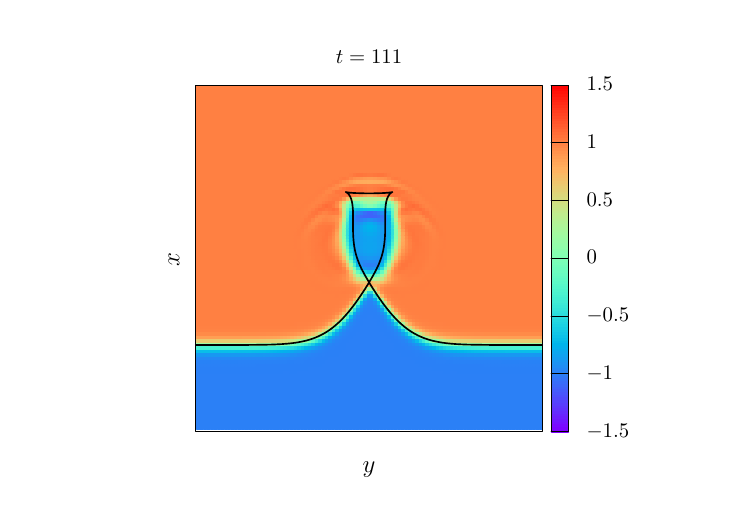}
\caption{Two snapshots of the field theory and Nambu-Goto evolutions for Gaussian wiggles of amplitude $A=1.6A_{\rm threshold}$.}
\label{fig:FTvsNG}
\end{figure}

\begin{figure}[h!]
\centering
\hspace*{-1.5cm}
\includegraphics[scale=0.72]{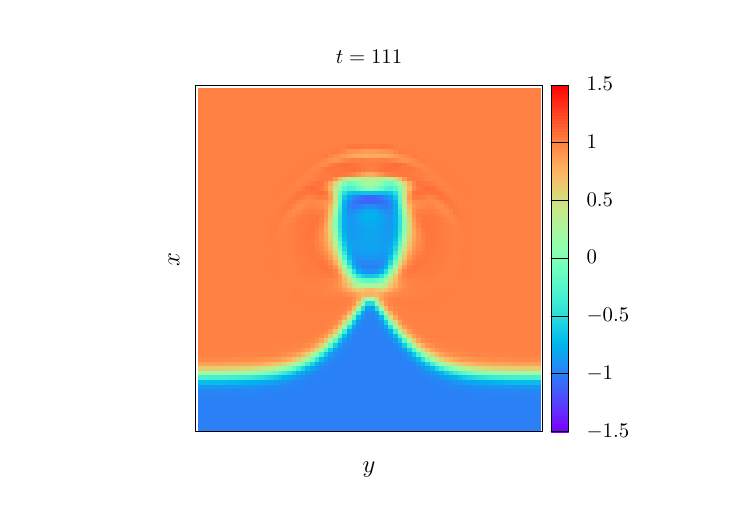}
\hspace*{-1.7cm}
\includegraphics[scale=0.72]{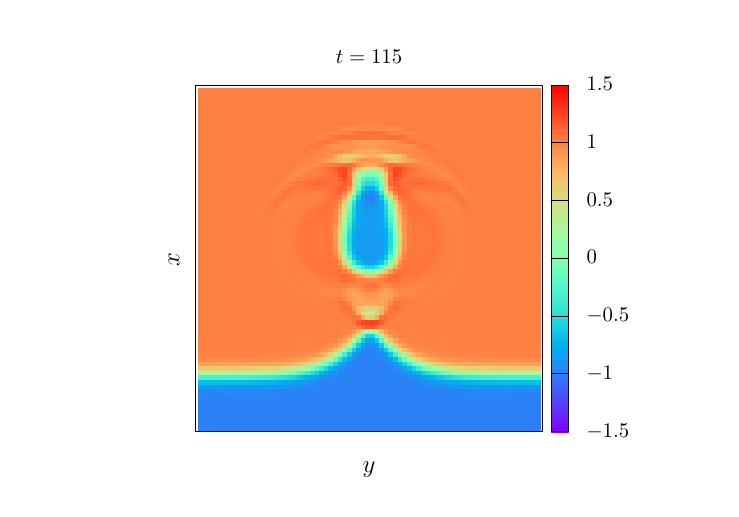}
\hspace*{-1.5cm}
\includegraphics[scale=0.72]{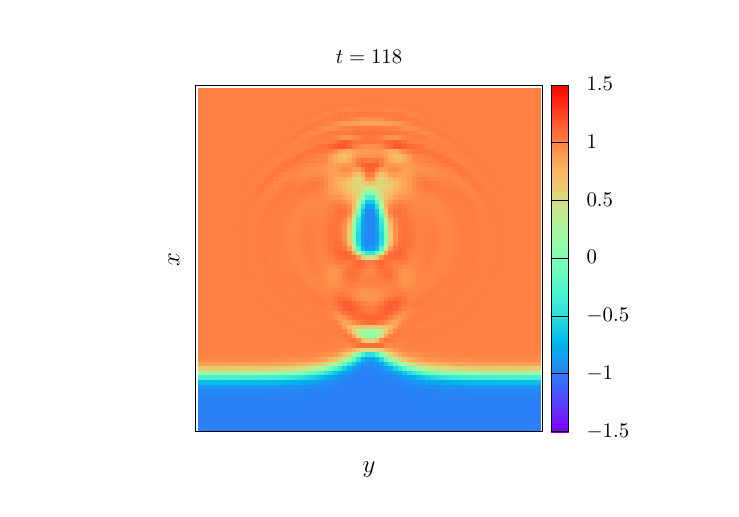}
\hspace*{-1.7cm}
\includegraphics[scale=0.72]{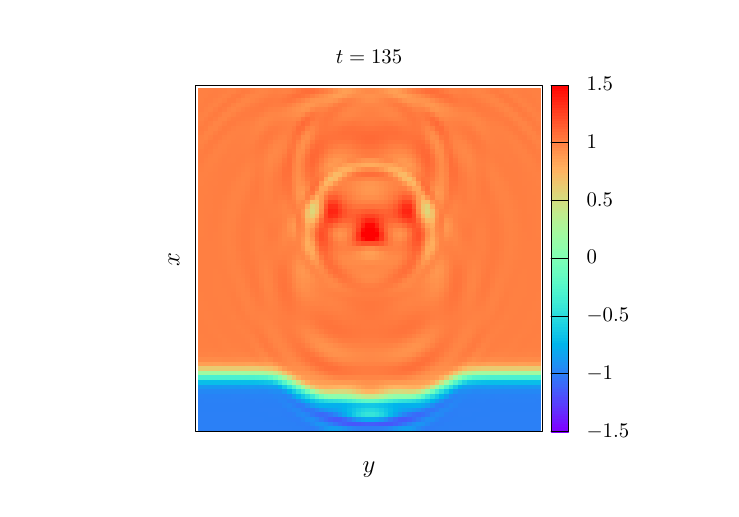}
\caption{Snapshots showing the formation of a loop in field theory and its subsequent decay into radiation.}
\label{fig:FT-loop}
\end{figure}

Finally, we consider configurations with very large initial amplitudes. In the NG framework, such conditions lead to secondary self-intersections of the domain wall string. Despite the expectation that the full field-theoretic dynamics might diverge significantly from the NG prediction in this regime, our simulations demonstrate a surprising level of agreement, both in the qualitative features and in the global behavior of the evolving configuration.
See a few snapshots of this type of simulations with $A=1.6 A_{\text{threshold}}$ in Figs. \ref{fig:FTvsNG} and \ref{fig:FT-loop}. The loop produced in this process exhibits two semi-persistent cusps and emits a substantial amount of radiation throughout its evolution, ultimately leading to its collapse.

\begin{figure}[h!]
\centering
\includegraphics[scale=0.22]{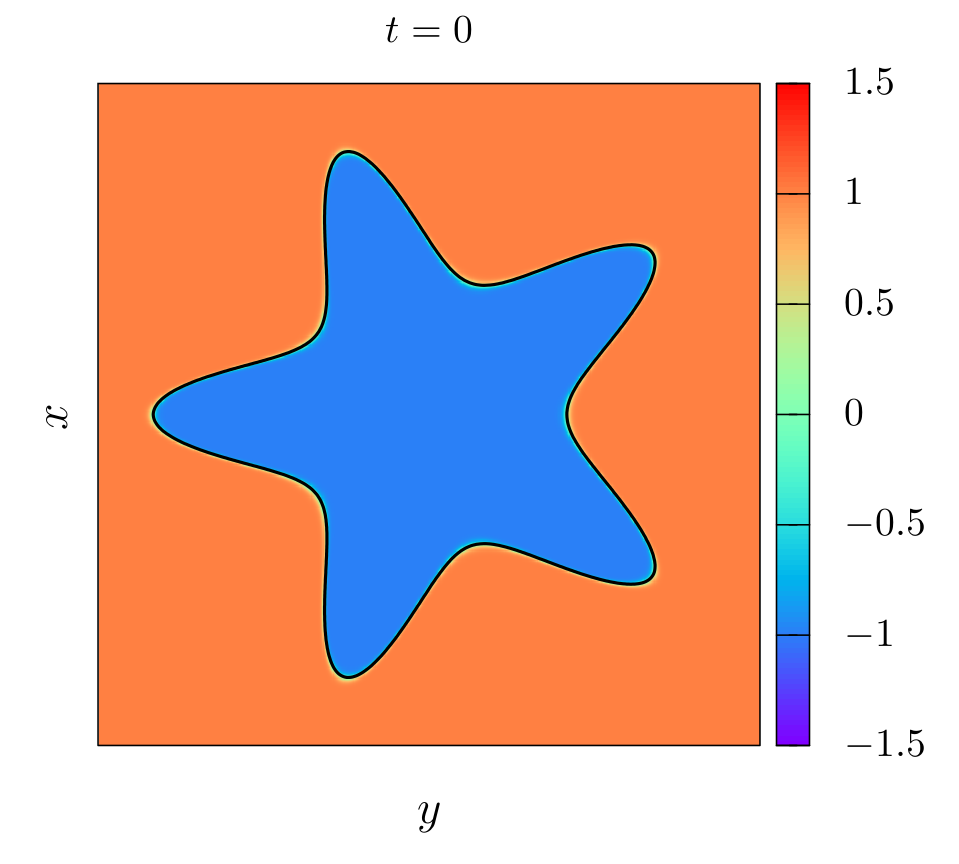}
\includegraphics[scale=0.22]{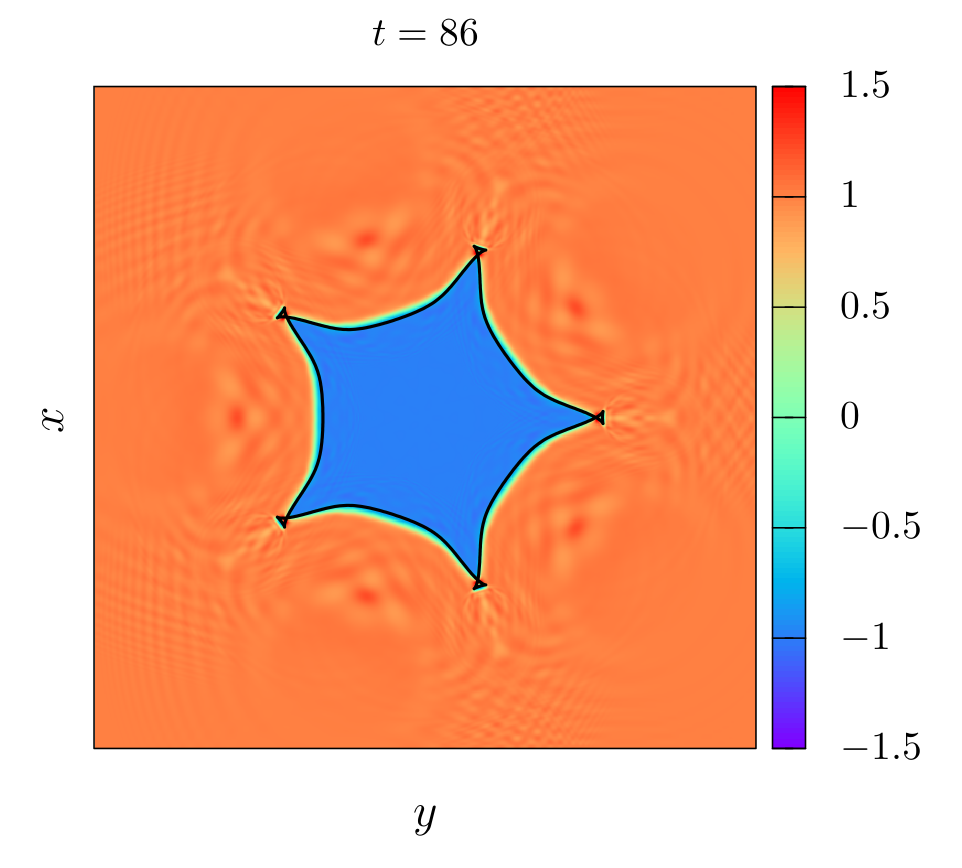}
\includegraphics[scale=0.22]{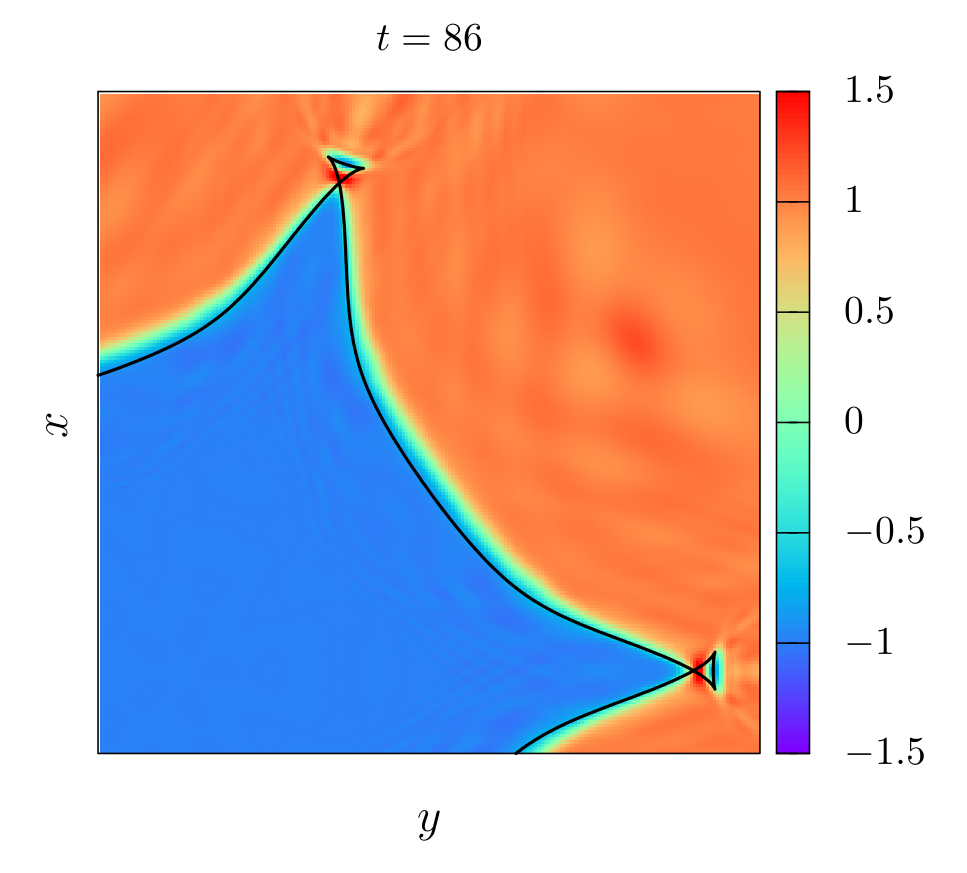}
\includegraphics[scale=0.22]{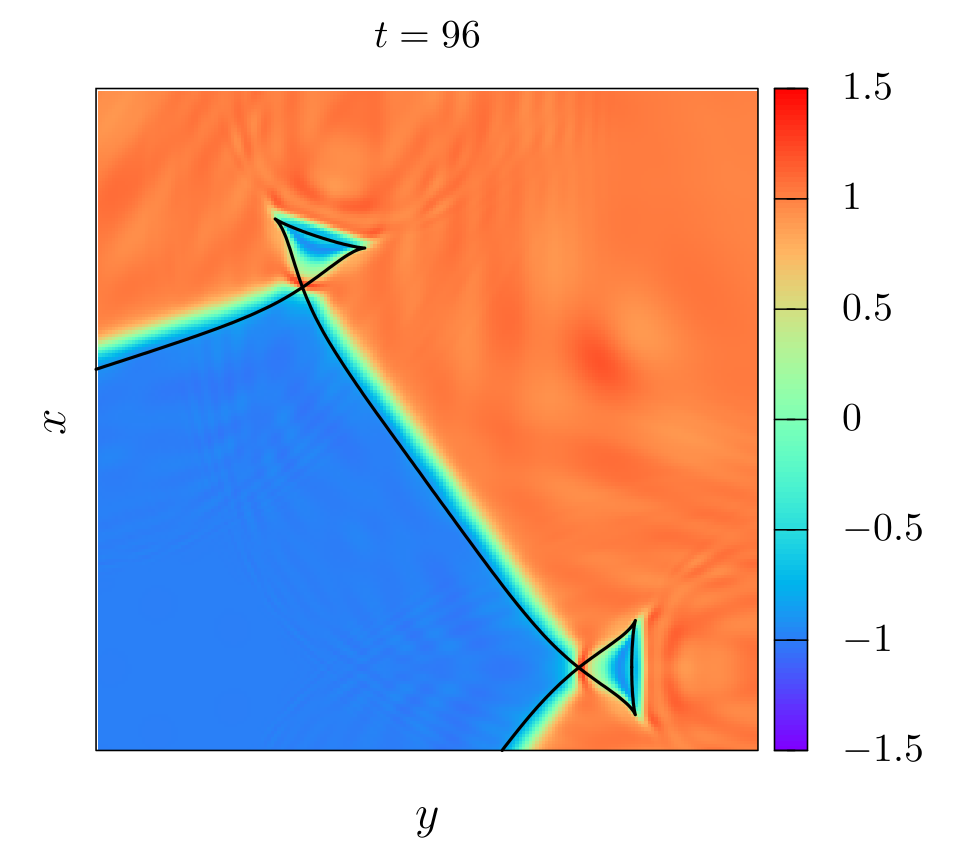}
\caption{Snapshots showing the evolution of a star-shaped loop.}
\label{fig:star-loop}
\end{figure}

As explained in detail in Appendix \ref{appendix loops}, swallowtail bifurcations can also take place during the collapse of closed domain wall strings. We illustrate this in Fig. \ref{fig:star-loop}, where we show the field theory evolution of a star-shaped loop initialized at rest with the profile (\ref{eq:star equation}) along with its NG trajectory. The shape of the loop has been chosen to be symmetric for ease of visualization, but let us emphasize that this behavior is generic. In fact, for loops that start out at rest, the appearance of swallowtail bifurcations is guaranteed if the string has inflection points at the initial time.

\clearpage

\section{Conclusions}

In this work, we have demonstrated that swallowtail bifurcations are a robust and generic
feature of the Nambu–Goto evolution equations for domain wall strings in (2+1) dimensions.
These structures emerge dynamically from smooth initial data, and their appearance is tied
to a special class of bifurcation cusps in the worldsheet geometry of the evolving string.
At such points, a single cusp undergoes a bifurcation, giving rise to two semi-persistent
cusps that subsequently propagate along the string for a while.

The local geometry of these cusps is different for each case. The bifurcation cusp exhibits
a distinctive scaling of the form $X\propto Y^{4/3}$, while the semi-persistent cusps follow
the canonical $X\propto Y^{3/2}$ behavior known from standard cusp dynamics. Importantly, we
have shown that such swallowtail catastrophes arise naturally in the collision of localized
NG wiggles on a straight domain wall, and that the local conditions on the initial
perturbations can be used to predict their formation.

Our analysis further reveals that many key features of these singular structures persist beyond
the effective NG description. Field theory simulations of scalar domain wall solitons, governed
by the full nonlinear equations of motion, exhibit clear signatures of swallowtail bifurcations
when the wall is sufficiently thin and the initial conditions are appropriately chosen reproducing the
NG initial conditions. These results provide strong evidence that the NG approximation captures
essential aspects of the nonlinear dynamics, including the emergence of cusp-driven caustics.

However, a notable distinction between the NG model and the full field theory lies in the presence
of radiation. In the field theory setting, the formation of bifurcation and semi-persistent cusps
is accompanied by the emission of scalar radiation, with the energy loss often being a large percentage
of the initial energy stored in the colliding wiggles. This radiative backreaction is absent in the
NG approximation but is crucial for understanding the physical evolution and eventual dissipation of
the string structures.

A faithful reproduction of these phenomena in field-theoretic simulations requires a substantial
dynamical range: the ratio between the scale of these structures and the width of the domain wall
must be sufficiently large to simultaneously capture the formation of swallowtails and the fine-scale dynamics at the cusps. Simulations with limited resolution may fail to resolve these features, potentially overestimating the radiative output of these models. Moreover, the resulting radiation spectrum in these scenarios may be significantly influenced by the limited numerical resolution of some numerical simulations, which prevents an accurate characterization of these small scale structures.

We have also extended our analysis to the case of loop configurations in $2+1$ dimensions,
demonstrating that swallowtail bifurcations can appear generically when loops evolve from rest under
smooth initial conditions. This suggests that such singular behavior is not limited to open or infinite
strings, but is a widespread feature of domain wall string dynamics in two spatial dimensions and it
should be widely present in domain wall network simulations in $2+1$ dimensions.

Finally, while our work here serves as a controlled toy model, it offers valuable insight into the
possible dynamics of domain walls in (3+1) dimensions, particularly those arising in cosmological
scenarios involving decaying networks. If similar singular structures form generically during the
collapse of such networks, they could have significant implications for the resulting gravitational
wave spectrum or the formation of primordial black holes. In particular, a population of high-curvature,
cusp-like events across the network may produce a hard component in the gravitational wave background,
distinct from the smoother contributions of large-scale coherent motion associated with the large
scale dynamics of the network.

It is therefore crucial to determine whether the caustic phenomena observed in the NG framework and
reproduced in field theory are genuine features of physical domain walls in the early universe, or
artifacts of idealized approximations. Further investigation in (3+1) dimensional field theory is
essential to resolve this question. We plan to report on the corresponding analysis in higher-dimensional
settings in forthcoming work \cite{3+1}.

%%%%%%%%%%%%%%%%%%%%%%%%%%%%%%%%%%%
\begin{acknowledgments}
%%%%%%%%%%%%%%%%%%%%%%%%%%%%%%%%%%%

We are grateful to Matt Elley, Francesc Ferrer, Alberto Garcia Martin-Caro, Ken D. Olum and Juan Sebastián Valbuena Bermúdez for stimulating discussions. This work has been supported in part by the PID2021-123703NB-C21 grant funded by
MCIN/AEI/10.13039/501100011033/ and by ERDF;“ A way of making Europe”; the Basque
Government grant (IT-1628-22) and the Basque Foundation for Science (IKERBASQUE).
D.~J.-A. is supported in part by National Science Foundation grant 
PHY-2419848. 
O.P. acknowledges support from the Spanish Ministry of Science and Innovation (MICINN) through the Spanish State Research Agency under the R\&D\&i project PID2023-146686NB-C31 funded by MICIU/AEI/10.13039/501100011033/ and by ERDF/EU, and under Severo Ochoa Centres of Excellence Programme 2025-2029 (CEX2024001442-S). IFAE is partially funded by the CERCA program of the Generalitat de Catalunya. 
\end{acknowledgments}

%%%%%%%%%%%%%%%%%%%%%%%%%%%%%%%%%%%%%%%%%%%%%%%%%%%%%%%%%%%%%%%%%%%%%%%%%%%%%%%
%%%%%%%%%%%%%%%%%%%%%%%%%%%%%%%%%%%%%%%%%%%%%%%%%%%%%%%%%%%%%%%%%%%%%%%%%%%%%%%

\newpage

\appendix

%%%%%%%%%%%%%%%%%%%%%%%%%%%%%%%%%%%%%%%%%%%%%%%%%%%%%%%%%%%%%%%%%%%%%%%%%%%%
\section{Swallowtail bifurcations in the collision of wiggles}
\label{appendix wiggles}

%%%%%%%%%%%%%%%%%%%%%%%%%%%%%%%%%%%%%%%%%%%%%%%%%%%%%%%%%%%%%%%%%%%%%%%%%%%%
In this Appendix we will find the conditions for cusp formation (and, consequently, for the appearance of swallowtail singularities) in the collision of two travelling waves on a straight domain wall string. For simplicity, we will consider wiggles with a single maximum (not necessarily symmetric), but we will not require them to be identical.

\begin{figure}[h!]
\begin{center}
\includegraphics[scale=1.0]{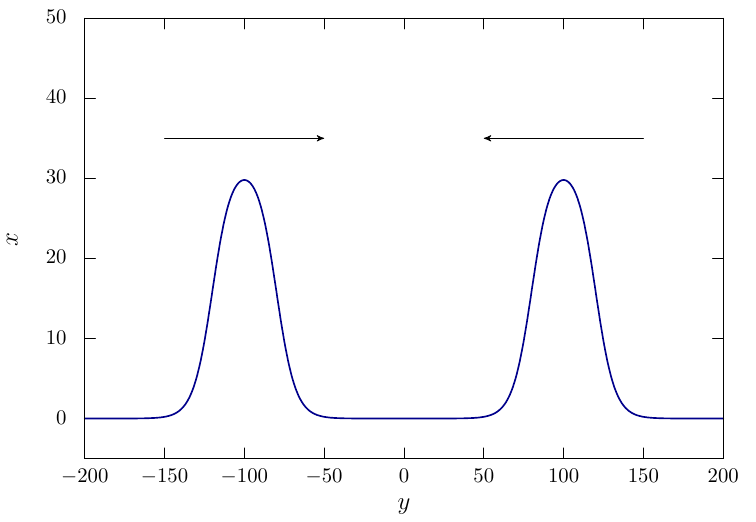}
\caption{Initial configuration for the collision of two wave packets traveling at the speed of light on a straight string.}
\label{fig:straight string 1}
\end{center}
\end{figure}

Consider the initial configuration shown in Figure \ref{fig:straight string 1}, and let the function $\psi\left(t=0,y\right)$ be the $x$ coordinate of the spatial position of the string at $t=0$. In order to find the Nambu-Goto evolution of the string, we need the right and left mover functions, $\vec{a}$ and $\vec{b}$ as functions of $\sigma$. Here they will be obtained from the unit vectors $\vec{a}\,'\left(\sigma\right)$ and $\vec{b}\,'\left(\sigma\right)$ by evaluating the derivatives of the spatial position $\vec{X}\left(t,\sigma\right)=\left(X\left(t,\sigma\right),Y\left(t,\sigma\right)\right)$ at $t=0$:

\begin{equation}
a'_{x}\left(\sigma\right)=\frac{\partial X}{\partial\sigma}\,\bigg\rvert_{t=0}-\,\,\,\frac{\partial X}{\partial t}\,\bigg\rvert_{t=0}\,,
\label{eq:ax prime straight string}
\end{equation}
\\
\begin{equation}
a'_{y}\left(\sigma\right)=\frac{\partial Y}{\partial\sigma}\,\bigg\rvert_{t=0}-\,\,\,\frac{\partial Y}{\partial t}\,\bigg\rvert_{t=0}\,,
\label{eq:ay prime straight string}
\end{equation}
\\
\begin{equation}
b'_{x}\left(\sigma\right)=\frac{\partial X}{\partial\sigma}\,\bigg\rvert_{t=0}+\,\,\,\frac{\partial X}{\partial t}\,\bigg\rvert_{t=0}\,,
\label{eq:bx prime straight string}
\end{equation}
\\
\begin{equation}
b'_{y}\left(\sigma\right)=\frac{\partial Y}{\partial\sigma}\,\bigg\rvert_{t=0}+\,\,\,\frac{\partial Y}{\partial t}\,\bigg\rvert_{t=0}\,.
\label{eq:by prime straight string}
\end{equation}
\\

\begin{figure}[h!]
\begin{center}
\includegraphics[scale=1.0]{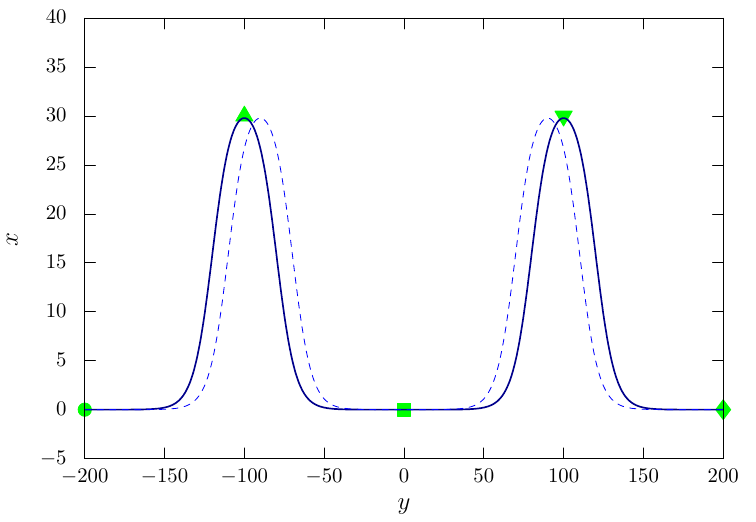}
\caption{String at $t=0$ (solid line) and $t=\Delta t$ (dashed line). The magnitude of $\Delta t$ has been exaggerated to easily extract the information we need from this plot, and the green symbols will be useful for the discussion in the main text.}
\label{fig:straight string 2}
\end{center}
\end{figure}

As detailed in the main text, it is instructive to derive the explicit relation between the functions that parametrize the position of the string in the conformal gauge and the single scalar degree of freedom employed in the static gauge, namely the function $\psi(t,y)$. To this end, we first recall that, in the conformal gauge, the spatial coordinate on the worldsheet is effectively identified with the energy per unit length. Employing the corresponding expression for the energy in the static gauge, we find that 
\begin{equation} 
Y'=\frac{\sqrt{1+(\partial_y \psi)^{2}-(\partial_t \psi)^{2}}}{1+(\partial_y \psi)^2}\,\,.
\label{eq:y prime}
\end{equation}
In order to find $X'$, we can use the conformal gauge constraint 
\begin{equation}
\left(X'\right)^{2}+\left(Y'\right)^{2}+\dot{\vec{X}}^{2}=1\,.
\label{eq:conformal gauge constraint one}
\end{equation}
Since the length of an infinitesimal string segment is $dl=dy\sqrt{1+(\partial_y \psi)^2}$, we have
\begin{equation}
d\sigma=dy\,\,\sqrt{\frac{1+(\partial_y \psi)^2}{1-\dot{\vec{X}}^{2}}}\,\,,
\label{eq:dsigma 2}
\end{equation}
so comparison of (\ref{eq:y prime}) and (\ref{eq:dsigma 2}) yields
\begin{equation}
\dot{\vec{X}}^{2}=\frac{(\partial_t \psi)^2}{1+(\partial_y \psi)^2}~.
\label{eq:velocity squared psi}
\end{equation}
Substituting (\ref{eq:y prime}) and (\ref{eq:velocity squared psi}) into (\ref{eq:conformal gauge constraint one}), we get
\begin{equation}
|X'|=|Y'|\,|\partial_y \psi|=Y\,'\,|\partial_y \psi|\,.
\label{eq:x prime}
\end{equation}
The other conformal gauge constraint is
\begin{equation}
\dot{X}X'+\dot{Y}Y'=0\,.
\label{eq:conformal gauge constraint two}
\end{equation}
Taking absolute values and using (\ref{eq:x prime}), we obtain $|\dot{Y}|=|\dot{X}|\,|\partial_y \psi|\,$, and if we combine this result with (\ref{eq:velocity squared psi}), we find
\begin{equation}
|\dot{X}|=\frac{|\partial_t \psi|}{1+(\partial_y \psi)^2}
\label{eq:x dot}
\end{equation}
and
\begin{equation}
|\dot{Y}|=\frac{|\partial_y \psi||\partial_t \psi|}{1+(\partial_y \psi)^2}\,.
\label{eq:y dot}
\end{equation}
Now we can use (\ref{eq:y prime}), (\ref{eq:x prime}), (\ref{eq:x dot}) and (\ref{eq:y dot}) to express the components of $\vec{a}\,'$ and $\vec{b}\,'$ in equations (\ref{eq:ax prime straight string}) to (\ref{eq:by prime straight string}) in terms of $\partial_y \psi$ and $\partial_t \psi$ evaluated at $t=0$. Note, however, that we need to specify the signs of $X\,'$, $\dot{X}$ and $\dot{Y}$. On the one hand, it is obvious that $X\,'$ must have the same sign as the slope of the string, which is $\partial_y \psi$. Therefore, one can safely remove the absolute values in (\ref{eq:x prime}). On the other hand, in order to determine the signs of $\dot{X}$ and $\dot{Y}$, it may be useful to visualize simultaneously the position of the string at $t=0$ and at some infinitesimal $t=\Delta t$. This is shown in Figure \ref{fig:straight string 2}. The velocity at each point on the string at $t=0$ is perpendicular to the tangent vector at that point. Then it is clear from the figure that the signs of $\dot{X}$ and $\dot{Y}$ are the following:
\begin{table}[h!]
\centering
\begin{tabular}{||c | c | c ||} 
 \hline
 & $\dot{X}$ & $\dot{Y}$ \\
 \hline\hline
$\bigcirc\rightarrow\bigtriangleup$ & - & + \\
\hline
$\bigtriangleup\rightarrow\square$ & + & + \\
\hline
$\square\rightarrow\bigtriangledown$ & + & - \\
\hline
$\bigtriangledown\rightarrow\Diamond$ & - & - \\
\hline
 \end{tabular}
\caption{Signs of $\dot{X}$ and $\dot{Y}$ in different segments of the string.}
\label{table:signs straight string}
\end{table} 

It is also clear that $\partial_t \psi$ has the same signs as $\dot{X}$, so we can also remove the absolute values in (\ref{eq:x dot}). Finally, with this information, it is straightforward to show that $\dot{Y}$ is the negative of the right-hand side of (\ref{eq:y dot}) without absolute values. All these observations allow us to rewrite equations (\ref{eq:ax prime straight string}) to (\ref{eq:by prime straight string}) as follows:\\\\ 
\begin{equation}
a'_{x}\left(\sigma\right)=\frac{(\partial_y \psi) \sqrt{1 + (\partial_y \psi)^2-(\partial_t \psi)^2}-\partial_t \psi}{1+(\partial_y \psi)^2}\,\,\Bigg\rvert_{t=0}\,,
\label{eq:ax prime psi}
\end{equation}
\\
\begin{equation}
a'_{y}\left(\sigma\right)=\frac{ \sqrt{1 + (\partial_y \psi)^2-(\partial_t \psi)^2}+(\partial_y \psi) (\partial_t \psi)}{1+(\partial_y \psi)^2}\,\,\Bigg\rvert_{t=0}\,,
\label{eq:ay prime psi}
\end{equation}
\\
\begin{equation}
b'_{x}\left(\sigma\right)=\frac{(\partial_y \psi) \sqrt{1 + (\partial_y \psi)^2-(\partial_t \psi)^2}+\partial_t \psi}{1+(\partial_y \psi)^2}\,\,\Bigg\rvert_{t=0}\,,
\label{eq:bx prime psi}
\end{equation}
\\
\begin{equation}
b'_{y}\left(\sigma\right)=\frac{ \sqrt{1 + (\partial_y \psi)^2-(\partial_t \psi)^2} - (\partial_y \psi) (\partial_t \psi)}{1+(\partial_y \psi)^2}\,\,\Bigg\rvert_{t=0}\,,
\label{eq:by prime psi}
\end{equation}
\\
where the right-hand sides depend on $\sigma$ through $y\left(\sigma\right)$, which should be found by integrating (\ref{eq:y prime}) evaluated at $t=0$. One can first find
\begin{equation}
\sigma(y)=\int_{-l/2}^{y}dy\frac{1+(\partial_{y}\psi)^{2}}{\sqrt{1+(\partial_{y}\psi)^{2}-(\partial_{t}\psi)^{2}}}\,\,\Bigg\rvert_{t=0}\,,
\label{eq:sigma of y}
\end{equation}
which can be inverted numerically to get $y(\sigma)$.

One can easily check with (\ref{eq:ax prime psi})-(\ref{eq:by prime psi}) that $\vec{a}\,'$ and $\vec{b}\,'$ are unit vectors, as required in the conformal gauge. Note that this is true for any $\psi$, so any amplitude is allowed for the wave packets.

One can ultimately use these results to find the Nambu-Goto position of the string at any time as
\begin{equation}
X\left(t,\sigma\right)=\frac{1}{2}\left[a_{x}\left(\sigma-t\right)+b_{x}\left(\sigma+t\right)\right]\,,
\label{eq:x}
\end{equation}
\begin{equation}
Y\left(t,\sigma\right)=\frac{1}{2}\left[a_{y}\left(\sigma-t\right)+b_{y}\left(\sigma+t\right)\right]\,,
\label{eq:x}
\end{equation}
with
\\
\begin{equation}
a_{x}\left(\sigma\right)=\psi\left(t=0,y\left(\sigma\right)\right)-\int_{-l/2}^{y\left(\sigma\right)} dy\frac{\partial_t \psi}{\sqrt{1+ (\partial_y \psi)^{\,2}-(\partial_t \psi)^{2}}}\,\,\Bigg\rvert_{t=0}\,,
\label{eq:ax}
\end{equation}
\begin{equation}
b_{x}\left(\sigma\right)=\psi\left(t=0,y\left(\sigma\right)\right)+\int_{-l/2}^{y\left(\sigma\right)} dy\frac{\partial_t \psi}{\sqrt{1+ (\partial_y \psi)^{\,2}-(\partial_t \psi)^{2}}}\,\,\Bigg\rvert_{t=0}\,,
\label{eq:bx}
\end{equation}
\begin{equation}
a_{y}\left(\sigma\right)=y\left(\sigma\right)+\int_{-l/2}^{y\left(\sigma\right)} dy\frac{(\partial_y \psi) (\partial_t \psi)}{\sqrt{1+ (\partial_y \psi)^{\,2}-(\partial_t \psi)^{2}}}\,\,\Bigg\rvert_{t=0}\,,
\label{eq:ay}
\end{equation}
\begin{equation}
b_{y}\left(\sigma\right)=y\left(\sigma\right)-\int_{-l/2}^{y\left(\sigma\right)} dy\frac{(\partial_y \psi) (\partial_t \psi)}{\sqrt{1+ (\partial_y \psi)^{\,2}-(\partial_t \psi)^{2}}}\,\,\Bigg\rvert_{t=0}\,.
\label{eq:by}
\end{equation}
\\
In these expressions, $l$ is the size of the box in the $y$ direction, namely, the length of the string in the absence of wave packets.

This analysis shows how one can always find the Nambu-Goto solution for any $\psi\left(t,y\right)$ giving the initial shape and velocity of the string. Note that this is true for any amplitude of the packets.

Let us find now the conditions for the formation of swallowtail bifurcation. To this end, it will be useful to take $|\partial_y \psi|=|\partial_t \psi|$ in the expressions for the components of $\vec{a}\,'$ and $\vec{b}\,'$ as an approximation, which is justified in the case of distant packets that travel at the speed of light. In this case, the components take the following form in each of the four segments of string indicated in Figure \ref{fig:straight string 2}:
\\
\begin{equation}
\bigcirc\rightarrow\bigtriangleup:
\begin{cases} 
a'_{x}=\frac{2\,|\partial_y \psi|}{1+(\partial_y \psi)^{\,2}}\,,&a'_{y}=\frac{1-(\partial_y \psi)^{\,2}}{1+(\partial_y \psi)^{\,2}}\,.\\\\
-b'_{x}=0\,,&-b'_{y}=-1\,.
\end{cases}
\label{eq:circle to triangleup}
\end{equation}
\\
\begin{equation}
\bigtriangleup\rightarrow\square:
\begin{cases} 
a'_{x}=-\frac{2\,|\partial_y \psi|}{1+(\partial_y \psi)^{\,2}}\,,&a'_{y}=\frac{1-(\partial_y \psi)^{\,2}}{1+(\partial_y \psi)^{\,2}}\,.\\\\
-b'_{x}=0\,,&-b'_{y}=-1\,.
\end{cases}
\label{eq:triangleup to square}
\end{equation}
\\
\begin{equation}
\square\rightarrow\bigtriangledown:
\begin{cases} 
a'_{x}=0\,,&a'_{y}=1\,.\\\\
-b'_{x}=-\frac{2\,|\partial_y \psi|}{1+(\partial_y \psi)^{\,2}}\,,&-b'_{y}=-\frac{1-(\partial_y \psi)^{\,2}}{1+(\partial_y \psi)^{\,2}}\,.\\\\
\end{cases}
\label{eq:square to triangledown}
\end{equation}
\\
\begin{equation}
\bigtriangledown\rightarrow\Diamond:
\begin{cases} 
a'_{x}=0\,,&a'_{y}=1\,.\\\\
-b'_{x}=\frac{2\,|\partial_y \psi|}{1+(\partial_y \psi)^{\,2}}\,,&-b'_{y}=-\frac{1-(\partial_y \psi)^{\,2}}{1+(\partial_y \psi)^{\,2}}\,.
\end{cases}
\label{eq:triangledown to diamond}
\end{equation}
\\

Since $\vec{a}\,'\left(\sigma\right)$ and $-\vec{b}\,'\left(\sigma\right)$ are unit vectors, one can draw them as curves on the surface of a unit-radius cylinder whose height corresponds to $\sigma$:
\begin{enumerate}
\item At $\bigcirc$, we have $(a'_{x},a'_{y})=\left(0,1\right)$ and $(-b'_{x},-b'_{y})=\left(0,-1\right)$, so the curves $\vec{a}\,'$ and $-\vec{b}\,'$ are at opposite points on the base of the cylinder. As we start moving away from $\bigcirc$, $a'_{x}$ increases and $a'_{y}$ decreases, while $-\vec{b}\,'$ climbs up as a straight line. At some point, $\vec{a}\,'$ turns around, because $a'_{x}$ is always positive in this part of the trajectory and at $\bigtriangleup$ we have $(a'_{x},a'_{y})=\left(0,1\right)$, which is the starting point. Whether this turnaround takes place at positive or negative $a'_{y}$ depends on the value of $\big|\partial_y \psi\big|$, i.e., on the slope of the string. If $\big|\partial_y \psi\big|>1$, $a'_{y}$ can make it to negative values. In this case, $\vec{a}\,'$ enters the semicylinder where $-\vec{b}\,'$ lies.
\item At $\bigtriangleup$, we have $(a'_{x},a'_{y})=\left(0,1\right)$. As we start moving away from $\bigtriangleup$, both $a'_{x}$ and $a'_{y}$ decrease, while $-\vec{b}\,'$ still climbs up as a straight line. As in the previous part, there must be another turnaround in $\vec{a}\,'$, because $a'_{x}$ is always negative in this part of the trajectory and at $\square$ we have $(a'_{x},a'_{y})=\left(0,1\right)$. Again, $\vec{a}\,'$ enters the semicylinder where $-\vec{b}\,'$ lies if $\big|\partial_y \psi\big|>1$ at some point in this part of the trajectory.\\\\
So far, $-\vec{b}\,'$ has been a straight line at $(-b'_{x},-b'_{y})=\left(0,-1\right)$, whereas $\vec{a}\,'$ has ``embraced'' the cylinder, with arms of equal length if the wiggle on the left is symmetric. In any case, it has touched $-\vec{b}\,'$'s semicylinder if $\big|\partial_y \psi\big|>1$ at some point. In the following parts of the trajectory, $\vec{a}\,'$ remains at $(a'_{x},a'_{y})=\left(0,1\right)$ and it is only $-\vec{b}\,'$ that moves.\\
\item As we start moving away from $\square$, $-b'_{x}$ decreases and $-b'_{y}$ increases. At some point, $-\vec{b}\,'$ turns around, because $-b'_{x}$ is always negative in this part of the trajectory and at $\bigtriangledown$ we have $(-b'_{x},-b'_{y})=\left(0,-1\right)$, which is the starting point. This arm has reached $\vec{a}\,'$'s semicylinder if $\big|\partial_y \psi\big|>1$ at some point, in which case $-b'_{y}$ has become positive.
\item At $\bigtriangledown$, we have $(-b'_{x},-b'_{y})=\left(0,-1\right)$. As we start moving away from $\bigtriangledown$, both $-b'_{x}$ and $-b'_{y}$ increase. Once again, $-\vec{b}\,'$ turns around at some point, because $-b'_{x}$ is always positive in this part of the trajectory and at $\Diamond$ we have $(-b'_{x},-b'_{y})=\left(0,-1\right)$. The curve $-\vec{b}\,'$ has entered the semicylinder where $\vec{a}\,'$ lies if $\big|\partial_y \psi\big|>1$ at some point in this part of the trajectory.\\
\end{enumerate}
The picture that emerges from this description is that $\vec{a}\,'$ embraces the lower part of the cylinder, while $-\vec{b}\,'$ embraces the upper part of it. Moreover, the right arm of both curves is higher than the left arm. As time passes, the curve $\vec{a}\,'$ rises and the curve $-\vec{b}\,'$ goes down. Swallowtail bifurcations will occur if the two curves intersect (recall that the condition for cusp formation is $\vec{a}\,'=-\vec{b}\,'$). This will happen in any of these three cases: (i) the left arm of $\vec{a}\,'$ and the right arm of $-\vec{b}\,'$ touch, (ii) the right arm of $\vec{a}\,'$ and the left arm of $-\vec{b}\,'$ touch, (iii) the two previous conditions are satisfied. Note that swallowtails cannot be formed if $\big|\partial_y \psi\big|$ is smaller than 1 for every $y$, because in this case the arms of $\vec{a}\,'$ and $-\vec{b}\,'$ are too short. 

Let $y_{1,2,3,4}$ be the $y$ coordinates corresponding to the inflection points of the string in the regions $\bigcirc\rightarrow\bigtriangleup$, $\bigtriangleup\rightarrow\square$, $\square\rightarrow\bigtriangledown$ and $\bigtriangledown\rightarrow\Diamond$, respectively. These are the points where the turnarounds take place ($a''_{y}=0$), so they determine the maximum extension of the arms. Using (\ref{eq:circle to triangleup})-(\ref{eq:triangledown to diamond}), conditions (i) and (ii) above can be expressed mathematically as follows:\\

(i) $a'_{y}(y_{1})<-b'_{y}(y_{4})$, which implies 
\begin{equation}
|\partial_y \psi(0,y_{1})|\times|\partial_y \psi(0,y_{4})|>1.\\
\label{eq:condition swallowtail formation 1}
\end{equation}

\begin{figure}[h!]
\begin{center}
\includegraphics[scale=1.1]{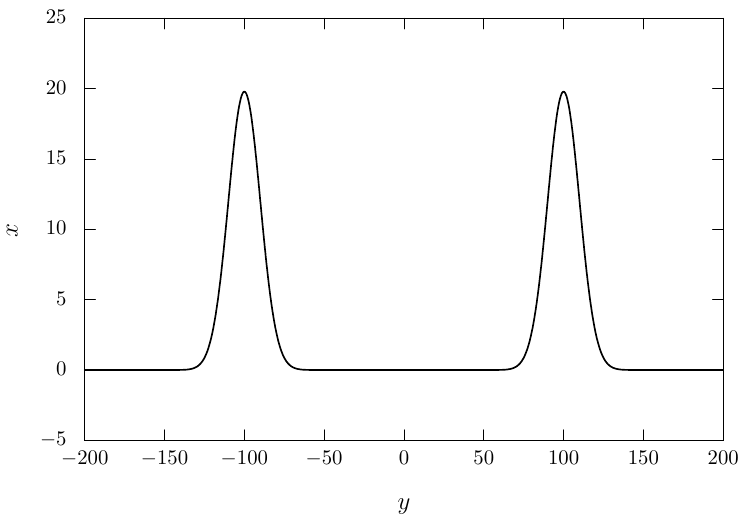}
\includegraphics[scale=1.4]{cylinder_Vachaspati_t0.pdf}
\caption{Upper panel: colliding  wiggles on a straight string at $t=0$. Bottom panel: curves defined by $\vec{a}\,'\left(\sigma-t\right)$ (red) and $-\vec{b}\,'\left(\sigma+t\right)$ (purple) on the surface of the unit-radius cylinder at $t=0$. $L$ denotes the invariant length of the string.}
\label{fig:straight example 1}
\end{center}
\end{figure}

\begin{figure}[h!]
\begin{center}
\includegraphics[scale=1.1]{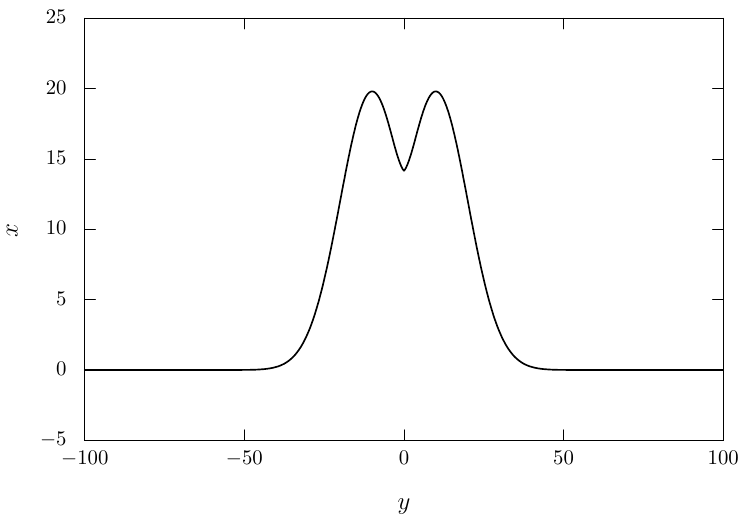}
\includegraphics[scale=1.4]{cylinder_Vachaspati_t89d5.pdf}
\caption{Upper panel: colliding wiggles on a straight string at $t\approx90$. Bottom panel: curves defined by $\vec{a}\,'\left(\sigma-t\right)$ (red) and $-\vec{b}\,'\left(\sigma+t\right)$ (purple) on the surface of the unit-radius cylinder at $t\approx90$. The right arm of $\vec{a}\,'\left(\sigma-t\right)$ and the left arm of $-\vec{b}\,'\left(\sigma+t\right)$ are tangent at a single point.}
\label{fig:straight example 2}
\end{center}
\end{figure}

\begin{figure}[h!]
\begin{center}
\includegraphics[scale=1.1]{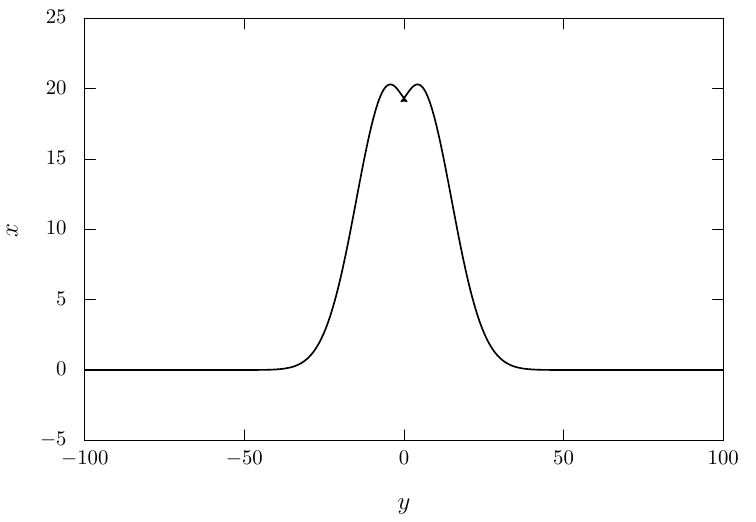}
\includegraphics[scale=1.4]{cylinder_Vachaspati_t95.pdf}
\caption{Upper panel: colliding  wiggles on a straight string at $t=95$. Bottom panel: curves defined by $\vec{a}\,'\left(\sigma-t\right)$ (red) and $-\vec{b}\,'\left(\sigma+t\right)$ (purple) on the surface of the unit-radius cylinder at $t=95$.}
\label{fig:straight example 3}
\end{center}
\end{figure}

(ii) $a'_{y}(y_{2})<-b'_{y}(y_{3})$, which implies
\begin{equation}
|\partial_y \psi(0,y_{2})|\times|\partial_y \psi(0,y_{3})|>1.\\\\
\label{eq:condition swallowtail formation 2}
\end{equation}
In Figures \ref{fig:straight example 1}, \ref{fig:straight example 2} and \ref{fig:straight example 3} we show some snapshots of a collision of Vachaspati waves where swallowtail bifurcations occur, along with the representation of $\vec{a}\,'$ and $-\vec{b}\,'$ on the surface of the unit-radius cylinder.

\clearpage

%%%%%%%%%%%%%%%%%%%%%%%%%%%%%%%%%%%%%%%%%%%%%%%%%%%%%%%%%%%%%%%%%%%%%%%%%%%
\section{Angular description of the swallowtail bifurcation}
\label{appendix angles}
%%%%%%%%%%%%%%%%%%%%%%%%%%%%%%%%%%%%%%%%%%%%%%%%%%%%%%%%%%%%%%%%%%%%%%%%%%%%

The swallowtail singularity arises as a direct consequence of the formation of a bifurcation cusp. In this Appendix, we demonstrate that, under the conditions required for cusp formation, the absence of a third spatial dimension renders the emergence of the swallowtail singularity inevitable. As shown in the main text, this result holds provided that the relation $\alpha_{2}=-\beta_{2}$ in Eq. (\ref{eq:ab prime prime}) is satisfied, corresponding to an initial cusp with vanishing second derivative. We further establish that this condition is generically fulfilled in collisions involving wiggles of sufficiently large amplitude. As discussed in Sec. \ref{sec:cusps}, this means that the resulting cusp is characterized by the scaling relation $X\propto Y^{4/3}$. In contrast, the daughter cusps formed subsequently satisfy $\alpha_{2}\neq-\beta_{2}$  , leading instead to the scaling behavior $X\propto Y^{3/2}$.

In order to prove these statements, let us start by assuming that the functions $\vec{a}\,'(\sigma)$ and $\vec{b}\,'(\sigma)$ are continuous (in other words, we do not consider here the presence of kinks). Then, since these are unit vectors, they can be represented in terms of a continuous angular variable as follows\footnote{Again, the $x$ axis is chosen to be aligned with the velocity of the cusp, assuming it forms.}:
\begin{equation}
\vec{a}\,'(\sigma)= \left(\sin[\theta_{a}(\sigma)],\cos[\theta_{a}(\sigma)]\right)\,,
\label{eq:aprime angle}
\end{equation}
\begin{equation}
\vec{b}\,'(\sigma)=(\sin[\theta_{b}(\sigma)],\cos[\theta_{b}(\sigma)])\,.
\label{eq:bprime angle}
\end{equation}
Let us now assume that we want to describe a straight string with propagating wiggles on it. In this case, the functions $\vec{a}\,'(\sigma)$ and $\vec{b}\,'(\sigma)$ should tend asymptotically to a constant (see equations (\ref{eq:ax prime straight string}) to (\ref{eq:by prime straight string})). Let $c_{1}$ and $c_{2}$ be the corresponding constant components, which depend on the angle between the asymptotic string and the velocity of the cusp:
\begin{equation}
\lim_{\sigma\rightarrow\pm\infty}\vec{a}\,'=\lim_{\sigma\rightarrow\pm\infty}\vec{b}\,'=(c_{1},c_{2})\,.
\label{eq:limit abprime}
\end{equation}
Then, according to (\ref{eq:aprime angle}) and (\ref{eq:bprime angle}), we need $\sin(\theta_{a})=\sin(\theta_{b})=c_{1}$ and $\cos(\theta_{a})=\cos(\theta_{b})=c_{2}$ in the asymptotic limit. This means that
\begin{equation}
\lim_{\sigma\rightarrow\pm\infty}\theta_{a}=c_{3}
\label{eq:limit theta a}
\end{equation}
and
\begin{equation}
\lim_{\sigma\rightarrow\pm\infty}\theta_{b}=c_{3}+2\pi
\label{eq:limit theta b}
\end{equation}
for some constant $c_{3}$. 
Now, combining the cusp condition (\ref{eq:general condition for cusp formation}) with (\ref{eq:aprime angle}) and (\ref{eq:bprime angle}), we see that
\begin{equation}
\theta_{a}(\sigma_{*}-t_{*})=\theta_{b}(\sigma_{*}+t_{*})-\pi\equiv\tilde{\theta}_{b}(\sigma_{*}+t_{*})\,.
\label{eq:theta a theta b}
\end{equation}
If we define the angle difference
\begin{equation}
\Delta=\tilde{\theta_{b}}-\theta_{a}\,,
\label{eq:Delta}
\end{equation}
the condition for cusp formation is $\Delta=0$. Since
\begin{equation}
\lim_{\sigma\rightarrow\pm\infty}\Delta=\pi\,,
\label{eq:limit Delta}
\end{equation}
this function will cross zero an even number of times. The only exception to this rule is the case where the function is tangent to zero. 

At the cusp event we have
\begin{equation}
\vec{a}\,'(\sigma_{*}-t_{*})=(-1,0)
\label{eq:aprime cusp event}
\end{equation}
and
\begin{equation}
\vec{b}\,'(\sigma_{*}+t_{*})=(1,0)\,,
\label{eq:aprime cusp event}
\end{equation}
and so $\theta_{a}=-\pi/2$ and $\theta_{b}=\pi/2$, which translates into
\begin{equation}
\tilde{\theta}_{b_{\,*}}=-\pi/2\,,
\label{eq:tilde theta b cusp}
\end{equation}
\begin{equation}
\Delta(t_{*},\sigma_{*})=0\,,
\label{eq:Delta cusp}
\end{equation}
as required in order to have a cusp.

\begin{figure}[h!]
\centering
\includegraphics[scale=0.7]{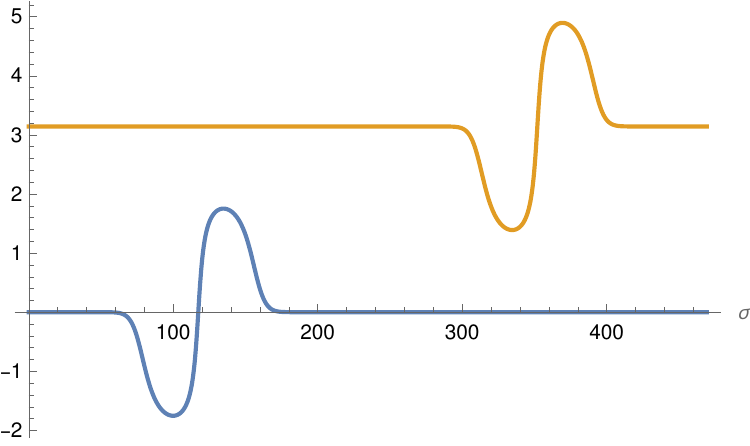}
%\vspace{1cm}
\includegraphics[scale=0.7]{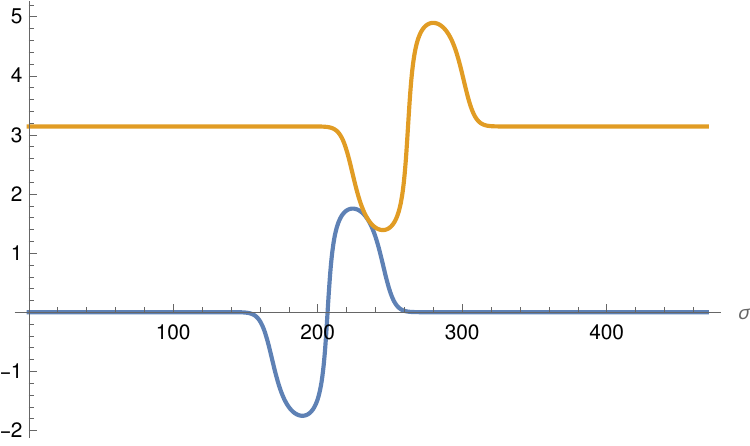}
%\vspace{1cm}
\includegraphics[scale=0.7]{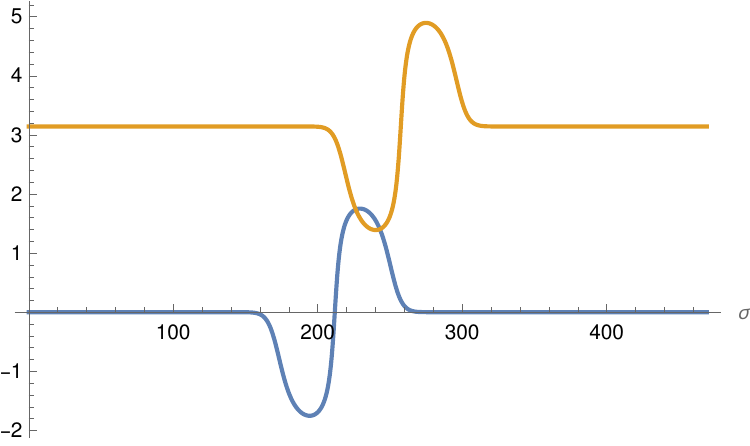}
\caption{Angular functions $\theta_{a}(\sigma)$ (in blue) and $\tilde{\theta}_{b}(\sigma)$ (in orange) at three different times for symmetric Gaussian wiggles colliding on a string. After meeting first at a single point (central panel) where the functions are tangent, the curves intersect at two different points for a while (lower panel). This corresponds to the birth and growth of the swallowtail.}
\label{fig:angles}
\end{figure}
On the other hand, also at the cusp event, the second derivatives of the right and left movers read
\begin{equation}
\vec{a}\,''_{*}=\theta_{a}'(\sigma_{*}-t_{*})(0,1)\,,
\label{eq:second derivative a cusp}
\end{equation}
\begin{equation}
\vec{b}\,''_{*}=\theta_{b}'(\sigma_{*}+t_{*})(0,-1)\,.
\label{eq:second derivative a cusp}
\end{equation}
We want to prove that $\alpha_{2}=-\beta_{2}$ in equation (\ref{eq:ab prime prime}), or in other words, that $\theta_{a}'(\sigma_{*}-t_{*})=\theta_{b}'(\sigma_{*}+t_{*})$. Note that this condition is equivalent to
\begin{equation}
\Delta'(t_{*},\sigma_{*})=0\,.
\label{eq:condition vanishing acceleration}
\end{equation}
According to (\ref{eq:limit theta a}), (\ref{eq:limit theta b}) and (\ref{eq:theta a theta b}), $\theta_{a}(\sigma)$ and $\tilde{\theta}_{b}(\sigma)$ are localized around some constant $c_{3}$ and $c_{3}+\pi$, respectively. As time goes by, $\theta_{a}$ will shift to the right and $\tilde{\theta}_{b}$ to the left, and if their amplitudes are big enough (at least $\pi/2$), the curves will meet at a single point $\sigma_{*}$ at time $t_{*}$. At this moment, $\Delta(t_{*},\sigma_{*})=0$ only at a single value of $\sigma$, which is $\sigma_{*}$. As justified above, the boundary conditions for $\Delta$ imply that this function must be zero an even number of times unless $\Delta$ touches zero tangentially. Since at this particular time $\Delta=0$ only at $\sigma=\sigma_{*}$, we necessarily have $\Delta'(t_{*},\sigma_{*})=0$, and so $\alpha_{2}=-\beta_{2}$. This proves that the acceleration of the string vanishes at the event of the first cusp. Immediately after, the curves $\theta_{a}$ and $\tilde{\theta_{b}}$ will unavoidably intersect at two values of $\sigma$, provided that their amplitudes are greater than $\pi/2$ (see Figure \ref{fig:angles}). This means that the initial cusp has split into two cusps, which will be permanent for a finite range of time. For these two daughter cusps, we have $\Delta'\neq 0$, which means that $\alpha_{2}\neq-\beta_{2}$.

This analysis confirms that, for generic perturbations of the straight domain-wall string, the criteria for the formation of bifurcation cusps are indeed fulfilled.

%%%%%%%%%%%%%%%%%%%%%%%%%%%%%%%%%%%%%%%%%%%%%%%%%%%%%%%%%%%%%%%%%%%%%%%%%%%%
\section{Computing the near cusp coefficients for symmetric wiggles}
\label{appendix identical-wiggles}
%%%%%%%%%%%%%%%%%%%%%%%%%%%%%%%%%%%%%%%%%%%%%%%%%%%%%%%%%%%%%%%%%%%%%%%%%%%%

In the Nambu-Goto approximation it is possible to trace the formation of singularities on the worldsheet to the initial  conditions. It is interesting then to examine which is the precise property of the initial condition that leads to singularities. This is the purpose of this Appendix, focusing on the example of the simplest type of wiggle collisions, assuming identical and shape-symmetric wiggles. 

This is readily done by assuming that the worldsheet has a singularity with a local expansion like in \ref{sec:cusps}. For identical and symmetric wiggles this pinpoints some special points in the initial condition. Moreover, it will be possible to express the local parameters in the singularity expansion to the wiggle properties on these pinpoints. To avoid clutter in this Appendix, we will use the following more compact notation:  
\begin{equation}
\psi_0(y)\equiv \psi(t=0,y)~\, ,
\qquad\psi_0'(y)\equiv \partial_y\psi(0,y)    \,
,\qquad\dot\psi_0(y)\equiv \partial_t\psi(t,y)|_{t=0}    \,.
\end{equation}

Recall that we label the worldsheet parameters on the singularity as $(\sigma_*,t_*)$, corresponding to Cartesian coordinate along the string $y_*$. With no loss of generality we set $\sigma_*=y_*=0$ so the only remaining parameter is the cusp formation time $t_{*}$, that scales like half the initial separation but does not exactly coincide with it. The symmetry in the collision implies $\psi_0(-y)=\psi_0(y)$, $\psi_0'(-y)=-\psi_0'(y)$, $\dot\psi_0(-y)=\dot\psi_0(y)$ etc. Asymptotically in the past of the collision the initial state is very well approximated by isolated left- or right-moving wave, so around either wiggle the initial condition satifies 
\begin{equation}\label{eq near wiggles}
    \dot\psi_0(y)= {\rm sign}(y)\;\psi_0'(y) 
\end{equation}
with arbitrarily high accuracy.

As shown in Appendix \ref{appendix wiggles} it is straightforward to obtain the full time evolution since $\psi_0$ and its derivatives already fix the form of the left- and right-mover functions $\vec{a}$, $\vec{b}$, see (\ref{eq:ax prime psi} - \ref{eq:by prime psi}). Time evolution proceeds by shifting the arguments of the functions as $\vec{a}(\sigma-t)$ and $\vec{b}(\sigma+t)$.
In order to identify the singular point(s) we only need to impose the cusp condition \eqref{eq:general condition for cusp formation}. This involves only $\vec{a}'$ and $\vec{b}'$ which are directly expressed in terms of $\dot\psi_0$ and $\psi_0'$. Since there is a one-to-one relation between the argument of $\vec{a},\,\vec{b}$ and the Cartesian coordinate (see \eqref{eq:sigma of y}), the cusp condition translates directly into an equation for some special points determined by the initial data, which we denote collectively as  $y_c$. 
These points are at $(\sigma_*,t_*)=(0,t_*)$ so they correspond to evaluating  the argument of $\vec{a}$ ($\vec{b}$) at $-t_*$ ($t_*$) respectively. This is equivalent to a shift in the initial condition by $\pm t_*$ from the center of symmetry; therefore, the points $y_c$ must be around the wiggle locations. Plugging (\ref{eq:ax prime psi} - \ref{eq:by prime psi}) into \eqref{eq:ab prime} and using \eqref{eq near wiggles}, one easily arrives at
\begin{equation}
\psi_0'\big|_{y=y_c}=\pm \, \dot\psi_0(y_c)=\pm1\, .
\label{eq:conditions for yc}
\end{equation}
In other words, points in the initial condition with slope $\pm 1$ will give rise to swallowtail bifurcations. Notice that in the asymptotic limit large initial separation $y_c$ essentially depends only on the wiggle shape $\psi_\text{wiggle}(z)$ and \eqref{eq:conditions for yc} reduces to
\begin{equation}
|\,\psi_\text{wiggle}'(z)\,|=1\, .
\label{eq: yc wiggle}
\end{equation}
This gives a sharp criterion for the formation of cusps in the collision of (symmetric) wiggles: the slope in the wiggle profile must exceed unity. 

\begin{figure}[t]
\begin{center}
\includegraphics[scale=0.37]{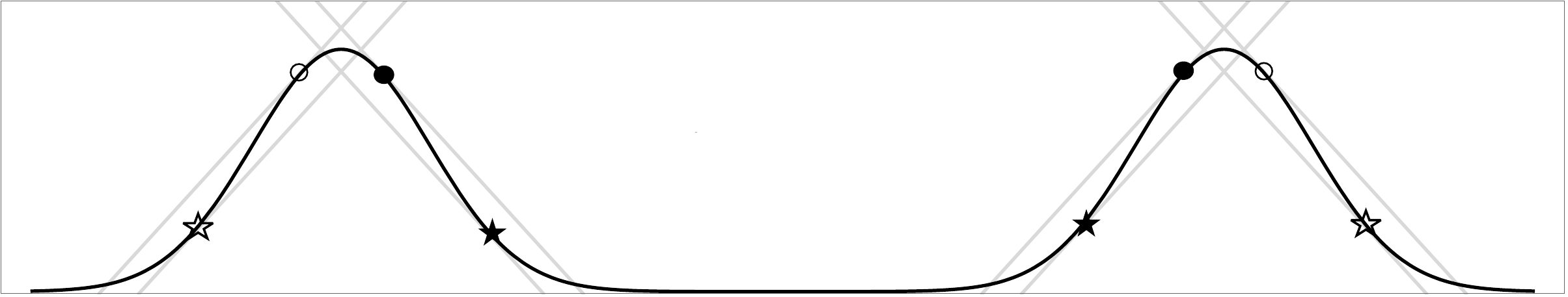}
\caption{Sketch of an initial condition $\psi_0(y)=\psi(t=0,y)$ for the collision of two identical wiggles with large amplitude: the slope is larger than $1$ in magnitude in two intervals, between star and circle, on either side of the peak. Gray lines have slope $\pm1$ and points tangent to them give rise to swallowtail singularities. The inner ones (filled star and circle) are resolvable in field theory simulations. }
\label{fig:caustic points}
\end{center}
\end{figure}

We show in Fig.~\ref{fig:caustic points} the caustic or bifurcation points $y_c$ resulting from \eqref{eq:conditions for yc} in a symmetric collision with large enough amplitude (slope) wiggles. (These are not to be confused with the $y_{1,\,2,\,3,\,4}$ introduced in Appendix \ref{appendix wiggles}, which correspond to the inflection points of $\psi_0$.)
There is a total of 8 roots satisfying  \eqref{eq: yc wiggle} for Gaussian-like wiggles and twice as many roots for \eqref{eq:conditions for yc}, related by $y\to-y$ symmetry. Let us denote collectively the 4 negative roots as $-y_c$. The inner ones (filled star and circle) are physical in the sense that are also reproduced in the field theory evolution. The star corresponds to appearance of the swallowtail bifurcation. The filled dot corresponds to an inverse swallowtail cusp where the bifurcation closes back into a smooth worldsheet. The white-filled circle and star correspond to a time-reversed repetition of the same which is forced by the symmetry in the collision and occurs only in the NG approximation.

Next, we can find the coefficients $\alpha_{n}$, $\beta_{n}$ of the near singularity expansion for the particular case of the  wiggle collisions \eqref{eq:Gaussian wiggles}. By definition, these coefficients reduce to the derivatives of $a'_{y}(\sigma)$, $b'_{y}(\sigma)$ evaluated at $(\sigma_*, t_*)$ (see {\em e.g.} \eqref{eq:ab prime prime}, \eqref{eq:ab prime prime prime} and \eqref{eq:ab prime prime prime prime}), and are readily expressed in terms of the (derivatives of the) initial condition $\psi_0$  using equations \eqref{eq:ax prime psi} to \eqref{eq:by prime psi}  
evaluated at $\pm y_c$.

Using  (\ref{eq:ax prime psi} - \ref{eq:by prime psi}), the chain rule, Eq. \eqref{eq:y prime}, \eqref{eq:conditions for yc} and \eqref{eq near wiggles}  we obtain
\begin{align}\label{eq:alpha 2}
\alpha_{2}=
a''_{y}(\sigma_*-t_*)=
\frac{dy}{d\sigma}\frac{da'_{y}}{dy}\Biggr|_{-y_{c}}&=\frac{\sqrt{1+(\psi_0')^{2}-(\dot\psi_0)^2}}{1+(\psi_0')^{2}}\frac{da'_{y}}{dy}\Biggr|_{-y_{c}}\\\nonumber
&=-\frac{1}{2}\psi_0'\;\psi_0''\Big|_{-y_{c}}
=\frac{1}{2}\psi_0''\Big|_{-y_{c}}\,.
\end{align}
where the last step holds for the inner roots (solid star and circle).
Similarly, 
\begin{equation}
\beta_{2}=b''_{y}(\sigma_*+t_*)=\frac{db'_{y}}{dy}\frac{dy}{d\sigma}\,\Bigg\rvert_{y_{c}}
=-\frac{1}{2}\psi_0'\,\psi_0''\Big|_{y_{c}} = -\alpha_2.
\label{eq:beta 2}
\end{equation}
where the last step following because $\psi_0''$ is even under $y\to- y$. Notice that this implies that the singularity must be of swallowtail bifurcation type. This might seem surprising since we have only imposed the condition of cusp \eqref{eq:ab prime}, however it is a consequence of the symmetry: semi-persistent cusps are formed but not at $\sigma=y=0$.

For the higher derivatives, one  finds
\begin{align}
\alpha_{3}&=a'''_{y}\big|_{-y_{c}}=-\frac{1}{4}
\left(\psi_0'\psi_0''' - \big(\psi_0''\big)^2 \right)\Big\rvert_{-y_{c}}=\frac{1}{4}
\left(\psi_0''' + \big(\psi_0''\big)^2 \right)\Big\rvert_{-y_{c}}\,,    \\
    \beta_{3}&=b'''_{y}\big|_{y_{c}}=-\frac{1}{4}
\left(\psi_0'\psi_0''' - \big(\psi_0''\big)^2 \right)\Big\rvert_{y=y_{c}}
=\frac{1}{4}
\left(-\psi_0''' + \big(\psi_0''\big)^2 \right)\Big\rvert_{y=y_{c}}= \alpha_3\;,
\label{eq:alpha beta 3}
\end{align}

and
\begin{align}
\alpha_{4}&=a^{IV}_{y}|_{-y_{c}}
=\frac{1}{8}\left( \psi_0'\,\psi_0^{IV} + 3 \,\psi_0''\,\psi_0''' \right)\Big|_{-y_{c}}
=\left( - \psi_0^{IV} + 3 \,\psi_0''\,\psi_0''' \right)\Big|_{-y_{c}}\,,\\
\beta_{4}&=b^{IV}_{y}(y_{c})
=\frac{1}{8}\left( \psi_0'\,\psi_0^{IV} + 3 \,\psi_0''\,\psi_0'''  \right)\Big|_{y_{c}} 
=\left( \psi_0^{IV} + 3 \,\psi_0''\,\psi_0'''  \right)\Big|_{y_{c}} = - \alpha_4\,.    
\label{eq:alpha beta 4}
\end{align}

For the Gaussian-shaped wiggles \eqref{eq:Gaussian wiggles}, 
$\psi_\text{wiggle}(z)=A \;e^{-z^2/2w^2}$, 
the  condition \eqref{eq near wiggles} reduces to 
\begin{equation}
    A \, z \, e^{-z^2/2w^2} =w^2\,,
\end{equation}
which has 2 roots for amplitude larger than the critical value
\begin{equation}
    A_{\rm threshold}=\sqrt e\,w~.
\end{equation}
Near (and above) $A_{\rm threshold}$, the roots appear like
$$
z=-w\pm w\sqrt{\frac{A-A_{\rm threshold}}{A_{\rm threshold}}}+O(A-A_{\rm threshold})
$$
and $\alpha_2$ scales as
$$
\alpha_2=\pm\frac{1}{w}\sqrt{\frac{A-A_{\rm threshold}}{A_{\rm threshold}}}+O(A-A_{\rm threshold})~. 
$$
Instead the higher coefficients are finite at $A_{\rm threshold}$, with $\alpha_3 = -1/2w^2$,  and $\alpha_4=-1/4w^3$.

Notice that this discussion can be easily extended to non-symmetric collisions, that is, with different left- and right-moving wiggles in the initial condition. In this case, the points on the wiggles that correspond to the singularity are not related by symmetry. Call them $y_{c_L}$ and $y_{c_R}$, which are understood to be located on/near each wiggle. The equations that define the swallowtail bifurcation cusp are \eqref{eq:ab prime} (or $\vec{X}'|_{(\sigma_*,t_*)}=\vec{0}$) and $\vec{X}''|_{(\sigma_*,t_*)}=\vec{0}$. These reduce to the following equations:
\begin{align}
    \psi_0'(y_{c_L}) \,\psi_0'(y_{c_R}) &=-1\,, \label{eq:sw general1}\\[2mm]
    \psi_0'(y_{c_R})\,\psi_0''(y_{c_L}) &= -\psi_0'(y_{c_L}) \,\psi_0''(y_{c_R})~,\label{eq:sw general2}
\end{align}
which for symmetric  collisions reduce to \eqref{eq:conditions for yc} because \eqref{eq:sw general2} becomes trivial. 

For non-identical wiggles, once their shapes are specified, Eqs. \eqref{eq:sw general1} and \eqref{eq:sw general2} determine a discrete set of solutions $(y_{c_L},y_{c_R})$, if any. As expected, the slope on one wiggle can be below $1$ provided the slope on the other one compensates enough. These correspond to the swallowtail cusps, that is where the  swallowtail bifurcation starts and ends. 
The points along the bifurcation (between the bifurcation cusps) then must obey $ - \psi_0'|_L \; \psi_0'|_R >1 $. The inflection points on the wiggles $\psi_0''|_{L,R}=0$ (denoted $y_{1,\,2,\,3,\,4}$ in Appendix \ref{appendix wiggles}) actually maximize (extremize) the product $\big|\psi_0'|_L \; \psi_0'|_R\big|$. Therefore, a sufficient condition to guarantee a swallowtail bifurcation can be simply expressed as $\big|\psi_0'|_L \; \psi_0'|_R\big|_{\rm inflection~points} >1$, see \eqref{eq:condition swallowtail formation 1}, \eqref{eq:condition swallowtail formation 2}. The expressions for the $\alpha_n$, $\beta_n$ coefficients, \eqref{eq:alpha 2} - \eqref{eq:alpha beta 4},  remain valid (before using symmetry and $|\psi_0'|=1$, and  with derivatives evaluated at $y_{c_{L,\,R}}$).

%%%%%%%%%%%%%%%%%%%%%%%%%%%%%%%%%%%%%%%%%%%
\section{Swallowtail bifurcations in the collapse of closed loops}
\label{appendix loops}
Here we will find the conditions for cusp formation (and, consequently, for the appearance of swallowtail singularities) for loops which are initially at rest.

Since $\vec{a}\,'\left(\sigma-t\right)$ and $-\vec{b}\,'\left(\sigma+t\right)$ are unit vectors, at a given time they are functions of $\sigma$ that can be represented as curves on the surface of a cylinder with unit radius. The basis of this cylinder is the plane defined by the $x$ and $y$ components of either $\vec{a}\,'\left(\sigma-t\right)$ or $-\vec{b}\,'\left(\sigma+t\right)$, while its longitudinal axis represents the value of $\sigma$. As time passes, the curve $\vec{a}\,'\left(\sigma-t\right)$ rises on the surface of the cylinder (i.e., it moves to higher values of $\sigma$) and $-\vec{b}\,'\left(\sigma+t\right)$ goes down. If the two curves intersect at some time, i.e., if $\vec{a}\,'\left(\sigma-t\right)=-\vec{b}\,'\left(\sigma+t\right)$, a cusp will form.

From (\ref{eq:velocity}) and (\ref{eq:r prime}) it follows that 
\begin{equation}
\vec{a}\,'\left(\sigma-t\right)=\vec{X}\,'\left(t,\sigma\right)-\dot{\vec{X}}\left(t,\sigma\right)
\label{eq:a prime}
\end{equation}
and 
\begin{equation}
\vec{b}\,'\left(\sigma-t\right)=\vec{X}\,'\left(t,\sigma\right)+\dot{\vec{X}}\left(t,\sigma\right)\,.
\label{eq:b prime}
\end{equation}
\\
Let $X\left(t,\sigma\right)$ and $Y\left(t,\sigma\right)$ be the $x$ and $y$ components of $\vec{X}\left(t,\sigma\right)$. Since the loop starts out at rest, at $t=0$ we have
\begin{equation}
a'_{x}\left(\sigma\right)=b'_{x}\left(\sigma\right)=\frac{\partial X}{\partial\sigma}\,\bigg\rvert_{t=0}\,,
\label{eq:ax prime and bx prime loops}
\end{equation}

\begin{equation}
a'_{y}\left(\sigma\right)=b'_{y}\left(\sigma\right)=\frac{\partial Y}{\partial\sigma}\,\bigg\rvert_{t=0}\,.
\label{eq:ay prime and by prime loops}
\end{equation}
\\
As we shall see in the following, in order to determine whether the swallowtail will appear or not, it suffices to draw these curves qualitatively on the surface of the unit-radius cylinder.

\subsection{Circular loop}
\begin{figure}[h!]
\begin{center}
\includegraphics[scale=0.9]{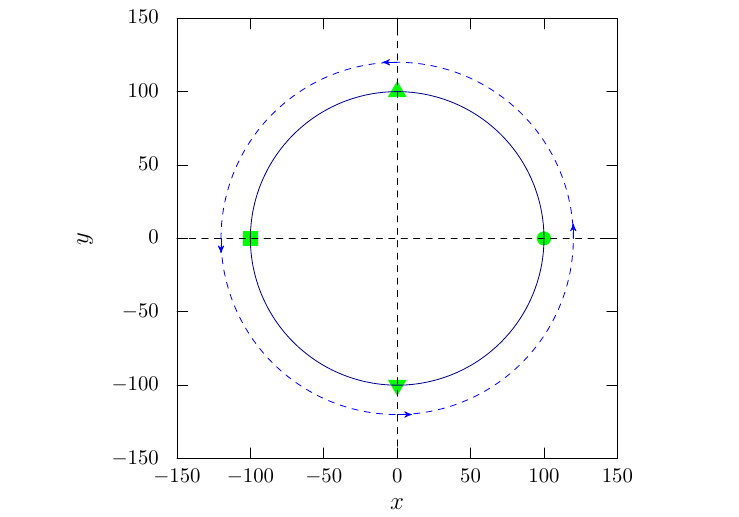}
\includegraphics[scale=1.3]{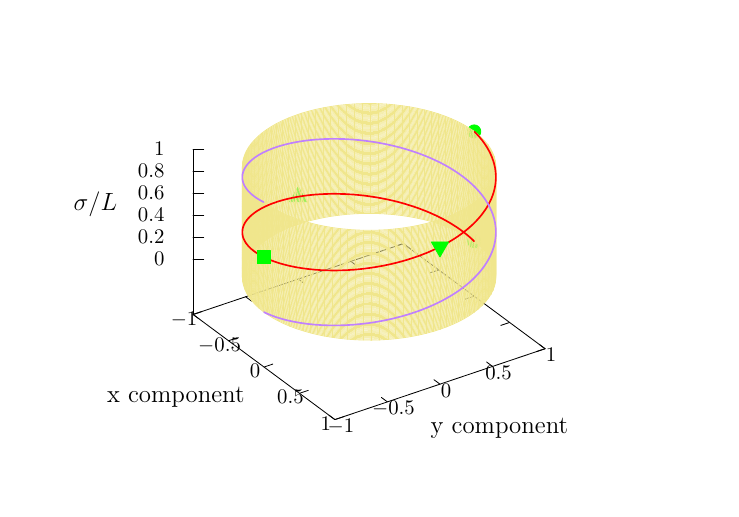}
\caption{Curves defined by $\vec{a}\,'\left(\sigma\right)$ (red) and $-\vec{b}\,'\left(\sigma\right)$ (purple) on the surface of the unit-radius cylinder for a circular loop.}
\label{fig:circle example}
\end{center}
\end{figure}

Although we know that the evolution of a circular loop that starts out at rest is only singular at the collapse time, let us consider this simple case in order to exemplify our procedure for the prediction of swallowtail bifurcations. The procedure is illustrated in Figure \ref{fig:circle example}. On the upper panel we have plotted a circular loop with radius $R=100$ in some units. On the bottom panel, we have drawn the curves $\vec{a}\,'\left(\sigma\right)$ (in red) and $-\vec{b}\,'\left(\sigma\right)$ (in purple) on the unit-radius cylinder as we go through the loop in the direction indicated by the blue arrow heads, taking the circle as the starting point: $\bigcirc\rightarrow\bigtriangleup\rightarrow\square\rightarrow\bigtriangledown\rightarrow\bigcirc$. The $\sigma$ parameter goes from 0 to $L$, which is the invariant length of the loop (the total amount of $\sigma$). Let us describe this trajectory in some more detail using (\ref{eq:ax prime and bx prime loops}) and (\ref{eq:ay prime and by prime loops}). 

Consider the first part of the trajectory, namely, $\bigcirc\rightarrow\bigtriangleup$. At the starting point, we have $\partial X/\partial\sigma=0$ and $\partial Y/\partial\sigma>0$. Therefore, $a'_{x}=0$ and $a'_{y}=1$ (green point at the base of the cylinder in the bottom panel of Figure \ref{fig:circle example}), and the components of $-\vec{b}\,'$ have opposite signs. All the way up to $\bigtriangleup$ we have $\partial X/\partial\sigma<0$ and $\partial Y/\partial\sigma>0$, so $a'_{x}<0$ and $a'_{y}>0$. This means that we have to climb up the cylinder counterclockwise until we get to $\bigtriangleup$, where $a'_{x}=\partial X/\partial\sigma<0$ and $a'_{y}=\partial Y/\partial\sigma=0$. We then make the same observations for the rest of the trajectory (see table \ref{table:signs circle}) to obtain the complete curves $\vec{a}\,'\left(\sigma\right)$ and $-\vec{b}\,'\left(\sigma\right)$ on the cylinder.

Now we just have to see how the red and purple curves intersect as they move in time (recall that the former moves upwards and the latter moves downwards). In this case, it is clear that the two curves will intersect only when they completely overlap. This corresponds to the moment of evolution at which the loop has collapsed to a point.
\begin{table}[h!]
\centering
\begin{tabular}{||c | c | c | c | c ||} 
 \hline
 & $a'_{x}=\partial X^{1}/\partial\sigma$ & $a'_{y}=\partial X^{2}/\partial\sigma$ & $-b'_{x}=-\partial X^{1}/\partial\sigma$ & $-b'_{y}=-\partial X^{2}/\partial\sigma$\\
 \hline\hline
$\bigcirc\rightarrow\bigtriangleup$ & - & + & + & -\\
\hline
$\bigtriangleup\rightarrow\square$ & - & - & + & +\\
\hline
$\square\rightarrow\bigtriangledown$ & + & - & - & +\\
\hline
$\bigtriangledown\rightarrow\bigcirc$ & + & + & - & -\\
\hline
 \end{tabular}
\caption{Signs of the components of $\vec{a}\,'\left(\sigma\right)$ and $-\vec{b}\,'\left(\sigma\right)$. This allows us to draw them qualitatively on the surface of the unit-radius cylinder (see Figure \ref{fig:circle example}).\\}
\label{table:signs circle}
\end{table}

%\vspace{6cm}

\subsection{Elliptical loop}

One can easily obtain the same conclusions for an elliptical loop. In particular, table \ref{table:signs circle} would be exactly the same, and the only difference in Figure \ref{fig:circle example} would be the slope of the curves, which depends on the eccentricity of the ellipse. Therefore, all the points on an elliptical loop that starts out at rest also reach the speed of light simultaneously. No swallowtail bifurcations are formed during its collapse.

As we will see in the next example, the feature we need in order for the swallowtail to appear is that the curves on the cylinder present turnarounds.

\subsection{Star-shaped loop}

\begin{figure}[h!]
\begin{center}
\includegraphics[scale=0.9]{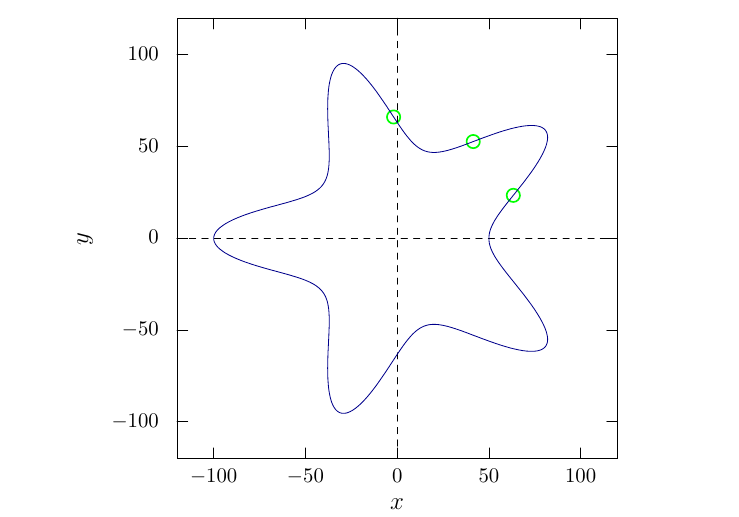}
\includegraphics[scale=1.3]{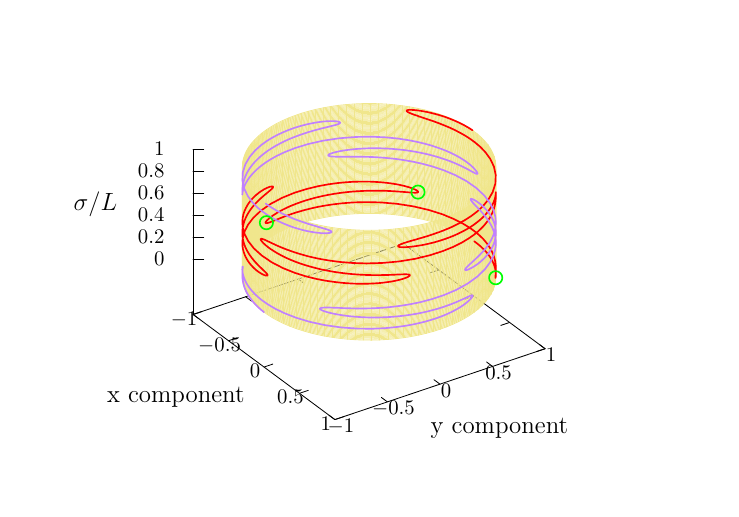}
\caption{Curves defined by $\vec{a}\,'\left(\sigma\right)$ (red) and $-\vec{b}\,'\left(\sigma\right)$ (purple) on the surface of the unit-radius cylinder for a star-shaped loop. The first three turnarounds are indicated in green.}
\label{fig:star example}
\end{center}
\end{figure}

\begin{figure}[h!]
\begin{center}
\includegraphics[scale=0.9]{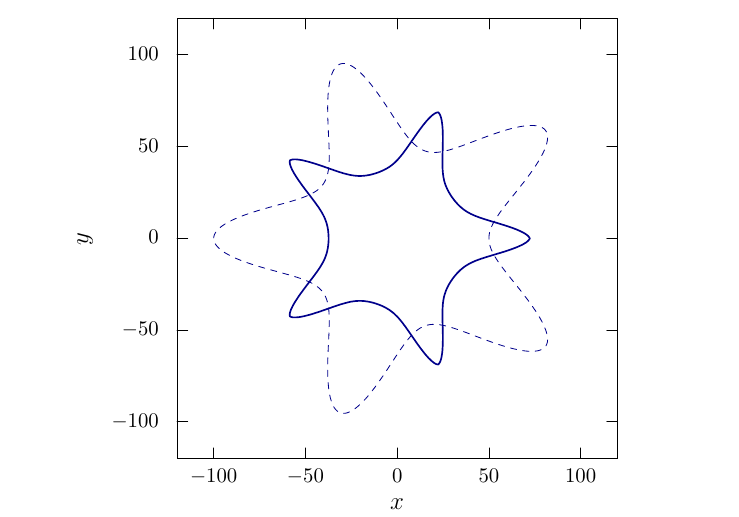}
\includegraphics[scale=1.3]{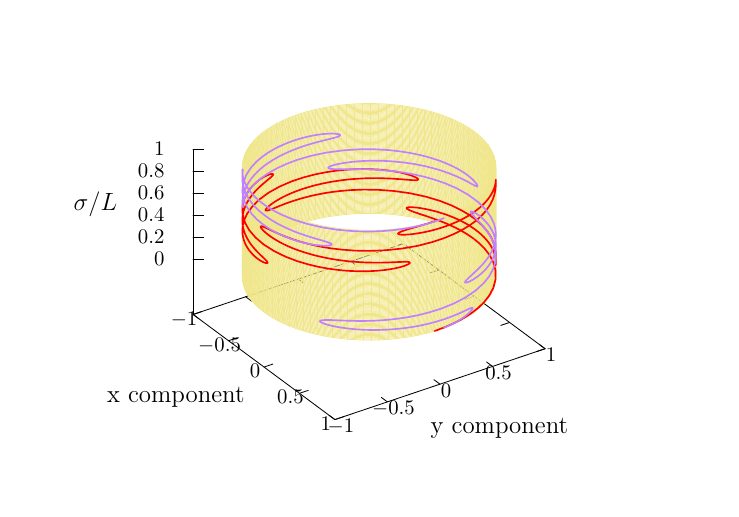}
\caption{Upper panel: star-shaped loop at $t=0$ (dashed line) and $t=78$ (solid line). Bottom panel: curves defined by $\vec{a}\,'\left(\sigma-t\right)$ (red) and $-\vec{b}\,'\left(\sigma-t\right)$ (purple) on the surface of the unit-radius cylinder at $t=78$. The curves intersect at five points, which are the tips of the solid loop in the upper panel. A couple of cusps is about to emerge from each of these tips.}
\label{fig:star example 2}
\end{center}
\end{figure}

\begin{figure}[h!]
\begin{center}
\includegraphics[scale=0.9]{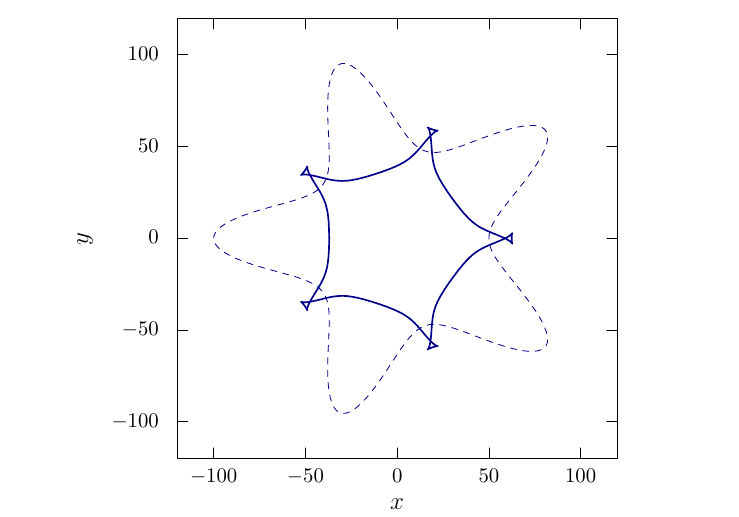}
\includegraphics[scale=1.3]{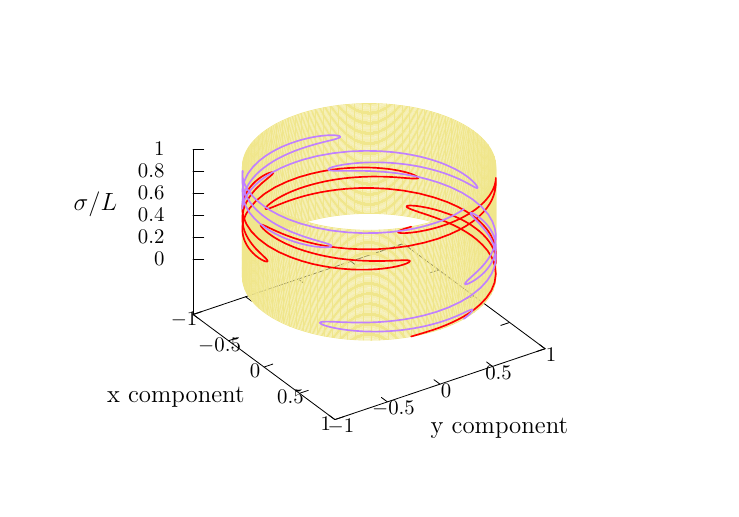}
\caption{Upper panel: star-shaped loop at $t=0$ (dashed line) and $t=88$ (solid line). Bottom panel: curves defined by $\vec{a}\,'\left(\sigma-t\right)$ (red) and $-\vec{b}\,'\left(\sigma+t\right)$ (purple) on the surface of the unit-radius cylinder at $t=88$. The curves intersect at ten points, which are easily identified as pairs of cusps in the solid loop in the upper panel. Each of these pairs constitutes a swallowtail.}
\label{fig:star example 3}
\end{center}
\end{figure}

An example of star-shaped loop can be obtained with the following radius as a function of the polar angle:
\begin{equation}
r\left(\theta\right)=\frac{A}{\left[\cos^{\alpha}\left(\frac{N\theta}{4}\right)+\sin^{\alpha}\left(\frac{N\theta}{4}\right)\right]^{1/k}}\,,
\label{eq:star equation}
\end{equation}
\\
where $A,\,\alpha,\,N$ and $k$ are parameters. For $A=50$, $\alpha=6$, $N=5$ and $k=2$, the analysis presented in the previous subsections yields what is shown in Figure \ref{fig:star example}. As one can check from this figure, turnarounds take place at the inflection points one finds along the loop, that is, whenever the center of curvature jumps from one side of the loop to the other. Clearly, this cannot happen in an elliptical loop. It is also clear from the bottom panel of Figure \ref{fig:star example} that the two curves will intersect at a few points before they completely overlap (the complete overlap corresponds to the collapse of the loop to a double line). We show in Figure \ref{fig:star example 2} the moment at which the two curves first intersect, and in Figure \ref{fig:star example 3} a later time when the curves intersect at ten points.

The conclusion of this appendix is the following: swallowtail bifurcations will appear in an initially static loop contained in a plane if the loop presents inflection points at $t=0$.

\newpage

\newpage

\newpage

\clearpage

%%%%%%%%%%%%%%%%%%%%%%%%%%%%%%%%%%%%%%%%%%%
\bibliography{biblio}
\bibliographystyle{JHEP.bst}

\end{document}